\documentstyle[aps]{revtex}

\begin{document}
\title{Renormalization of the $\sigma -\omega $ model within the framework of U(1)
gauge symmetry }
\author{Jun-Chen Su$^{1,2}$ and Hai-Jun Wang$^2$}
\address{1. Department of Physics, Harbin Institute of Technology , Harbin 150006,\\
People's Republic of China\\
2. Center for Theoretical Physics, College of Physics, Jilin University,\\
Changchun 130023, People's Republic of China}
\maketitle

\begin{abstract}
It is shown that the $\sigma -\omega $ model which is widely used in the
study of nuclear relativistic many-body problem can exactly be treated as an
Abelian massive gauge field theory. The quantization of this theory can
perfectly be performed by means of the general methods described in the
quantum gauge field theory. Especially, the local U(1) gauge symmetry of the
theory leads to a series of Ward-Takahashi identities satisfied by Green's
functions and proper vertices. These identities form an uniquely correct
basis for the renormalization of the theory. The renormalization is carried
out in the mass-dependent momentum space subtraction scheme and by the
renormalization group approach. With the aid of the renormalization boundary
conditions, the solutions to the renormalization group equations are given
in definite expressions without any ambiguity and renormalized S-matrix
elememts are exactly formulated in forms as given in a series of tree
diagrams provided that the physical parameters are replaced by the running
ones. As an illustration of the renormalization procedure, the one-loop
renormalization is concretely carried out and the results are given in
rigorous forms which are suitable in the whole energy region. The effect of
the one-loop renormalization is examined by the two-nucleon elastic
scattering.

\end{abstract}

\section{Introduction}

The quantum hadrodynamics (QHD), as a relativistic quantum field theory for
baryons and mesons, has been widely applied to studying various nuclear
phenomena including the hadron-hadron interaction, the hadron-nucleus
scattering, the bulk and single-particle properties of nuclei, etc. [1-5].
It is commonly recognized that although the quantum chromodynamics is a
fundamental theory for strong interaction, the QHD, as an effective field
theory formulated in terms of hadronic degrees of freedom, provides a simple
and reliable approach to produce the nuclear observables that are
insensitive to the short-range dynamics. There are various QHD models,
renormalizable and nonrenormalizable, which were tested in the past to
reproduce the empirical nuclear properties and the experimental data. Among
these models, the $\sigma -\omega $ model proposed by Walecka [1] has been
raising particular interest. This model contains proton, neutron and
isoscalar, Lorentz scalar and vector mesons $\sigma $ and $\omega $ and in
the tree diagram and nonrelativistic approximations leads to a
nucleon-nucleon interaction potential which behaves as short-range repulsion
and medium-range attraction. The early development of this model is based on
the relativistic mean-field and Hartree approximation and shows that the
model is quite successful in the applications to the infinite nuclear matter
and atomic nuclei. Since the model is renormalizable, it is necessary to
consider higher order perturbative corrections to the results given in the
mean field approximation by a certain renormalization procedure. Along this
line, a number of efforts were made previously [6-19]. Especially, the
efforts were mostly concentrated on the renormalization of the model in the
study of the nuclear matter at finite temperature and density. In this
renormalization, the loop expansion and spectral function methods were
applied to evaluate the loop corrections. However, there are various
difficulties to occur in the renormalization [8-16]. For example, in Ref.
[12], the authors calculated the nuclear matter energy density up to the two
loop level and found enormous contributions arising from the loop terms that
alter the description of the nuclear bound state qualitatively. Therefore,
it was concluded that ''the loop expansion does not provide a reliable
approximation scheme in renormalizable QHD'' [19]. To this end, one may ask
what is the correct procedure of performing the renormalization for a model
of QHD? and how to assess the applicability of a renormalizable model of QHD
for which the renormalization is carried out? To answer these questions, it
is meaningful to examine the renormalization of a QHD model from different
angles and, as suggested in Ref. [19], ''to develop and apply systematic and
consistent ''power counting'' schemes that lead to more general conserving
approximations and to study renormalization group methods that could
determine the analytic structure of the ground-state energy functional''.

In this paper, we confine ourself to discussing the renormalization of the $%
\sigma -\omega $ model by the renormalization group method in the case of
zero temperature. The procedure is very similar to that described in our
previous work on the QED and QCD renormalizations [20]. The main features of
the renormalization given in this paper contain two aspects: (1) the
renormalization is based on the U(1) gauge symmetry because the $\sigma
-\omega $ model, as argued in the next section, is exactly of the U(1) local
gauge symmetry; (2) The renormalization is carried out by a mass-dependent
momentum space subtraction [21-24] which will lead to rigorous renormalized
results by the renormalization group method [25--28]. Ordinarily, the
massive vector fields such as the $\omega $ meson field, the $\rho $ meson
field and so on are not viewed as gauge fields because the mass term in the
Lagrangian is not gauge-invariant [29-31]. On the contrary, it was pointed
out in Refs. [32-34] that a massive vector field must be viewed as a
constrained system in the whole space of the vector potential $A_\mu (x)$.
This is because a massive vector meson has only three polarization states
which need only three spatial components of the vector potential $A_\mu (x)$
to describe them. While, the remaining component of the $A_\mu (x)$ appears
to be a redundant degree of freedom which must be eliminated by introducing
the Lorentz condition. According to the general principle for constrained
systems, the gauge-invariance of a massive Abelian or non-Abelian gauge
field should be seen from its action given in the physical space defined by
the Lorentz condition. This viewpoint will be explained in more detail in
the next section. From this viewpoint, it is easy to see that the $\sigma
-\omega $ model is surely of U(1) local gauge symmetry. Therefore, the model
may be quantized by the method as used in the gauge field theory. In this
paper, we will describe the Lorentz-covariant quantization performed in the
both of Hamiltonian and Lagrangian path-integral formalisms by following the
procedure proposed in Refs. [32--34]. From this quantization, we obtain an
effective action which contains a gauge-fixing term and a ghost term in it
and manifests itself to be invariant under a set of BRST transformations
[35]. It should be mentioned that the quantum theory of the $\sigma -\omega $
model was set up previously by the method of canonical quantization and in
the path-integral formalism [29-31, 36, 37 ]. Especially, with the time
paths being generalized to a manifestly covariant form, a covariant
path-integral formulation for the model at finite temperature was achieved
in Ref. [37] and led to manifestly covariant Feynman rules for both real and
imaginary times. Nevertheless, owing to lack of the gauge-fixing term and
ghost term in the effective action, the generating functional given in these
quantizations would not exhibit the BRST-symmetry.

As emphasized in Ref. [20], a correct renormalization procedure for a gauge
field theory must respects the gauge-symmetry (the Ward-Takahashi identities
[38, 39]), the Lorentz-invariance (the energy-momentum conservation) and the
mathematical convergence principles. Otherwise, the renormalization would be
incorrect. From the gauge-invariance (or say, the BRST-symmetry) of the $%
\sigma -\omega $ model, we derive a set of Ward-Takahashi (W-T) identities
satisfied by the generating functionals, Green's functions and vertices
which provide a firm basis for the renormalization of the model. As
mentioned before, in this paper, the renormalization of the $\sigma -\omega $
model will be performed in the mass-dependent momentum space subtraction.
The prominent advantage of such a subtraction is that it naturally provides
boundary conditions satisfied by the renormalied wave functions, propagators
and proper vertices for the quantum $\sigma -\omega $ model. These boundary
conditions enable us to uniquely determine the solutions to the
renormalization group equations for those renormalized quantities. With the
solutions of the renormalization group equations, a $S$-matrix element can
be expressed in the form as given in the tree diagrams provided that the
physical parameters in the $S$-matrix element are replaced by the effective
(running) ones. To specify the procedure of the renormalization group
method, the one-loop effective physical parameters are concretely calculated
and given exact and analytical expressions.

The remainder of this paper is arranged as follows. In Sec. II, we present
arguments for the gauge-invariance of the $\sigma -\omega $ model. In Sec.
III, the $\sigma -\omega $ model will be respectively quantized in the
Hamiltonian and Lagrangian path-integral formalisms. In Sec. IV, we will
derive a set of W-T identities obeyed by the generating functionals. In Sec.
V, a W-T identity satisfied by the $\omega $ meson propagator will be
derived and the renormalization of the propagator will be discussed. In Sec.
VI, we will derive a W-T identity satisfied by the vectorial vertex
(nucleon--nucleon-$\omega $ meson vertex) and discuss the renormalizations
of the vertex and the nucleon propagator. In Sec. VII, the renormalizations
of the $\sigma $ meson propagator and the scalar coupling vertex
(nucleon--nucleon-$\sigma $ meson vertex) will be derived and discussed
.Sec. VIII is used to sketch the renormalization group method and the
renormalized S-matrix elements. Sec. IX serves to derive the one-loop
effective coupling constants and masses. In the last section, summary and
discussions will be made. In Appendix A, the gauge-independence of the
S-matrix elements given in the one-loop level will be proved. In Appendix B,
we will show the differential cross section of the two-nucleon elastic
scattering in the approximation of order $g^2$ and examine the effect of the
one-loop renormalization on it..

\section{Argument of gauge-invariance for the $\sigma -\omega $ model}

The $\sigma -\omega $ model is described by the following Lagrangian density
[1] 
\begin{equation}
{\cal L}=\overline{\psi }(i\gamma ^\mu D_\mu -M)\psi -\frac 14F^{\mu \nu
}F_{\mu \nu }+\frac 12m_\omega ^2A^\mu A_\mu +\frac 12\partial ^\mu \varphi
\partial _\mu \varphi -\frac 12m_\sigma ^2\varphi ^2  \eqnum{2.1}
\end{equation}
where 
\begin{equation}
\psi =\left( 
\begin{array}{c}
\psi _p \\ 
\psi _n
\end{array}
\right)  \eqnum{2.2}
\end{equation}
is the nucleon isospin doublet in which $\psi _p$ and $\psi _n$ are the
proton and neutron field functions respectively, 
\begin{equation}
D_\mu =\partial _\mu -ig_vA_\mu -\frac i4g_s\gamma _\mu \varphi  \eqnum{2.3}
\end{equation}
is the covariant derivative in which $A_\mu $ and $\varphi $ stand for the $%
\omega $ and $\sigma $ meson fields, $g_v$ and $g_s$ designate the vectorial
and scalar coupling constants, 
\begin{equation}
F_{\mu \nu }=\partial _\mu A_\nu -\partial _\nu A_\mu  \eqnum{2.4}
\end{equation}
is the vector field strength and $M$, $m_\omega $ and $m_\sigma $ are the
masses of nucleon, $\omega $ meson and $\sigma $ meson respectively. In the
above Lagrangian, the scalar self-couplings are ignored as was done
originally in the Walecka model [1].

In the previous, the $\sigma -\omega $ model was considered to be
gauge-non-invariant with respect to the following local U(1) gauge
transformations [29-31] 
\begin{equation}
\begin{array}{c}
\psi ^{\prime }(x)=e^{ig_v\theta (x)}\psi (x), \\ 
\overline{\psi }^{\prime }(x)=e^{-ig_v\theta (x)}\overline{\psi }(x), \\ 
A_\mu ^{\prime }(x)=A_\mu (x)+\partial _\mu \theta (x), \\ 
\varphi ^{\prime }(x)=\varphi (x)
\end{array}
\eqnum{2.5}
\end{equation}
where $\theta (x)$ is the scalar parametric function of U(1) group since the
mass term of the $\omega $ meson in the Lagrangian is not gauge-invariant.
But, this does not mean that the dynamics of the $\omega $ meson system is
not gauge-invariant. As mentioned in the Introduction, the $\omega $ meson
field must be viewed as a constrained system in the space spanned by the
four-dimensional vector potential $A_\mu (x)$. As we know, a massive gauge
field has three polarization states which need only three spatial components
of the four-dimensional vector potential $A_\mu $ to describe them. In the
Lorentz-covariant formulation, a full vector potential $A^\mu (x)$ can be
split into two Lorentz-covariant parts: the transverse vector potential $%
A_T^\mu (x)$ and the longitudinal vector potential $A_L^\mu (x)$, 
\begin{equation}
A^\mu (x)=A_T^\mu (x)+A_L^\mu (x)  \eqnum{2.6}
\end{equation}
where 
\begin{equation}
A_T^\mu (x)=(g^{\mu \nu }-\frac 1{\Box }\partial ^\mu \partial ^\nu )A_\nu
(x),  \eqnum{2.7}
\end{equation}
\begin{equation}
A_L^\mu (x)=\frac 1{\Box }\partial ^\mu \partial ^\nu A_\nu (x)  \eqnum{2.8}
\end{equation}
with $\Box =\partial ^\mu \partial _\mu $ being the D'Alembertian operator.
The vector potentials $A_T^\mu (x)$ and $A_L^\mu (x)$ satisfy the following
transverse and longitudinal field conditions (identities): 
\begin{equation}
\partial _\mu A_T^\mu (x)=0,  \eqnum{2.9}
\end{equation}
\begin{equation}
(g_{\mu \nu }-\frac 1{\Box }\partial _\mu \partial _\nu )A_L^\nu (x)=0 
\eqnum{2.10}
\end{equation}
and the orthogonality relation 
\begin{equation}
\int d^4xA_T^\mu (x)A_{L\mu }(x)=0  \eqnum{2.11}
\end{equation}
which characterizes the linear independence of the two field variables.
Since the Lorentz-covariant transverse vector potential $A_T^\mu (x)$
contains three-independent spatial components, it is sufficient to represent
the polarization states of a massive vector boson. Whereas, the
Lorentz-covariant longitudinal vector potential $A_L^\mu $ appears to be a
redundant unphysical variable which must be constrained by introducing the
Lorentz condition: 
\begin{equation}
\chi \equiv \partial ^\mu A_\mu =0  \eqnum{2.12}
\end{equation}
whose solution is 
\begin{equation}
A_L^\mu =0.  \eqnum{2.13}
\end{equation}
With this solution, the $\sigma -\omega $ model Lagrangian may be expressed
in terms of the independent dynamical variables $A_T^\mu (x),$%
\begin{equation}
\begin{tabular}{l}
${\cal L}=\overline{\psi }[\gamma ^\mu (i\partial _\mu +g_vA_{T\mu }+\frac 14%
g_s\gamma _\mu \varphi )-M]\psi -\frac 14F_T^{\mu \nu }F_{T\mu \nu }+\frac 12%
m_\omega ^2A_T^\mu A_{T\mu }$ \\ 
$+\frac 12\partial ^\mu \varphi \partial _\mu \varphi -\frac 12m_\sigma
^2\varphi ^2$%
\end{tabular}
\eqnum{2.14}
\end{equation}
where $F_T^{\mu \nu }$ is defined as in Eq. (2.4) with replacing the $A^\mu
(x)$ by $A_T^\mu (x).$ The Lagrangian represented above gives a complete
description of the dynamics of the $\sigma -\omega $ model. If we want to
represent the dynamics in the whole space of the full vector potential as
described by the Lagrangian in Eq. (2.1), the $\omega $ field must be
treated as a constrained system. In this case, according to the general
procedure for constrained systems as formulated in Mechanics, the Lorentz
condition in Eq. (2.12), as a constraint, must be introduced from the onset
and imposed on the Lagrangian in Eq. (2.1) so as to guarantee the redundant
degree of freedom to be eliminated from the Lagrangian. Otherwise, the
Lagrangian in Eq. (2.1) itself can not give a complete description for the $%
\omega $ field system. From the Lagrangian in Eq. (2.14), one may derive an
equation of motion satisfied by the $\omega $ meson field as follows: 
\begin{equation}
\partial _\mu F_T^{\mu \nu }+m_\omega ^2A_T^\nu =-j^\nu  \eqnum{2.15}
\end{equation}
where 
\begin{equation}
j^\nu =g_v\overline{\psi }\gamma ^\nu \psi  \eqnum{2.16}
\end{equation}
is the current generated from the nucleon field. The above equation
describes the evolution of the independent variable $A_T^\mu $ with time. In
particular, when we take divergence of the both sides of Eq. (2.15),
considering the identities in Eq. (2.9) and $\partial _\nu \partial _\mu
F_T^{\mu \nu }\equiv 0$, we immediately obtain the current conservation 
\begin{equation}
\partial ^\mu j_\mu =0  \eqnum{2.17}
\end{equation}
which shows that the current is transverse.

Ordinarily, the Lorentz condition is viewed as a consequence of the
following $\omega $ field equation of motion which is derived from the
Lagrangian in Eq. (2.1) [29-30, 36] 
\begin{equation}
\partial _\mu F^{\mu \nu }+m_\omega ^2A^\nu =-j^\nu  \eqnum{2.18}
\end{equation}
The argument of this viewpoint is as follows. When we take divergence of the
equation (2.18) and notice the current conservation, it is found that 
\begin{equation}
m_\omega ^2\partial ^\mu A_\mu =0  \eqnum{2.19}
\end{equation}
Since $m_\omega \neq 0$, the above equation leads to the Lorentz condition
which implies that one component of the vector potential is not independent.
It is pointed out here that the above viewpoint actually is an ill-concept
and the procedure leading to the Lorentz condition logically is not
consistent with the principle established well in the mechanics for
constrained systems. In fact, the aforementioned derivation seems to imply
that the Lorentz condition has already been included in the Lagrangian
denoted in Eq. (2.1). If so, when the Lagrangian is written in the first
order form, we should see a term in the Lagrangian which is given by
incorporating the Lorentz condition with the aid of the Lagrange multiplier
method. Nevertheless, as will be shown in the next section, there is no such
a term to appear in the Lagrangian. Moreover, as we know, equations of
motion should describe the evolution of the independent variables with time
as the equation given in Eq. (2.15) does and should not lead to a constraint
condition which implies some variable in the equation is not independent.
Therefore, the viewpoint stated above is not reasonable. In accordance with
the general principle for constrained systems, the correct procedure is to
treat the Lorentz condition as a primary constraint and to impose this
condition on the Lagrangian in Eq. (2.1) from the beginning. The necessity
of introducing the Lorentz condition can also be seen from the derivation
mentioned in Eqs. (2.18) and (2.19). The equation (2.19) can be understood
in such a way that if the Lorentz condition is not introduced, there would
appear a contradiction that the right hand side of the equation is zero, but
the left hand side is not. Only when the Lorentz condition is introduced,
the contradiction disappears. In this case, due to the Lorentz condition,
the equation (2.19), as a trivial identity, naturally holds and the equation
of motion (2.18) can naturally go over to the equation (2.15), exhibiting
the self-consistency of the theory. Particularly, in the latter case, when
the divergence of the equation (2.18) is taken and the Lorentz condition is
employed, one immediately obtains the current conservation in Eq. (2.17). In
addition, we would like to note that for the quantum theory, in the
zero-mass limit: $m_\omega \rightarrow 0$, the vector field part of the
Lagrangian in Eq. (2.1) naturally goes over to the one for the massless
vector meson, but, as shown in Sec. V, the vector meson propagator does not
and occurs a worse singularity, revealing a severe inconsistence of the
theory. Only when the Lorentz condition is introduced initially and
incorporated into the Lagrangian by the Lagrange multiplier method, a
consistent quantum theory can be constructed.

Now, let us turn to address the gauge-invariance of the $\sigma -\omega $
model. Usually, the gauge-invariance is required to the Lagrangian. From the
dynamical viewpoint, as pointed out in Refs. [32-34], the action is of more
essential significance than the Lagrangian. This is why in Mechanics and
Field Theory, to investigate the dynamical and symmetric properties of a
system, one always starts from the action of the system. Similarly, when we
examine the gauge-symmetric property of a field system, in more general, we
should also see whether the action for the system is gauge-invariant or not.
In particular, for a constrained system such as the massive vector field, we
should see whether or not the action represented in terms of the independent
dynamical variables is gauge-invariant. This point of view is easy to
understand from the mechanics for constrained systems. Suppose a mechanical
system is described by a Hamiltonian

\begin{equation}
H(p_i,q_i)(i=1,2,\cdots ,n)  \eqnum{2.20}
\end{equation}
which is given in the 2n-dimensional phase space and constraint conditions 
\begin{equation}
\varphi _a(p_i,q_i)=0(\alpha =1,2,\cdots ,2m<2n)  \eqnum{2.21}
\end{equation}
which define a physical phase space of dimension $2(n-m)$ where the system
exists and moves only. If the constrained variables can be solved out from
the constraint conditions, we may write a Hamiltonian

\begin{equation}
H^{*}(p_j^{*},q_j^{*})(j=1,2,\cdots ,n-m)  \eqnum{2.22}
\end{equation}
which is expressed via the independent variables and gives a complete
formulation of the constrained system. Obviously, to examine some symmetry
of the constrained system, it is only necessary to see if the Hamiltonian $%
H^{*}(p_j^{*},q_j^{*})$ other than the Hamiltonian $H(p_i,q_i)$ to have the
desired symmetry because in contrast to the $H^{*}(p_j^{*},q_j^{*})$, the $%
H(p_i,q_i)$ is not complete for describing the system.

Certainly, in some special cases, the Lagrangian given in the physical space
itself is locally gauge-invariant so that the gauge-invariance of the
corresponding action is ensured. This situation happens for the massless
gauge fields and the massive Abelian gauge field. The gauge transformation
of an Abelian gauge field was shown in the third equality in Eq. (2.5).
Since $\partial _\mu \theta (x)$ acts as a longitudinal field, according to
the decomposition denoted in Eq. (2.6) and considering the independence of
the fields $A_T^\mu (x)$ and $A_L^\mu (x)$, the gauge transformation of the $%
\omega $ field can be equivalently divided into two transformations: 
\begin{equation}
A^{\prime }{}_T^\mu (x)=A_T^\mu (x)  \eqnum{2.23}
\end{equation}
\begin{equation}
A^{\prime }{}_L^\mu (x)=A_L^\mu (x)+\partial ^\mu \theta (x)  \eqnum{2.24}
\end{equation}
Eqs. (2.23) and (2.24) clearly express the fact that the gauge
transformation only changes the unphysical longitudinal part of the vector
potential, while, the physical transverse vector potential is a
gauge-invariant quantity. Furthermore, it is easy to verify that the
longitudinal vector potential $A_L^\mu (x)$, which may be expressed as $%
A_L^\mu (x)=\partial ^\mu \varphi (x)$ where $\varphi (x)$ is a scalar
function, is cancelled in the field strength tensor so that 
\begin{equation}
F^{\mu \nu }=\partial ^\mu A^\nu -\partial ^\nu A^\mu =\partial ^\mu A_T^\nu
-\partial ^\nu A_T^\mu =F_T^{\mu \nu }  \eqnum{2.25}
\end{equation}
This indicates that the longitudinal part of the vector potential has no
kinetic energy term in the Lagrangian and hence has no any dynamical
meaning. Such a vector potential can only be viewed as a constrained
variable. Since the transverse field variable $A_T^\mu $ is gauge-invariant,
the Lagrangian (2.14) which is written in the physical space is manifestly
gauge-invariant. Therefore, the action given by this Lagrangian is
gauge-invariant. Alternatively, the gauge-invariance may also be seen from
the action given by the Lagrangian in Eq. (2.1) which is now constrained by
the Lorentz condition. Under the gauge transformation written in Eq. (2.5)
and the Lorentz condition denoted in Eq. (2.12), it is easy to find that 
\begin{equation}
\delta S=-m_\omega ^2\int d^4x\theta \partial ^\mu A_\mu =0  \eqnum{2.26}
\end{equation}
This indicates that the $\sigma -\omega $ model can surely be set up on the
basis of gauge-invariance principle.

\section{Path-integral quantization of the $\sigma -\omega $ model}

\subsection{Quantization in the Hamiltonian path-integral formalism}

According to the general procedure of dealing with constrained systems, the
Lorentz condition (2.12) may be incorporated into the Lagrangian (2.1) by
the Lagrange undetermined multiplier method to give a generalized Lagrangian
[32, 40]. In the first order formalism [32, 40, 41] , this Lagrangian can be
written as 
\begin{equation}
\begin{tabular}{l}
${\cal L}_\lambda =\overline{\psi }(i\gamma ^\mu D_\mu -M)\psi +\frac 12%
\partial ^\mu \varphi \partial _\mu \varphi -\frac 12m_\sigma ^2\varphi ^2+%
\frac 14F^{\mu \nu }F_{\mu \nu }$ \\ 
$-\frac 12F^{\mu \nu }(\partial _\mu A_\nu -\partial _\nu A_\mu )+\frac 12%
m_\omega ^2A^\mu A_\mu +\lambda \partial ^\mu A_\mu $%
\end{tabular}
\eqnum{3.1}
\end{equation}
where $A_\mu $ and $F_{\mu \nu }$ are now treated as the mutually
independent variables and $\lambda $ is chosen to represent the Lagrange
multiplier. Using the canonically conjugate variables defined by 
\begin{equation}
\Pi _\psi =\frac{\partial {\cal L}}{\partial \dot \psi }=i\overline{\psi }%
\gamma ^0,  \eqnum{3.2}
\end{equation}
\begin{equation}
\Pi _{\overline{\psi }}=\frac{\partial {\cal L}}{\partial \stackrel{.}{%
\overline{\psi }}}=0,  \eqnum{3.3}
\end{equation}
\begin{equation}
\Pi _\varphi =\frac{\partial {\cal L}}{\partial \dot \varphi }=\dot \varphi 
\eqnum{3.4}
\end{equation}
and 
\begin{equation}
\Pi _\mu (x)=\frac{\partial {\cal L}}{\partial \dot A^\mu }=F_{\mu
0}+\lambda \delta _{\mu 0}={\cal \{}
\begin{tabular}{l}
$F_{k0}=E_k,$ if $\mu =k=1,2,3;$ \\ 
$\lambda =-E_0,$ if $\mu =0,$%
\end{tabular}
\eqnum{3.5}
\end{equation}
the Lagrangian in Eq. (3.1) may be rewritten in the canonical form 
\begin{equation}
{\cal L}=E^\mu \dot A_\mu +\Pi _\psi \dot \psi +\Pi _\varphi \dot \varphi
+A_0C-E_0\chi -{\cal H}  \eqnum{3.6}
\end{equation}
where $E_\mu =(E_0,E_k)$ is a Lorentz vector, 
\begin{equation}
C\equiv \partial ^\mu E_\mu +m^2A_0+g_\omega \overline{\psi }\gamma ^0\psi ,
\eqnum{3.7}
\end{equation}
$\chi $ was defined in (2.12) and ${\cal H}$ is the Hamiltonian density
expressed by 
\begin{equation}
\begin{tabular}{l}
${\cal H}=\frac 12(E_k)^2+\frac 14(F_{ij})^2+\frac 12m_\omega
^2[(A_0)^2+(A_k)^2]+\frac 12[\Pi _\varphi ^2+(\nabla \varphi )^2$ \\ 
$+m_\sigma ^2\varphi ^2]-i\overline{\psi }\overrightarrow{\gamma }\cdot
\nabla \psi +M\overline{\psi }\psi -g_v\overline{\psi }\gamma ^k\psi A_k-g_s%
\overline{\psi }\psi \varphi $%
\end{tabular}
\eqnum{3.8}
\end{equation}
in which $F_{ij}$ was defined in Eq. (2.4). In the above, the
four-dimensional and the spatial indices are respectively denoted by the
Greek and Latin letters. Eq. (3.6) clearly shows that the terms $A_0C$ and $%
E_0\chi $ are respectively given by incorporating the constraint condition 
\begin{equation}
C=0  \eqnum{3.9}
\end{equation}
and the Lorentz condition into the Lagrangian by the Lagrange multiplier
method and the Lagrange multipliers $A_0$ and $E_0$ are just the constrained
variables themselves in this case. Since the $A_0$ and $E_0$ are a pair of
the canonically conjugate unphysical variables, their constraint conditions
in Eqs. (2.12) and (3.9) should simultaneously occur in the Lagrangian
(3.6). Otherwise, if the Lorentz condition is not introduced, the term $%
E_0\chi $ does not appear in the Lagrangian shown in Eq. (3.6) or in Eq.
(2.1). In this case, the Lagrangian could not be complete for describing the
constrained system under consideration.

From the stationary condition of the action constructed by the Lagrangian
(3.6), one may derive the following first-order canonical equations of
motion: 
\begin{equation}
\dot A_k=\partial _kA_0-E_k,  \eqnum{3.10}
\end{equation}
\begin{equation}
\dot E_k=\partial ^iF_{ik}+m_\omega ^2A_k+\partial _kE_0+g_v\overline{\psi }%
\gamma _k\psi ,  \eqnum{3.11}
\end{equation}
\begin{equation}
\dot \varphi =\Pi _\varphi ,  \eqnum{3.12}
\end{equation}
\begin{equation}
\dot \Pi _\varphi =\nabla ^2\varphi -m_\sigma ^2\varphi +g_s\overline{\psi }%
\psi ,  \eqnum{3.13}
\end{equation}
\begin{equation}
(i\gamma ^\mu \partial _\mu -M+g_v\gamma ^\mu A_\mu +g_s\varphi )\psi =0, 
\eqnum{3.14}
\end{equation}
\begin{equation}
\overline{\psi }(i\gamma ^\mu \overleftarrow{\partial }_\mu +M-g_v\gamma
^\mu A_\mu -g_s\varphi )=0  \eqnum{3.15}
\end{equation}
as well as the constraint equations written in Eqs. (2.12) and (3.9). Eqs
(3.10) and (3.11) act as the equations of motion satisfied by the
independent canonical variables $A_k$ and $E_k(k=1,2,3)$ which precisely
describe the three degrees of freedom of polarization for the massive $%
\omega $ field, while, Eqs. (2.12) and (3.9) can only be regarded as the
constraint equations obeyed by the constrained variables $A_0$ and $E_0$
because in these equations, there are no time-derivatives of the dynamical
variables $A_k$ and $E_k$. It is clear to see that in Eqs. (3.10)-(3.15),
(2.12) and (3.9), there are altogether twelve equations. They are sufficient
to determine the twelve variables including the dynamical canonical
variables $\psi $, $\overline{\psi }$, $\Pi _\varphi $, $\varphi $, $A_k$
and $E_k(k=1,2.3)$ and one pair constrained variables $A_0$ and $E_0$,
showing the completeness of the equations.

Now, we turn to formulate the quantization performed in the Hamiltonian
path-integral formalism for the $\sigma -\omega $ model. In accordance with
the general procedure of the quantization, we should first write a
generating functional of Green's functions in terms of the independent
canonical variables which are $\psi $, $\overline{\psi }$, $\Pi _\varphi $, $%
\varphi $ and the transverse parts of the vectors $A_\mu $ and $E_\mu $ for
the $\omega $ meson field [32, 40, 41] 
\begin{equation}
\begin{array}{c}
Z[J^\mu ,J,\overline{\eta },\eta ]=\frac 1N\int D(A_T^\mu ,E_T^\mu ,\psi ,%
\overline{\psi },\Pi _\varphi ,\varphi )exp\{i\int d^4x[E_T^\mu \dot A_{T\mu
}+\Pi _\psi \dot \psi \\ 
+\Pi _\varphi \dot \varphi -{\cal H}^{*}(A_T^\mu ,E_T^\mu ,\psi ,\overline{%
\psi },\Pi _\varphi ,\varphi )+J_T^\mu A_{T\mu }+J\varphi +\overline{\eta }%
\psi +\overline{\psi }\eta ]\}
\end{array}
\eqnum{3.16}
\end{equation}
where ${\cal H}^{*}(A_T^\mu ,E_T^\mu ,\psi ,\overline{\psi },\Pi _\varphi
,\varphi )$ is the Hamiltonian which is obtained from the Hamiltonian (3.8)
by replacing the constrained variables $A_L^\mu $ and $E_L^\mu $ with the
solutions of the equations (2.12) and (3.9) 
\begin{equation}
{\cal H}^{*}(A_T^\mu ,E_T^\mu ,\cdot \cdot \cdot )={\cal H}(A^\mu ,E^\mu
,\cdot \cdot \cdot )\mid _{\chi =0,C=0}  \eqnum{3.17}
\end{equation}
and $J_\mu ,J,\eta $ and $\overline{\eta }$ are the external sources coupled
to the $\omega $ meson, $\sigma $ meson and nucleon fields respectively. As
mentioned before, Eq. (2.12) leads to $A_L^\mu =0$. Noticing this solution
and the decomposition 
\begin{equation}
E^\mu (x)=E_T^\mu (x)+E_L^\mu (x),  \eqnum{3.18}
\end{equation}
when setting 
\begin{equation}
E_L^\mu (x)=\partial _x^\mu Q(x)  \eqnum{3.19}
\end{equation}
where $Q(x)$ is a scalar function, one may get from Eq. (3.9) an equation
obeyed by the scalar function $Q(x)$ 
\begin{equation}
\Box _xQ(x)=W(x)  \eqnum{3.20}
\end{equation}
where 
\begin{equation}
W(x)=-g_v\overline{\psi }(x)\gamma ^0\psi (x)-m_\omega ^2A_T^0(x). 
\eqnum{3.21}
\end{equation}
With the aid of the Green's function $G(x-y)$ (the ghost particle
propagator) which satisfies the equation: 
\begin{equation}
\Box _xG(x-y)=\delta ^4(x-y),  \eqnum{3.22}
\end{equation}
one may find the solution to the equation (3.20) as follows: 
\begin{equation}
Q(x)=\int d^4yG(x-y)W(y).  \eqnum{3.23}
\end{equation}
From the expressions given in Eqs. (3.19), (3.21) and (3.23), we see that
the $E_L^\mu (x)$ is a complicated functional of the variables $A_T^\mu $
and $E_T^\mu $ so that the Hamiltonian ${\cal H}^{*}(A_T^\mu ,E_T^\mu ,\cdot
\cdot \cdot )$ is of a much more complicated functional structure which is
not convenient for constructing the diagram technique in perturbation
theory. Therefore, it is better to express the generating functional in Eq.
(3.16) in terms of the variables $A_\mu $ and $E_\mu $. For this purpose, it
is necessary to insert the following delta-functional into Eq. (3.16) [32,
40, 41] 
\begin{equation}
\delta [A_L^\mu ]\delta [E_L^\mu -E_L^\mu (A_T^0,\psi ,\overline{\psi }%
)]=detM\delta [C]\delta [\chi ]  \eqnum{3.24}
\end{equation}
where $M$ is the matrix whose elements are 
\begin{equation}
\begin{tabular}{l}
$M(x,y)=\{C(x),\chi (y)\}\equiv \int d^4z\{\frac{\delta C(x)}{\delta A_\mu
(z)}\frac{\delta \chi (y)}{\delta E^\mu (z)}-\frac{\delta \chi (y)}{\delta
E_\mu (z)}\frac{\delta C(x)}{\delta A^\mu (z)}\}$ \\ 
$=\Box _x\delta ^4(x-y)$%
\end{tabular}
\eqnum{3.25}
\end{equation}
where $\{C(x),\chi (y)\}$ is the Poisson bracket as defined in the second
equality in Eq. (3.25). The relation in Eq. (3.24) is easily derived from
Eq. (2.12) and (3.9) by applying the property of delta-functional. Upon
inserting Eq. (3.24) into Eq. (3.16) and utilizing the Fourier
representation of the delta-functional 
\begin{equation}
\delta [C]=\int D(\rho /2\pi )e^{i\int d^4x\rho (x)C(x)},  \eqnum{3.26}
\end{equation}
we have 
\begin{equation}
\begin{array}{c}
Z[J^\mu ,J,\overline{\eta },\eta ]=\frac 1N\int D(A_\mu ,E_\mu ,\psi ,%
\overline{\psi },\Pi _\varphi ,\varphi ,\rho /2\pi )detM\delta [\chi
]exp\{i\int d^4x[E^\mu \dot A_\mu \\ 
+\Pi _\psi \dot \psi +\Pi _\varphi \dot \varphi +\rho C-{\cal H}(A_\mu
,E_\mu ,\psi ,\overline{\psi },\Pi _\varphi ,\varphi )+J^\mu A_\mu +J\varphi
+\overline{\eta }\psi +\overline{\psi }\eta ]\}.
\end{array}
\eqnum{3.27}
\end{equation}
In the above exponent, there is a $E_0$-related term $E_0(\partial
_0A_0-\partial _0\rho )$ which permits us to perform the integration over $%
E_0$, giving a delta-functional 
\begin{equation}
\delta [\partial _0A_0-\partial _0\rho ]=det|\partial _0|^{-1}\delta
[A_0-\rho ].  \eqnum{3.28}
\end{equation}
The determinant $det|\partial _0|^{-1}$, as a constant, may be put in the
normalization constant $N$ and the delta-functional $\delta [A_0-\rho ]$
will disappear when the integration over $\rho $ is carried out. The
integrals over $E_k$ $\Pi _\varphi $ are of Gaussian-type and hence easily
calculated. After these computations and noticing the expression in Eq.
(3.2), we arrive at 
\begin{equation}
\begin{array}{c}
Z[J^\mu ,J,\overline{\eta },\eta ]=\frac 1N\int D(A_\mu ,\psi ,\overline{%
\psi },\varphi ,)detM\delta [\partial ^\mu A_\mu ]exp\{i\int d^4x[{\cal L}
\\ 
+J^\mu A_\mu +J\varphi +\overline{\eta }\psi +\overline{\psi }\eta ]\}
\end{array}
\eqnum{3.29}
\end{equation}
where ${\cal L}$ was written in Eq. (2.1). When employing the familiar
expression [41, 42] 
\begin{equation}
detM=\int D(\bar C,C)e^{i\int d^4xd^4y\bar C(x)M(x,y)C(y)}=\int D(\bar C%
,C)e^{i\int d^4x\bar C(x))\Box C(x)}  \eqnum{3.30}
\end{equation}
where $\bar C(x)$ and $C(x)$ are the mutually conjugate ghost field
variables and the following limit for the Fresnel functional 
\begin{equation}
\delta [\partial ^\mu A_\mu ]=\lim_{\alpha \to 0}C[\alpha ]e^{-\frac i{%
2\alpha }\int d^4x(\partial ^\mu A_\mu )^2}  \eqnum{3.31}
\end{equation}
where $C[\alpha ]\sim \prod\limits_x(\frac i{2\pi \alpha })^{1/2}$ and
supplementing the external source terms for the ghost fields, the generating
functional in equation (3.29) is finally given in the form 
\begin{equation}
\begin{tabular}{l}
$Z[J^\mu ,J,\overline{\eta },\eta ,\overline{\xi },\xi ]=\frac 1N\int
D(A_\mu ,\psi ,\overline{\psi },\varphi ,\bar C,C)exp\{i\int d^4x[{\cal L}%
_{eff}$ \\ 
$+J^\mu A_\mu +J\varphi +\overline{\eta }\psi +\overline{\psi }\eta +%
\overline{\xi }C+\bar C\xi ]\}$%
\end{tabular}
\eqnum{3.32}
\end{equation}
where 
\begin{equation}
{\cal L}_{eff}={\cal L}-\frac 1{2\alpha }(\partial ^\mu A_\mu )^2+\overline{C%
}\Box C  \eqnum{3.33}
\end{equation}
which is the effective Lagrangian for the quantized $\sigma -\omega $ model
in which the last two terms are the so-called gauge-fixing term and the
ghost term, respectively. In Eq. (3.32), the limit $\alpha \to 0$ is
implied. Certainly, the theory may be given in arbitrary gauges $(\alpha \ne
0)$. In this case, as will be seen shortly, the ghost particle will acquire
a spurious mass $\nu =\sqrt{\alpha }m_\omega $.

\subsection{Quantization in the Lagrangian path-integral formalism}

Now let us to quantize the $\sigma -\omega $ model in the (second order)
Lagrangian path-integral formalism following the procedure proposed in Ref.
[32-34, 40]. For later convenience, the Lagrangian in Eq. (2.1) and the
Lorentz constraint condition in Eq. (2.12) are respectively generalized to
the following forms: 
\begin{equation}
{\cal L}_\lambda ={\cal L}-\frac \alpha 2\lambda ^2  \eqnum{3.34}
\end{equation}
and 
\begin{equation}
\partial ^\mu A_\mu +\alpha \lambda =0  \eqnum{3.35}
\end{equation}
where $\lambda (x)$ is an extra function which will be identified with the
Lagrange multiplier and $\alpha $ is an arbitrary constant playing the role
of gauge parameter. According to the general procedure for constrained
systems, Eq. (3.35) may be incorporated into Eq. (3.34) by the Lagrange
multiplier method to give a generalized Lagrangian 
\begin{equation}
{\cal L}_\lambda ={\cal L}+\lambda \partial ^\mu A_\mu +\frac 12\alpha
\lambda ^2.  \eqnum{3.36}
\end{equation}
This Lagrangian is obviously not gauge-invariant. However, for building up a
correct gauge field theory, it is necessary to require the dynamics of the
gauge field to be gauge-invariant. In other words, the action given by the
Lagrangian (3.36) is required to be invariant under the gauge
transformations shown in Eq. (2.5). By this requirement and applying the
constraint condition (3.35), we have 
\begin{equation}
\delta S_\lambda =-\frac 1\alpha \int d^4x\partial ^\nu A_\nu (x)(\Box
_x+\nu ^2)\theta (x)=0  \eqnum{3.37}
\end{equation}
where $\nu ^2=\alpha m_\omega ^2$. From equation (3.35) we see $\frac 1\alpha
\partial ^\nu A_\nu =-\lambda \ne 0$. Therefore, to ensure the action to be
gauge-invariant, the following constraint condition on the gauge group is
necessary to be required 
\begin{equation}
(\Box _x+\nu ^2)\theta (x)=0.  \eqnum{3.38}
\end{equation}
The constraint condition in Eq. (3.38) may also be incorporated into the
Lagrangian in Eq. (3.36) by the Lagrange undetermined multiplier method. In
doing this, it is convenient, as usually done, to introduce the ghost field
variable $C(x)$ in such a fashion 
\begin{equation}
\theta (x)=\varsigma C(x)  \eqnum{3.39}
\end{equation}
where $\varsigma $ is an infinitesimal Grassmann's number. Based on the
above definition, the constraint condition (3.38) can be rewritten as 
\begin{equation}
(\Box _x+\nu ^2)C=0  \eqnum{3.40}
\end{equation}
where the number $\varsigma $ has been dropped. This constraint condition
usually is called ghost equation. When the condition (3.40) is incorporated
into the Lagrangian (3.36) by the Lagrange multiplier method, we obtain a
more generalized Lagrangian as follows: 
\begin{equation}
{\cal L}_\lambda ={\cal L}+\lambda \partial ^\mu A_\mu +\frac 12\alpha
\lambda ^2+\bar C(\Box _x+\nu ^2)C  \eqnum{3.41}
\end{equation}
where $\bar C(x)$, acting as a Lagrange undetermined multiplier, is the new
scalar variable conjugate to the ghost variable $C(x).$ At present, we are
ready to formulate the quantization of the $\sigma -\omega $ model. As we
learn from the Lagrange undetermined multiplier method, the dynamical and
constrained variables as well as the Lagrange multiplier in the Lagrangian
(3.41) can all be treated as free ones, varying arbitrarily. Therefore, we
are allowed to use this kind of Lagrangian to construct the generating
functional of Green's functions 
\begin{equation}
\begin{tabular}{l}
$Z[J^\mu ,J,\overline{\eta },\eta ,\overline{\xi },\xi ]=\frac 1N\int
D(A_\mu ,\psi ,\overline{\psi },\varphi ,\bar C,C,\lambda )exp\{i\int d^4x[%
{\cal L}_\lambda (x)$ \\ 
$+J^\mu A_\mu +J\varphi +\overline{\eta }\psi +\overline{\psi }\eta +%
\overline{\xi }C+\bar C\xi ]\}.$%
\end{tabular}
\eqnum{3.42}
\end{equation}
Looking at the expression of the Lagrangian in Eq. (3.41), we see, the
integral over $\lambda (x)$ is of Gaussian-type. Upon completing the
calculation of this integral, we finally obtain 
\begin{equation}
\begin{tabular}{l}
$Z[J^\mu ,J,\overline{\eta },\eta ,\overline{\xi },\xi ]=\frac 1N\int
D(A_\mu ,\psi ,\overline{\psi },\varphi ,\bar C,C)exp\{i\int d^4x[{\cal L}%
_{eff}(x)$ \\ 
$+J^\mu A_\mu +J\varphi +\overline{\eta }\psi +\overline{\psi }\eta +%
\overline{\xi }C+\bar C\xi ]\}$%
\end{tabular}
\eqnum{3.43}
\end{equation}
where 
\begin{equation}
{\cal L}_{eff}={\cal L}-\frac 1{2\alpha }(\partial ^\mu A_\mu )^2+\bar C%
(\Box _x+\nu ^2)C  \eqnum{3.44}
\end{equation}
is the effective Lagrangian given in the general gauges. In the Landau gauge
($\alpha \rightarrow 0$), the Lagrangian (3.44) just goes over to the one
given in Eq. (3.33). As proved in Ref. [32], the quantization described in
Eqs. (3.34)-(3.44) is equivalent to the quantization performed by the
Faddeev-Popov approach [42]. At last of this section, we would like to
emphasize that the ghost term in the ${\cal L}_{eff}$ does not couple to the
other fields. But, we do not integrate it out in the generating functional.
Keeping this term in the effective action and in the generating functional
is helpful to later derivations of W-T identities.

\section{Ward-Takahashi identities for generating functionals}

\subsection{BRST-transformation}

In this section, we show that the action and the generating functional in
Eq. (3.43) are invariant with respect to a set of BRST transformations which
include the infinitesimal gauge transformations of the nucleon, $\sigma $
meson and $\omega $ meson fields as well as the transformations for the
ghost fields [29-31, 35]. The BRST transformations can be written in the
form 
\begin{equation}
\begin{array}{c}
\delta \psi =\varsigma \triangle \psi , \\ 
\delta \overline{\psi }=\varsigma \triangle \overline{\psi }, \\ 
\delta A_\mu =\varsigma \triangle A_\mu , \\ 
\delta \overline{C}=\varsigma \triangle \overline{C}, \\ 
\delta C=0, \\ 
\delta \varphi =0
\end{array}
\eqnum{4.1}
\end{equation}
where 
\begin{equation}
\begin{array}{c}
\triangle \psi =ig_vC\psi , \\ 
\triangle \overline{\psi }=-ig_vC\overline{\psi }, \\ 
\triangle A_\mu =\partial _\mu C, \\ 
\triangle \overline{C}=\frac 1\alpha \partial ^\mu A_\mu .
\end{array}
\eqnum{4.2}
\end{equation}
The above transformations for the nucleon, $\sigma $ meson and $\omega $
meson fields can directly be written out from Eqs. (2.5) and (3.39). The
transformations for the ghost fields may be found from the stationary
condition of the effective action under the BRST transformations for the
nucleon, $\sigma $ meson and $\omega $ meson fields, 
\begin{equation}
\delta S_{eff}=\int d^4x\delta {\cal L}_{eff}=\int d^4x\{(\delta \bar C-%
\frac \varsigma \alpha \partial ^\nu A_\nu )(\Box _x+\nu ^2)C+\bar C(\Box
_x+\nu ^2)\delta C)\}=0.  \eqnum{4.3}
\end{equation}
This expression suggests that when the ghost fields undergo the
transformations shown in Eqs. (4.1) and(4.2), the effective action is
invariant. It is easy to prove that the integration measure in Eq. (3.43) is
also invariant under the BRST-transformations owing to the Jacobian of the
transformations being unity.

\subsection{W-T identity satisfied by the generating functionals for Green's
functions}

When we make the BRST-transformations shown in Eqs. (4.1) to the generating
functional in Eq. (3.43) and consider the invariance of the generating
functional, the action and the integration measure under the
transformations, we obtain an identity such that [29-31] 
\begin{equation}
\begin{tabular}{l}
$\frac 1N\int {\cal D}(A_\mu ,\psi ,\overline{\psi },\varphi .\bar C,C)\int
d^4x\{J^\mu (x)\delta A_\mu (x)+\overline{\eta }(x)\delta \psi (x)+$ \\ 
$\delta \overline{\psi }(x)\eta (x)+\delta \overline{C}(x)\xi
(x)\}e^{iS_{eff}+iE\cdot \Phi }=0$%
\end{tabular}
\eqnum{4.4}
\end{equation}
where $E\cdot \Phi $ with $E=(J_\mu ,J,\overline{\eta },\eta ,\overline{\xi }%
,\xi )$ and $\Phi =(A_\mu ,\varphi ,\psi ,\overline{\psi },C,\overline{C})$
stands for the external source terms appearing in Eq. (3.43). The Grassmann
number $\varsigma $ contained in the BRST-transformations in Eq. (4.1) may
be eliminated by performing a partial differentiation of Eq. (4.4) with
respect to $\varsigma $. As a result, we get a W-T identity as follows: 
\begin{equation}
\begin{tabular}{l}
$\frac 1N\int {\cal D}(A_\mu ,\psi ,\overline{\psi },\varphi .\bar C,C)\int
d^4x\{J^\mu (x)\triangle A_\mu (x)-\overline{\eta }(x)\triangle \psi (x)$ \\ 
$+\triangle \overline{\psi }(x)\eta (x)+\triangle \overline{C}(x)\xi
(x)\}e^{iS_{eff}+iE\cdot \Phi }=0.$%
\end{tabular}
\eqnum{4.5}
\end{equation}

In order to represent the composite field functions $\Delta A_\mu ,\Delta 
\bar \psi $ and $\Delta \psi $ in Eq. (4.5) in terms of derivatives of the
functional $Z$ with respect to external sources, we may, as usual, construct
a generalized generating functional by introducing new external sources
(called BRST-sources later on) into the generating functional written in Eq.
(3.43) 
\begin{equation}
\begin{tabular}{l}
$Z[J^\mu ,J,\overline{\eta },\eta ,\overline{\xi },\xi ;u^\mu ,\overline{v}%
,v]=\frac 1N\int D(A_\mu ,\psi ,\overline{\psi },\varphi ,\bar C%
,C)exp\{iS_{eff}+iE\cdot \Phi $ \\ 
$+i\int d^4x[u^\mu \triangle A_\mu +\overline{v}\triangle \psi +\triangle 
\overline{\psi }v]$%
\end{tabular}
\eqnum{4.6}
\end{equation}
where $u^\mu ,\overline{v}$ and $v$ are the sources coupled to the functions 
$\Delta A_\mu $, $\Delta \Psi $ and $\Delta \overline{\Psi \text{ }}$
respectively. Obviously, $u^\mu $ and $\Delta A_\mu $ are anticommuting
quantities, while, $\overline{v},v,\Delta \bar \psi $ and $\Delta \psi $ are
commuting ones. It is easy to verify that the BRST-source terms are
invariant under the BRST- transformation because the functions $\Delta A_\mu 
$, $\Delta \bar \psi $ and $\Delta \psi $ are nilpotent with respect to the
BRST- transformations. Thus, we may start from the above generating
functional to re-derive the W-T identity. The result is the same as given in
Eq. (4.5) except that the external source terms are now extended to include
the BRST-sources 
\begin{equation}
\begin{tabular}{l}
$\frac 1N\int D(A_\mu ,\psi ,\overline{\psi },\varphi .\bar C,C)\int
d^4x\{J^\mu (x)\triangle A_\mu (x)-\overline{\eta }(x)\triangle \psi
(x)+\triangle \overline{\psi }(x)\eta (x)$ \\ 
$+\triangle \overline{C}(x)\xi (x)\}\exp \{iS_{eff}+iE\cdot \Phi +i\int
d^4x[u^\mu \triangle A_\mu +\overline{v}\triangle \psi +\triangle \overline{%
\psi }v]=0.$%
\end{tabular}
\eqnum{4.7}
\end{equation}
Clearly, Eq. (4.7) may be represented as [29-31] 
\begin{equation}
\begin{tabular}{l}
$\int d^4x[J^\mu (x)\frac \delta {\delta u^\mu (x)}-\bar \eta (x)\frac \delta
{\delta \overline{v}(x)}+\eta (x)\frac \delta {\delta v(x)}+\frac 1\alpha
\xi (x)\partial _x^\mu \frac \delta {\delta J^\mu (x)}]Z[J^\mu ,\cdot \cdot
\cdot ,v]$ \\ 
$=0.$%
\end{tabular}
\eqnum{4.8}
\end{equation}
This is the W-T identity satisfied by the generating functional of full
Green's functions.

Apart from the identity in Eq. (4.8), there is another identity called ghost
equation. The ghost equation may easily be derived by first making the
translation transformation: $\bar C\rightarrow \bar C+\bar \lambda $ in Eq.
(4.6) where $\bar \lambda $ is an arbitrary Grassmann variable, then
differentiating Eq. (4.6) with respect to $\bar \lambda $ and finally
setting $\overline{\lambda }=0$. The result is 
\begin{equation}
\begin{tabular}{l}
$\frac 1N\int D(A_\mu ,\bar C,C,\bar \psi ,\psi )\{\xi (x)+(\Box _x+\nu
^2)C(x))\}\exp \{iS_{eff}+iE\cdot \Phi $ \\ 
$+i\int d^4x[u^\mu \triangle A_\mu +\overline{v}\triangle \psi +\triangle 
\overline{\psi }v]=0$%
\end{tabular}
\eqnum{4.9}
\end{equation}
which may be represented in the form [29-31] 
\begin{equation}
\lbrack \xi (x)+(\Box _x+{\nu }^2)\frac \delta {i\delta \overline{\xi }(x)}%
]Z[J_\mu ,\cdots ,v]=0.  \eqnum{4.10}
\end{equation}

On substituting into Eqs. (4.8) and (4.10) the relation $Z=e^{iW}$ where $W$
denotes the generating functional of connected Green's functions, one may
obtain a W-T identity and a ghost equation satisfied by the functional $W$
such that 
\begin{equation}
\begin{tabular}{l}
$\int d^4x[J^\mu (x)\frac \delta {\delta u^\mu (x)}-\bar \eta (x)\frac \delta
{\delta \overline{v}(x)}+\eta (x)\frac \delta {\delta v(x)}+\frac 1\alpha
\xi (x)\partial _x^\mu \frac \delta {\delta J^\mu (x)}]W[J^\mu ,\cdot \cdot
\cdot ,v]$ \\ 
$=0$%
\end{tabular}
\eqnum{4.11}
\end{equation}
and 
\begin{equation}
\xi (x)+(\Box _x+{\nu }^2)\frac \delta {\delta \overline{\xi }(x)}W[J_\mu
,\cdots ,v]=0.  \eqnum{4.12}
\end{equation}

\subsection{W-T identity obeyed by the generating functional for proper
vertex functions}

The W-T identity in Eq. (4.11) and the ghost equation in Eq. (4.12) may be
represented in terms of the generating functional $\Gamma $ for proper
(one-particle-irreducible) vertex functions. The functional $\Gamma $ is
usually defined by the following Legendre transformation [29-31] 
\begin{equation}
\begin{tabular}{l}
$\Gamma [A^\mu ,\bar C,C,\varphi ,\bar \psi ,\psi ;u_\mu ,\overline{v}%
,v]=W[J_\mu ,\overline{\xi },\xi ,J,\bar \eta ,\eta ;u_\mu ,\overline{v},v]$
\\ 
$-\int d^4x[J_\mu A^\mu +\overline{\xi }C+\bar C\xi +J\varphi +\bar \eta
\psi +\bar \psi \eta ]$%
\end{tabular}
\eqnum{4.13}
\end{equation}
where $A_\mu ,\bar C,C,\varphi ,\bar \psi $ and $\psi $ are field variables
defined by the following functional derivatives: 
\begin{equation}
\begin{tabular}{l}
$A_\mu (x)=\frac{\delta W}{\delta J^\mu (x)},\bar C(x)=-\frac{\delta W}{%
\delta \xi (x)},C(x)=\frac{\delta W}{\delta \overline{\xi }(x)}$ \\ 
$\bar \psi (x)=-\frac{\delta W}{\delta \eta (x)},\psi (x)=\frac{\delta W}{%
\delta \bar \eta (x)},\varphi (x)=\frac{\delta W}{\delta J(x)}$%
\end{tabular}
\eqnum{4.14}
\end{equation}
From Eq. (4.13), it is not difficult to get the inverse transformations 
\begin{equation}
\begin{tabular}{l}
$J^\mu (x)=-\frac{\delta \Gamma }{\delta A_\mu (x)},\overline{\xi }(x)=\frac{%
\delta \Gamma }{\delta C(x)},\xi (x)=-\frac{\delta \Gamma }{\delta \bar C(x)}%
,$ \\ 
$\bar \eta (x)=\frac{\delta \Gamma }{\delta \psi (x)},\eta (x)=-\frac{\delta
\Gamma }{\delta \bar \psi (x)},J(x)=-\frac{\delta \Gamma }{\delta \varphi (x)%
}.$%
\end{tabular}
\eqnum{4.15}
\end{equation}
It is obvious that 
\begin{eqnarray}
\frac{\delta W}{\delta u_\mu }=\frac{\delta \Gamma }{\delta u_\mu },\frac{%
\delta W}{\delta v}=\frac{\delta \Gamma }{\delta v},\frac{\delta W}{\delta 
\overline{v}}=\frac{\delta \Gamma }{\delta \overline{v}}.  \eqnum{4.16}
\end{eqnarray}
Employing Eqs. (4.15) and (4.16), Eqs. (4.11) and (4.12) will be represented
as 
\begin{equation}
\begin{tabular}{l}
$\int d^4x\{\frac{\delta \Gamma }{\delta A_\mu (x)}\frac{\delta \Gamma }{%
\delta u^\mu (x)}+\frac{\delta \Gamma }{\delta \psi (x)}\frac{\delta \Gamma 
}{\delta \overline{v}(x)}+\frac{\delta \Gamma }{\delta \bar \psi (x)}\frac{%
\delta \Gamma }{\delta v(x)}+\frac 1\alpha \partial _x^\mu A_\mu (x)\frac{%
\delta \Gamma }{\delta \bar C(x)}\}$ \\ 
$=0$%
\end{tabular}
\eqnum{4.17}
\end{equation}
and 
\begin{equation}
\frac{\delta \Gamma }{\delta \bar C(x)}-(\Box _x+\nu ^2)C(x)=0.  \eqnum{4.18}
\end{equation}

When we define a new functional $\hat \Gamma $ in such a manner 
\begin{equation}
\hat \Gamma =\Gamma +\frac 1{2\alpha }\int d^4x(\partial ^\mu A_\mu )^2, 
\eqnum{4.19}
\end{equation}
it follows that 
\begin{equation}
\frac{\delta \Gamma }{\delta A_\mu }=\frac{\delta \hat \Gamma }{\delta A_\mu 
}+\frac 1\alpha \partial ^\mu \partial ^\nu A_\nu  \eqnum{4.20}
\end{equation}
and 
\begin{equation}
\frac{\delta \Gamma }{\delta \Phi }=\frac{\delta \hat \Gamma }{\delta \Phi }
\eqnum{4.21}
\end{equation}
where $\Phi =\psi ,\overline{\psi },u^\mu ,v$ and $\overline{v}$. Upon
inserting Eqs. (4.18)-(4.21) into Eq. (4.17) and noticing $\frac{\delta
\Gamma }{\delta u_\mu }=\partial ^\mu C$, we arrive at 
\begin{equation}
\int d^4x\{\frac{\delta \hat \Gamma }{\delta A_\mu }\frac{\delta \hat \Gamma 
}{\delta u^\mu }+\frac{\delta \hat \Gamma }{\delta \psi }\frac{\delta \hat 
\Gamma }{\delta \overline{v}}+\frac{\delta \hat \Gamma }{\delta \bar \psi }%
\frac{\delta \hat \Gamma }{\delta v}+m^2\partial ^\nu A_\nu C\}=0. 
\eqnum{4.22}
\end{equation}
The ghost equation represented through the functional ${\hat \Gamma }$ may
be written as 
\begin{equation}
\frac{\delta \hat \Gamma }{\delta \bar C(x)}-\partial _x^\mu \frac{\delta 
\hat \Gamma }{\delta u^\mu (x)}-{\nu }^2C(x)=0  \eqnum{4.23}
\end{equation}
In the Landau gauge, since $\nu =0$ and ${\partial ^\nu A_\nu =0}$, Eqs.
(4.22) and (4.23) are respectively reduced to 
\begin{equation}
\int d^4x\{\frac{\delta \hat \Gamma }{\delta A_\mu }\frac{\delta \hat \Gamma 
}{\delta u^\mu }+\frac{\delta \hat \Gamma }{\delta \psi }\frac{\delta \hat 
\Gamma }{\delta \overline{v}}+\frac{\delta \hat \Gamma }{\delta \bar \psi }%
\frac{\delta \hat \Gamma }{\delta v}\}=0  \eqnum{4.24}
\end{equation}
and 
\begin{equation}
\frac{\delta \hat \Gamma }{\delta \bar C}-\partial ^\mu \frac{\delta \hat 
\Gamma }{\delta u^\mu }=0.  \eqnum{4.25}
\end{equation}
These equations formally are the same as those for the massless Abelian
gauge field theory [29-31].

From the W-T identities formulated in this section, we may derive various
W-T identities obeyed by Green's functions and vertices, as will be
illustrated soon later.

\section{W-T identity for $\omega $ meson propagator and renormalization of
the propagator}

The W-T identity satisfied by the $\omega $ meson propagator can be derived
from the identities shown in Eqs. (4.8) and (4.10). By successive
differentiations of the identity in Eq. (4.8) with respect to the sources $%
J^\nu (y)$ and $\xi (x)$ and then setting all the sources to be zero, one
may obtain 
\begin{equation}
\partial _x^\mu \frac{\delta ^2Z}{\delta J^\mu (x)\delta J^\nu (y)}\mid
_{J^\nu =\xi =\cdot \cdot \cdot =v=0}=-\alpha \frac{\delta ^2Z}{\delta \xi
(x)\delta u^\nu (y)}\mid _{J^\nu =\xi =\cdot \cdot \cdot =v=0}.  \eqnum{5.1}
\end{equation}
Noticing the definitions of the $\omega $ meson and ghost particle
propagators 
\begin{equation}
iD_{\mu \nu }(x-y)=\frac{\delta ^2Z}{i^2\delta J^\mu (x)\delta J^\nu (y)}%
\mid _{J^\nu =\xi =\cdot \cdot \cdot =v=0}=\left\langle 0^{+}\left| T\{{\bf A%
}_\mu (x){\bf A}_\nu (y)\}\right| 0^{-}\right\rangle ,  \eqnum{5.2}
\end{equation}
\begin{equation}
i\bigtriangleup (x-y)=\frac{\delta ^2Z}{\delta \overline{\xi }(x)\delta \xi
(y)}\mid _{J^\nu =\xi =\cdot \cdot \cdot =v=0}=\left\langle 0^{+}\left| T\{%
{\bf C}(x)\overline{{\bf C}}(y)\}\right| 0^{-}\right\rangle  \eqnum{5.3}
\end{equation}
(Here and afterward the bold letters represent the operators) and
interchanging the coordinate variables and Lorentz indices, Eq. (5.1) may be
written as 
\begin{equation}
\begin{tabular}{l}
$\partial _y^\nu D_{\mu \nu }(x-y)=i\alpha \left\langle 0^{+}\left|
T^{*}\{\triangle {\bf A}_\mu (x)\overline{{\bf C}}(y)\}\right|
0^{-}\right\rangle $ \\ 
$=-\alpha \partial _\mu ^x\bigtriangleup (x-y)$%
\end{tabular}
\eqnum{5.4}
\end{equation}
where $T^{*}$ symbolizes the covariant time-ordering product and the
definition of $\triangle A_\mu $ given in Eq. (4.2) has been considered.
Similarly, when taking the derivative of Eq. (4.10) with respect to the
source $\xi (y)$ and then letting all the sources to vanish, we get 
\begin{equation}
(\Box _x+{\nu }^2)\bigtriangleup (x-y)=\delta ^4(x-y).  \eqnum{5.5}
\end{equation}
This is the equation obeyed by the ghost particle propagator.
Differentiating Eq. (5.4) with respect to $x$ and utilizing Eq. (5.5), we
find 
\begin{equation}
\partial _x^\mu \partial _y^\nu D_{\mu \nu }(x-y)=-\alpha \Box _x(\Box _x+{%
\nu }^2)^{-1}\delta ^4(x-y).  \eqnum{5.6}
\end{equation}
This just is the W-T identity satisfied by the full $\omega $ meson
propagator.

By the Fourier transformation 
\begin{equation}
D_{\mu \nu }(x-y)=\int d^4xD_{\mu \nu }(k)e^{-ik(x-y)},  \eqnum{5.7}
\end{equation}
Eq. (5.6) becomes

\begin{equation}
k^\mu k^\nu D_{\mu \nu }(k)=-\frac{\alpha k^2}{k^2-\nu ^2}.  \eqnum{5.8}
\end{equation}
The propagator $D_{\mu \nu }(k)$ may be decomposed into a transverse part
and a longitudinal part: 
\begin{equation}
D_{\mu \nu }(k)=D_T(k^2)(g_{\mu \nu }-\frac{k_\mu k_\nu }{k^2})+D_L(k^2)%
\frac{k_\mu k_\nu }{k^2}.  \eqnum{5.9}
\end{equation}
Substitution of Eq. (5.9) into Eq. (5.8) gives rise to 
\begin{equation}
D_L(k^2)=-\frac \alpha {k^2-\nu ^2}.  \eqnum{5.10}
\end{equation}
In comparison of the above expressions with the free propagator which was
given in the indefinite-metric approach previously [29] and may easily be
derived from the generating functional in Eq. (3.43) by the perturbation
method [32] 
\begin{equation}
D_{\mu \nu }^{(0)}(k)=-\{\frac{g_{\mu \nu }-k_\mu k_\nu /k^2}{k^2-m_\omega
^2+i\epsilon }+\frac{\alpha k_\mu k_\nu /k^2}{k^2-\nu ^2+i\epsilon }\}, 
\eqnum{5.11}
\end{equation}
one can see that the longitudinal parts in Eqs. (5.9) and (5.11) are the
same, implying that the longitudinal part of the $\omega $ meson propagator
does not undergo renormalization.

To derive the expression of the function $D_T(k^2)$, it is convenient to
start from the Dyson equation satisfied by the full $\omega $ meson
propagator [29-31, 36] 
\begin{equation}
D_{\mu \nu }(k)=D_{\mu \nu }^{(0)}(k)+D_{\mu \lambda }^{(0)}(k)\Pi ^{\lambda
\tau }(k)D_{\tau \nu }(k)  \eqnum{5.12}
\end{equation}
where $\Pi ^{\lambda \tau }(k)$ stands for the vacuum polarization operator,
or say, the self-energy operator of the $\omega $ meson. Contraction of Eq.
(5.12) with $k^\mu $ and use of the expressions in Eqs. (5.9)-(5.11) yield
[1] 
\begin{equation}
k_\lambda \Pi ^{\lambda \tau }(k)=0.  \eqnum{5.13}
\end{equation}
This is the W-T identity obeyed by the vacuum polarization operator which is
a consequence of the gauge symmetry of the theory. The above identity
indicates that the operator $\Pi ^{\mu \nu }(k)$ is transverse and therefore
can be written in the form 
\begin{equation}
\Pi ^{\mu \nu }(k)=(k^2g_{\mu \nu }-k_\mu k_\nu )\Pi (k^2)  \eqnum{5.14}
\end{equation}
where $\Pi (k^2)$ is a scalar function characterizing the vacuum
polarization. With the above representation, it is easy to find from Eq.
(5.12) that

\begin{equation}
D_T(k^2)=-\frac 1{k^2[1+\Pi (k^2)]-m_\omega ^2+i\epsilon }.  \eqnum{5.15}
\end{equation}
Thus, the full propagator in Eq. (5.9) can be written as 
\begin{equation}
D_{\mu \nu }(k)=-\{\frac{g_{\mu \nu }-k_\mu k_\nu /k^2}{k^2[1+\Pi
(k^2)]-m_\omega ^2+i\epsilon }+\frac{\alpha k_\mu k_\nu /k^2}{k^2-\nu
^2+i\epsilon }\}.  \eqnum{5.16}
\end{equation}
When the gauge parameter $\alpha $ is taken to be 0 and 1, we obtain the
propagators given in the Landau gauge and in the Feynman gauge respectively.
When the $\alpha $ tends to infinity, we have the propagator given in the
so-called unitary gauge. In the lowest order perturbative approximation, the
latter propagator is of the form [1-5, 29-31] 
\begin{equation}
D_{\mu \nu }^{(0)}(k)=-\frac{g_{\mu \nu }-k_\mu k_\nu /m_\omega ^2}{%
k^2-m_\omega ^2+i\epsilon }.  \eqnum{5.17}
\end{equation}
This propagator was originally derived in the canonical quantization from
the vacuum expectation value of the time-ordered product of the transverse
field operators, $iD_{\mu \nu }^{(0)}(x-y)=\left\langle 0\left| T\{{\bf A}%
_{T\mu }(x){\bf A}_{T\nu }(y)\}\right| 0\right\rangle $ and by making use of
the Fourier representation of the transverse field operator ${\bf A}_{T\mu
}(x)$ in which the $\omega $ meson momentum $k$ is put on the mass-shell, $%
k^2=m_\omega ^2$, so that the propagator in Eq. (5.17) is transverse only
for this momentum [29, 36]. However, due to the on-shell property of the
momentum in Eq. (5.17), when evaluating the contraction $k^\mu D_{\mu \nu
}^{(0)}(k)$, as we see, there appears an indefinite result since the
numerator and the denominator in Eq. (5.17) all come to zero. Especially, in
the zero-mass limit, there is a serious contradiction that the vector field
part of the Lagrangian in Eq. (2.1) is converted to the massless one, but
the propagator in Eq. (5.17) does not and is of an awful singularity. In
contrast, for the propagator in Eq. (5.11), the momentum is off-shell, $%
k^2\neq m_\omega ^2.$ Therefore, for the transverse part of the propagator
(or say, the propagator given in the Laudau gauge), we have $k^\mu D_{\mu
\nu }^T(k)=0$, showing a definite result. Moreover, in the calculation of a
loop diagram involving internal $\omega $ meson lines in which the momentum
of the $\omega $ meson line is off-shell, it is necessary to use the
propagator in Eq. (5.11). In particular, the good ultraviolet property of
the propagator allows us to perform the renormalization safely (In spite of
whether the current conservation holds or not). In the zero-mass limit, the
propagator in Eq. (5.11) and the vector field part of the Lagrangian in Eq.
(3.44) simultaneously go over to the massless ones, exhibiting the logical
consistency of the theory.

Now let us discuss renormalization of the $\omega $ meson propagator.
According to the conventional procedure of renormalization, the divergence
included in the functions $\Pi (k^2)$ may be subtracted at a renormalization
point, say, $k^2=\mu ^2$ where $\mu $ may be real or imaginary,
corresponding to the subtraction point being timelike or spacelike, 
\begin{equation}
\Pi (k^2)=\Pi (\mu ^2)+\Pi _c(k^2)  \eqnum{5.18}
\end{equation}
where $\Pi (\mu ^2)$ and $\Pi _c(k^2)$ are respectively the divergent part
and the finite part of the functions $\Pi (k^2)$. The divergent part can be
absorbed in the renormalization constant $Z_3$ which is defined as 
\begin{equation}
\;Z_3^{-1}=1+\Pi (\mu ^2).  \eqnum{5.19}
\end{equation}
With this definition, on inserting Eq. (5.18) into Eq. (5.16), the $\omega $
meson propagator will be renormalized as 
\begin{equation}
D_{\mu \nu }(k)=Z_3D_{R\mu \nu }(k)  \eqnum{5.20}
\end{equation}
where 
\begin{equation}
D_{R\mu \nu }(k)=-\{\frac{g_{\mu \nu }-k_\mu k_\nu /k^2}{k^2-(m_\omega
^R)^2+\Pi _R(k^2)+i\varepsilon }+\frac{\alpha _Rk_\mu k_\nu /k^2}{(k^2-\nu
^2+i\varepsilon )}\}  \eqnum{5.21}
\end{equation}
is the renormalized propagator in which $m_\omega ^R$ is the renormalized
mass, $\alpha _R$ the renormalized gauge parameter and $\Pi _R(k^2)$ denotes
the finite correction coming from the loop diagrams. They are defined as 
\begin{equation}
m_\omega ^R=\sqrt{Z_3}m_\omega ,\alpha _R=Z_3^{-1}\alpha ,\Pi
_R(k^2)=Z_3k^2\Pi _c(k^2).  \eqnum{5.22}
\end{equation}
It is noted that the spurious mass $\nu $ is a renomalization-invariant
quantity, $\nu ^2=\alpha m_\omega ^2=\alpha _Rm_{\omega R}^2=\nu _R^2$.
Especially, at the renormalization point, $\Pi _R(\mu ^2)=0,$ as seen from
Eqs. (5.18) and (5.22). In this case, we have a renornalization boundary
condition such that 
\begin{equation}
D_{R\mu \nu }(k)\mid _{k^2=\mu ^2}=-\{\frac{g_{\mu \nu }-k_\mu k_\nu /k^2}{%
k^2-(m_\omega ^R)^2+i\epsilon }+\frac{\alpha _Rk_\mu k_\nu /k^2}{k^2-\nu
^2+i\epsilon }\}  \eqnum{5.23}
\end{equation}
which is of the form of free propagator except that the parameters are
replaced by the renormalized ones.

\section{W-T identity for the vectorial vertex function and renormalization
of the vertex and the nucleon propagator}

\subsection{W-T identity for the vectorial vertex function}

The W-T identity for the vectorial vertex (nucleon-nucleon-$\omega $ meson
vertex) can be derived by differentiating the identity in Eq. (4.8) with
respect to the sources $\xi (x),\overline{\eta }(y)$ and $\eta (z)$ and then
turning off all the sources. The result derived, written in the operator
form, is 
\begin{equation}
\begin{tabular}{l}
$\frac 1{\alpha g_\omega }\partial ^\mu \left\langle 0^{+}\left| T\{{\bf A}%
_\mu (x){\bf \psi }(y)\overline{{\bf \psi }}(z)\}\right| 0^{-}\right\rangle $
\\ 
$=i\left\langle 0^{+}\left| T\{\overline{{\bf C}}(x){\bf \psi }(y){\bf C}(z)%
\overline{{\bf \psi }}(z)\}\right| 0^{-}\right\rangle $ \\ 
$+i\left\langle 0^{+}\left| T\{\overline{{\bf C}}(x){\bf C}(y){\bf \psi }(y)%
\overline{{\bf \psi }}(z)\}\right| 0^{-}\right\rangle $%
\end{tabular}
\eqnum{6.1}
\end{equation}
where the definitions written in Eq. (4.2) have been used. Similarly, by
differentiating the ghost equation in Eq. (4.10) with respect to the sources 
$\xi (y),\overline{\eta }(y)$ and $\eta (z)$ and then letting the sources
vanishing, one may derive the following equation: 
\begin{equation}
\begin{tabular}{l}
$\delta ^4(x-y)\left\langle 0^{+}\left| T\{{\bf \psi }(y)\overline{{\bf \psi 
}}(z)\}\right| 0^{-}\right\rangle $ \\ 
$=-i(\Box _x+{\nu }^2)\left\langle 0^{+}\left| T\{{\bf C}(x)\overline{{\bf C}%
}(y){\bf \psi (y)}\overline{{\bf \psi }}(z)\}\right| 0^{-}\right\rangle .$%
\end{tabular}
\eqnum{6.2}
\end{equation}
Here it is noted that since there is no coupling between the ghost field and
the fermion field, the two fields can not construct a connected Green's
function. Therefore, we can write 
\begin{equation}
\begin{tabular}{l}
$\left\langle 0^{+}\left| T\{{\bf C}(x)\overline{{\bf C}}(y){\bf \psi }(y)%
\overline{{\bf \psi }}(z)\}\right| 0^{-}\right\rangle $ \\ 
$=\left\langle 0^{+}\left| T\{{\bf C}(x)\overline{{\bf C}}(y)\}\right|
0^{-}\right\rangle \left\langle 0^{+}\left| T\{{\bf \psi }(y)\overline{{\bf %
\psi }}(z)\}\right| 0^{-}\right\rangle $ \\ 
$=-\bigtriangleup (x-y)S_F(y-z)$%
\end{tabular}
\eqnum{6.3}
\end{equation}
where 
\begin{equation}
\left\langle 0^{+}\left| T\{{\bf \psi }(x)\overline{{\bf \psi }}(y)\}\right|
0^{-}\right\rangle =iS_F(x-y)  \eqnum{6.4}
\end{equation}
is the nucleon propagator. It is easy to verify that once Eq. (6.3) is
substituted into the right hand side of Eq. (6.2) and applying Eq. (5.5), we
just obtain the expression on the left hand side of Eq. (6.2). Acting on
both sides of Eq. (6.1) with the operator $\Box _x+{\nu }^2$ and employing
the decomposition in Eq. (6.3) and the ghost equation in Eq. (5.5), we
obtain the following W-T identity: 
\begin{equation}
\frac 1{\alpha g_v}(\Box _x+{\nu }^2)\partial _x^\mu G_\mu (x,y,z)=i[\delta
^4(x-y)-\delta ^4(x-z)]S_F(y-z)  \eqnum{6.5}
\end{equation}
where 
\begin{equation}
G_\mu (x,y,z)=\left\langle 0^{+}\left| T\{{\bf A}_\mu (x){\bf \psi }(y)%
\overline{{\bf \psi }}(z)\}\right| 0^{-}\right\rangle  \eqnum{6.6}
\end{equation}
is the three-point Green's function which is connected. This Green' function
has the following one-particle irreducible decomposition [29-31] 
\begin{equation}
G_\mu (x,y,z)=\int d^4x^{\prime }d^4y^{\prime }d^4z^{\prime }iD_{\mu \nu
}(x-x^{\prime })iS_F(y-y^{\prime })i\Gamma ^\nu (x^{\prime },y^{\prime
},z^{\prime })iS_F(z^{\prime }-z)  \eqnum{6.7}
\end{equation}
in which $\Gamma ^\nu (x^{\prime },y^{\prime },z^{\prime })$ is the
vectorial proper vertex. On inserting Eq. (6.7) into Eq. (6.5), through the
Fourier transformation, we get in the momentum space that 
\begin{equation}
\begin{tabular}{l}
$\frac 1{\alpha g_v}(k^2-\nu ^2)k^\mu D_{\mu \nu }(k)S_F(p)\Gamma ^\nu
(p,q,k)S_F(q)$ \\ 
$=(2\pi )^4\delta ^4(k+p-q)[S_F(q)-S_F(p)].$%
\end{tabular}
\eqnum{6.8}
\end{equation}
Considering that the energy-momentum conservation holds at the vertex, we
can write 
\begin{equation}
\Gamma _\mu (p,q,k)=(2\pi )^4\delta ^4(k+p-q)ig_v[\gamma _\mu +\Lambda _\mu
(p,q)]  \eqnum{6.9}
\end{equation}
With this representation and noticing 
\begin{equation}
(k^2-\nu ^2)k^\mu D_{\mu \nu }(k)=-\alpha k_\nu =\alpha (p-q)_\nu , 
\eqnum{6.10}
\end{equation}
one may obtain from Eq. (6.8) that 
\begin{equation}
(p-q)^\mu [\gamma _\mu +\Lambda _\mu (p,q)]=S_F^{-1}(p)-S_F^{-1}(q). 
\eqnum{6.11}
\end{equation}
It is well-known that the general expression of the nucleon propagator $%
S_F(p)$ can be found from the Dyson equation satisfied by the propagator, as
was similarly done in Eqs. (5.12)-(5.16) for the $\omega $ meson propagator.
The inverse of the propagator can be written as 
\begin{equation}
S_F^{-1}(p)={\bf p}-M-\Sigma (p)  \eqnum{6.12}
\end{equation}
where ${\bf p=}\gamma ^\mu p_\mu $ and $\Sigma (p)$ is the nucleon
self-energy. Noticing this expression, when we differentiate the both sides
of Eq. (6.11) with respect to $p^\mu $ and then set $q=p$, it is found that
[1] 
\begin{equation}
\Lambda _\mu (p,p)=-\frac{\partial \Sigma (p)}{\partial p^\mu }. 
\eqnum{6.13}
\end{equation}
This is the W-T identity which establishes the relation between the
vectorial proper vertex and the nucleon self-energy.

It is interesting to note that the above identity determines the subtraction
fashion of the nucleon self-energy. As one knows, the divergence in the
vertex $\Lambda _\mu (p,q)$ may be subtracted at the renormalization point $%
\mu $ in such a way 
\begin{equation}
\Lambda _\mu (p,p)=L\gamma _\mu +\Lambda _\mu ^c(p)  \eqnum{6.14}
\end{equation}
where 
\begin{equation}
L=\Lambda _\mu (p,p)\mid _{{\bf p=}\mu }  \eqnum{6.15}
\end{equation}
is a divergent constant. Substituting Eq. (6.14) into Eq. (6.13) and then
integrating Eq. (6.13) over $p_\mu $ from $p_\mu ^0$ to $p_\mu $, we get 
\begin{equation}
\Sigma (p)=\Sigma (\mu )-L({\bf p-}\mu )-({\bf p-}\mu )C(p^2)  \eqnum{6.16}
\end{equation}
where we have chosen the $p_\mu ^0$ to meet $\gamma ^\mu p_\mu ^0=\mu $ and
set the integral $\int_{p^0}^pdp^\mu \Lambda _\mu ^c(p^2)=({\bf p-}\mu
)C(p^2)$ with the consideration that when $p=p^0$, the integral vanishes and 
$\Lambda _\mu ^c(p)$ is finite, satisfying the boundary condition $\Lambda
_\mu ^c(p)\mid _{_{{\bf p=}\mu }}=0$ so that the $C(p^2)$ is also finite,
having the boundary condition $C(\mu ^2)=0$. When the divergent constants $%
\Sigma (\mu )$ and $L$ are set to be 
\begin{equation}
\Sigma (\mu )=A  \eqnum{6.17}
\end{equation}
and 
\begin{equation}
L=-B,  \eqnum{6.18}
\end{equation}
Eq. (6.16) will be written in the form 
\begin{equation}
\Sigma (p)=A+({\bf p-}\mu )[B-C(p^2)]  \eqnum{6.19}
\end{equation}
This is the formula that gives the uniquely correct way for the subtraction
of the nucleon self-energy.

\subsection{Renormalization of the nucleon propagator and the vectorial
vertex}

Based on the representation of the self-energy in Eq. (6.19), the full
nucleon propagator may be written as 
\begin{equation}
S_F(p)=\frac 1{{\bf p(}1-B)-(M+A-\mu B)+({\bf p-}\mu )C(p^2)}.  \eqnum{6.20}
\end{equation}
With the renormalization constant of the nucleon propagator defined by 
\begin{equation}
Z_2^{-1}=1-B,  \eqnum{6.21}
\end{equation}
the nucleon propagator will be renormalized as 
\begin{equation}
S_F(p)=Z_2S_F^R(p)  \eqnum{6.22}
\end{equation}
where 
\begin{equation}
S_F^R(p)=\frac 1{{\bf p-}M_R-\Sigma _R(p)+i\epsilon }  \eqnum{6.23}
\end{equation}
is the renormalized propagator in which $M_R$ and $\Sigma _R(p)$ are the
renormalized mass and the finite renormalization correction respectively.
They are separately represented in the following: 
\begin{equation}
M_R=Z_M^{-1}M  \eqnum{6.24}
\end{equation}
where $Z_M$ is the nucleon mass renormalization constant defined by 
\begin{equation}
Z_M^{-1}=1+Z_2[\frac AM+(1-\frac \mu M)B]  \eqnum{6.25}
\end{equation}
and 
\begin{equation}
\Sigma _R(p)=-Z_2({\bf p-}\mu )C(p^2)  \eqnum{6.26}
\end{equation}
with the boundary condition $\Sigma _R(p)\mid _{_{{\bf p=}\mu }}=0$ which
leads to the boundary condition of nucleon propagator like this 
\begin{equation}
S_F^R(p)\mid _{{\bf p=}\mu }=\frac 1{{\bf p-}M_R+i\epsilon }.  \eqnum{6.27}
\end{equation}
Clearly, this propagator is formally the same as the free propagator.

We would like to mention here the renormalization of the vertex function
defined by $\widehat{\Gamma }_\mu (p,q)=\gamma _\mu +\Lambda _\mu (p,q).$ In
view of the subtraction in Eq. (6.14) and the following definition of vertex
renormalization constant $Z_1$: 
\begin{equation}
Z_1^{-1}=1+L,  \eqnum{6.28}
\end{equation}
the vertex function will be renormalized as 
\begin{equation}
\widehat{\Gamma }_\mu (p,q)=Z_1^{-1}\widehat{\Gamma }_\mu
^R(p,q)=Z_1^{-1}[\gamma _\mu +\Lambda _\mu ^R(p,q)]  \eqnum{6.29}
\end{equation}
where 
\begin{equation}
\Lambda _\mu ^R(p,q)=Z_1\Lambda _\mu ^c(p,q).  \eqnum{6.30}
\end{equation}
is the finite renormalization correction to the vertex. From the boundary
condition $\Lambda _\mu ^c(p)\mid _{_{{\bf p=}\mu }}=0$ mentioned before, it
follows $\Lambda _\mu ^R(p,q)\mid _{{\bf p=q=}\mu }=0$ by which we have 
\begin{equation}
\widehat{\Gamma }_\mu ^R(p,q)\mid _{{\bf p=q}=\mu }=\gamma _\mu . 
\eqnum{6.31}
\end{equation}
This just is the boundary condition for the renormalized vertex function $%
\widehat{\Gamma }_\mu ^R(p,q)$ under which the vertex is of the form of the
bare vertex. In particular, from Eqs. (6.18), (6.21) and (6.28), it is clear
to see that 
\begin{equation}
Z_2=Z_1.  \eqnum{6.32}
\end{equation}
This is the Ward identity satisfied by the nucleon propagator
renormalization constant and the vertex one.

At last, it is pointed out that the identities shown in Eqs. (6.13) and
(6.32) and the subtraction represented in Eq. (6.19) formally are the same
as those in QED because they are all the consequence of U(1) gauge symmetry.
Originally, the identities mentioned above follow from the current
conservation. This result is natural because the current conservation, as
generally proved in the Gauge Field Theory [29-31], can be derived from the
global gauge symmetry or the local gauge symmetry. It would be emphasized
that the aforementioned identities hold not only for the case where the $%
\omega $ meson is considered only, but also for the general case that the $%
\omega $ meson and the $\sigma $ meson are taken into account together. We
take one-loop diagrams to illustrate this point. The one-loop nucleon
self-energy in the $\sigma -\omega $ model is represented in Fig. (1a) and
Fig. (1b). The one-loop vectorial vertex is shown in Fig. (2a) and Fig.
(2b). In the figures, the solid line designates the free nucleon propagator $%
iS_F^{(0)}(p)$ represented in Eq. (6.27), the wavy line stands for the free $%
\omega $ meson propagator $iD_{\mu \nu }^{(0)}(k)$ written in Eq. (5.11),
the dashed line denotes the free $\sigma $ meson propagator which is of the
form 
\begin{equation}
i\bigtriangleup (q)=\frac i{q^2-m_\sigma ^2+i\varepsilon },  \eqnum{6.33}
\end{equation}
the bare vectorial vertex $\Gamma _\mu ^{(0)}$ and the bare scalar vertex $%
\Gamma ^{(0)}$ are 
\begin{equation}
\begin{tabular}{l}
$\Gamma _\mu ^{(0)}=ig_v\gamma _\mu ,$ \\ 
$\Gamma ^{(0)}=ig_s.$%
\end{tabular}
\eqnum{6.34}
\end{equation}
The above Feynman rules are easily derived from the generating functional in
Eq. (3.43) by the perturbation method. Applying the Feynman rules, for the
one-loop nucleon self-energy defined by $-i\Sigma (p),$ we have the
following expression 
\begin{equation}
\Sigma (p)=i\int d^4k[g_v^2\gamma ^\mu S_F^{(0)}(p-k)\gamma ^\nu D_{\mu \nu
}^{(0)}(k)+g_s^2S_F^{(0)}(p-k)\bigtriangleup ^{(0)}(k)]  \eqnum{6.35}
\end{equation}
where the first term and the second one are respectively given by Fig. (1a)
and Fig. (1b). While, for the one-loop vectorial vertex, in accordance with
the definition in Eq. (6.9), we have 
\begin{equation}
\begin{tabular}{l}
$\Lambda _\mu (p,q)=i\int d^4k[g_v^2\gamma ^\nu S_F^{(0)}(q-k)\gamma _\mu
S_F^{(0)}(p-k)\gamma ^\lambda D_{\nu \lambda }^{(0)}(k)$ \\ 
$+g_s^2S_F^{(0)}(q-k)\gamma _\mu S_F^{(0)}(p-k)\bigtriangleup ^{(0)}(k)].$%
\end{tabular}
\eqnum{6.36}
\end{equation}
where the first and second terms are given by Fig. (2a) and Fig. (2b)
respectively. By making use of the derivative 
\begin{equation}
\frac \partial {\partial p^\mu }S_F(p-k)=-S_F(p-k)\gamma _\mu S_F(p-k), 
\eqnum{6.37}
\end{equation}
it is easy to find that the identity in Eq. (6.13) holds. Thus, the
correctness of the W-T identity in Eq. (6.13) which follows from the U(1)
gauge symmetry of the model is verified by the perturbative calculation. The
identities in Eqs. (6.13) and (6.32) will be helpful to facilitate
calculations of the renomalization of the $\sigma -\omega $ model.

\section{Renormalization of the $\sigma $ meson propagator and the scalar
vertex}

For later convenience, it is necessary to give a general description for the
renormalization of the $\sigma $ meson propagator and the scalar vertex. We
start from the Dyson equation satisfied by the $\sigma $ meson full
propagator $i\bigtriangleup (q)$ 
\begin{equation}
\bigtriangleup (q)=\bigtriangleup ^{^{(0)}}(q)+\bigtriangleup
^{^{(0)}}(q)\Omega (q)\bigtriangleup (q)  \eqnum{7.1}
\end{equation}
where $\bigtriangleup ^{^{(0)}}(q)$ is the $\sigma $ meson free propagator
shown in Eq. (6.33) and $-i\Omega (q)$ represents the $\sigma $ meson
self-energy. From Eq. (7.1) it may be solved that 
\begin{equation}
\bigtriangleup (q)=\frac 1{q^2-m_\sigma ^2-\Omega (q)+i\varepsilon }. 
\eqnum{7.2}
\end{equation}
The self-energy can be Lorentz-covariantly decomposed into 
\begin{equation}
\Omega (q)=\Omega _1(q^2)q^2+\Omega _2(q^2)m_\sigma ^2.  \eqnum{7.3}
\end{equation}
The divergence in the $\Omega (q)$ can be subtracted at the renormalization
point $\mu $ in such a way 
\begin{equation}
\begin{tabular}{l}
$\Omega _1(q^2)=\Omega _1(\mu ^2)+\Omega _1^c(q^2)$ \\ 
$\Omega _2(q^2)=\Omega _2(\mu ^2)+\Omega _2^c(q^2).$%
\end{tabular}
\eqnum{7.4}
\end{equation}
On substituting Eq. (7.4) in Eq. (7.2), the propagator $\bigtriangleup (q)$
will be renormalized as 
\begin{equation}
\bigtriangleup (q)=Z_3^{\prime }\bigtriangleup _R(q)  \eqnum{7.5}
\end{equation}
where 
\begin{equation}
Z_3^{\prime -1}=1-\Omega _1(\mu ^2)  \eqnum{7.6}
\end{equation}
is the renormalization constant of the $\sigma $ meson propagator, 
\begin{equation}
\bigtriangleup _R(q)=\frac 1{q^2-(m_\sigma ^R)^2-\Omega _R(q)+i\varepsilon }
\eqnum{7.7}
\end{equation}
is the renormalized propagator in which 
\begin{equation}
m_\sigma ^R=Z_m^{\sigma -1}m_\sigma   \eqnum{7.8}
\end{equation}
is the renormalized $\sigma $ meson mass with 
\begin{equation}
Z_m^\sigma =[Z_3^{\prime }(1+\Omega _2(\mu ^2))]^{-\frac 12}  \eqnum{7.9}
\end{equation}
being the renormalization constant of the $\sigma $ meson mass and 
\begin{equation}
\Omega _R(q)=Z_3^{\prime }[q^2\Omega _1^c(q^2)+m_\sigma ^2\Omega _2^c(q^2)] 
\eqnum{7.10}
\end{equation}
is the finite correction to the renormalized propagator. Obviously, the $%
\Omega _R(q)$ has the boundary condition $\Omega _R(q)\mid _{q^2=\mu ^2}=0$
which leads to the boundary condition of the propagator as follows: 
\begin{equation}
\bigtriangleup _R(q)\mid _{q^2=\mu ^2}=\frac 1{q^2-(m_\sigma
^R)^2+i\varepsilon }.  \eqnum{7.11}
\end{equation}
This propagator formally is the same as the free propagator in Eq. (6.33).

Analogous to Eq. (6.9 ) for the vectorial vertex, the scalar vertex can be
written as 
\begin{equation}
\Gamma (p,q,k)=(2\pi )^4\delta ^4(k+p-q)ig_s\widehat{\Gamma }(p,q) 
\eqnum{7.12}
\end{equation}
where 
\begin{equation}
\widehat{\Gamma }(p,q)=1+\Lambda (p,q)  \eqnum{7.13}
\end{equation}
in which $\Lambda (p,q)$ denotes the contribution of all higher order
diagrams. When the divergence in the $\Lambda (p,q)$ is subtracted at the
renormalization point $\mu $ , we have

\begin{equation}
\Lambda (p,q)=L^{\prime }+\Lambda _c(p,q)  \eqnum{7.14}
\end{equation}
where 
\begin{equation}
L^{\prime }=\Lambda (p,p)\mid _{{\bf p=}\mu }  \eqnum{7.15}
\end{equation}
is the divergent constant and $\Lambda _c(p,q)$ is the finite correction of
the $\Lambda (p,q)$. With the above subtraction, the vertex in Eq. (7.13)
will be renormalized as 
\begin{equation}
\widehat{\Gamma }(p,q)=Z_1^{\prime -1}\widehat{\Gamma }_R(p,q)=Z_1^{\prime
-1}[1+\Lambda _R(p,q)]  \eqnum{7.16}
\end{equation}
where $Z_1^{\prime }$ is the renormalization constant of the scalar vertex
defined by

\begin{equation}
Z_1^{\prime -1}=1+L^{\prime }.  \eqnum{7.17}
\end{equation}
and $\Lambda _R(p,q)=Z^{\prime }\Lambda _c(p,q)$ is the finite correction of
the $\widehat{\Gamma }_R(p,q)$ with the boundary condition $\Lambda
_R(p,q)\mid _{{\bf p=q=}\mu }=0$ which yields the boundary condition for the
renormalized vertex $\widehat{\Gamma }_R(p,q)$ as follows: 
\begin{equation}
\widehat{\Gamma }_R(p,q)\mid _{{\bf p=q=}\mu }=1.  \eqnum{7.18}
\end{equation}
This shows that at the renormalization point, the renormalized vertex is
reduced to the form of bare vertex.

It is interesting to note that there is an identity which holds between the
nucleon self-energy and the scalar vertex. For example, from the expression
of the one-loop scalar vertex $\Lambda (p,q)$ shown in Fig. (3a) and Fig.
(3b) 
\begin{equation}
\begin{tabular}{l}
$\Lambda (p,q)=i\int d^4k[g_s^2S_F^{(0)}(q-k)\bigtriangleup
^{(0)}(k)S_F(p-k) $ \\ 
$+g_v^2\gamma ^\mu S_F^{(0)}(q-k)S_F^{(0)}(p-k)\gamma ^\nu D_{\mu \nu
}^{(0)}(k)]$%
\end{tabular}
\eqnum{7.19}
\end{equation}
and the following derivative 
\begin{equation}
\frac \partial {\partial M}S_F(p-k)=S_F(p-k)S_F(p-k),  \eqnum{7.20}
\end{equation}
we find 
\begin{equation}
\Lambda (p,p)=\frac{\partial \Sigma (p)}{\partial M}.  \eqnum{7.21}
\end{equation}
Based on this identity and the expression in Eq. (6.19), the constant
defined in Eq. (7.15) can be computed by 
\begin{equation}
L^{\prime }=\frac{\partial \Sigma (p)}{\partial M}\mid _{_{{\bf p=}\mu }}=%
\frac{\partial A}{\partial M}.  \eqnum{7.22}
\end{equation}
This relation will be used to simplify the calculation of renormalization of
the $\sigma -\omega $ model.

\section{Renormalization group equation and renormalized S-matrix elements}

Suppose $F_R$ is a renormalized quantity. In the multiplicative
renormalization, it is related to the unrenormalized one $F$ in such a way 
\begin{eqnarray}
F=Z_FF_R  \eqnum{8.1}
\end{eqnarray}
where $Z_F$ is the renormalization constant of $F$. The $Z_F$ and $F_R$ are
all functions of the renormalization point $\mu =\mu _0e^t$ where $\mu _0$
is a fixed renormalization point corresponding the zero value of the group
parameter $t$. Differentiating Eq. (8.1) with respect to $\mu $ and noticing
that the $F$ is independent of $\mu $, we immediately obtain a
renormalization group equation (RGE) satisfied by the function $F_R$
[27--30] 
\begin{eqnarray}
\mu \frac{dF_R}{d\mu }+\gamma _FF_R=0  \eqnum{8.2}
\end{eqnarray}
where $\gamma _F$ is the anomalous dimension defined by 
\begin{eqnarray}
\gamma _F=\mu \frac d{d\mu }\ln Z_F.  \eqnum{8.3}
\end{eqnarray}
We first note here that because the renormalization constant is
dimensionless, the anomalous dimension can only depends on the ratio ${%
\sigma =\frac{m_R}\mu }$, ${\gamma }_F{=\gamma }_F{(g}_R{,\sigma ),}$ where $%
m_R$ and $g_R$ are the renormalized mass and coupling constant respectively.
Next, we note that Eq. (8.2) is suitable for a physical parameter (mass or
coupling constant), a propagator, a vertex, a wave function or some other
Green function. If the function ${F_R}$ stands for a renormalized Green
function, vertex or wave function, in general, it depends explicitly not
only on the scale $\mu $, but also on the renormalized coupling constant $%
g_R $, mass $m_R$ and gauge parameter $\alpha _R$ which are all functions of 
$\mu $, $F_R=F_R(p,g_R(\mu ),m_R(\mu ),\alpha _R(\mu );\mu )$ where $p$
symbolizes all the momenta. Considering that the function ${F_R}$ is
homogeneous in the momentum and mass, it may be written, under the scaling
transformation of momentum $p=\lambda p_0$, as follows: 
\begin{eqnarray}
F_R(p;g_R,m_R,\alpha _R;\mu )=\lambda ^{D_F}F_R(p_0;g_R,\frac{m_R}\lambda
,\alpha _R;\frac \mu \lambda )  \eqnum{8.4}
\end{eqnarray}
where $D_F$ is the canonical dimension of $F$. Since the renormalization
point is a momentum taken to subtract the divergence, we may set $\mu =\mu
_0\lambda $ where $\lambda =e^t$ which is taken to be the same as in $%
p=p_0\lambda $. Noticing the above transformation, the solution to the RGE
in Eq. (8.2) can be expressed as 
\begin{equation}
F_R(p;g_R,m_R,\alpha _R,\mu _0)=\lambda ^{D_F}e^{\int_1^\lambda \frac{%
d\lambda }\lambda \gamma _F(\lambda )}F_R(p_0;g_R(\lambda ),m_R(\lambda
)\lambda ^{-1},\alpha _R(\lambda );\mu _0)  \eqnum{8.5}
\end{equation}
where $g_R(\lambda ),m_R(\lambda )$ and $\alpha _R(\lambda )$ are the
effective (running) coupling constant, mass and gauge parameter,
respectively. The solution written above describes the behavior of the
function $F_R$ under the scaling of momenta.

How to determine the function $F_R(p_0;\cdots ,\mu _0)$ on the right-hand
side of Eq. (8.5) when the $F_R(p_0,...)$ stands for a wave function, a
propagator or a vertex? This question can be unambiguously answered in the
momentum space subtraction scheme. Noticing that the momentum $p_0$ and the
renormalization point $\mu _0$ are fixed, but may be chosen arbitrarily, we
can, certainly, set $\mu _0^2=p_0^2$. With this choice, by making use of the
boundary condition satisfied by the propagator, the vertex or the wave
function as denoted in Eqs. (5.23), (6.27). (6.31), (7.11) and (7.18), we
may write 
\begin{equation}
F_R(p_0;g_R,m_R,\alpha _R,\mu )\mid _{P_0^2=\mu
^2}=F_R^{(0)}(p_0;g_R,m_R,\alpha _R)  \eqnum{8.6}
\end{equation}
where the function $F_R^{(0)}(p;g_R,m_R,\alpha _R)$ is of the form of a free
propagator, a bare vertex (if the vertex is fundamental, i.e., follows
directly from the interaction Lagrangian) or a free wave function and
independent of the renormalization point (see the examples given in the
preceding sections). In light of the boundary condition in Eq. (8.6) and
considering the homogeneity of the function $F_R$ as mentioned in Eq. (8.4),
one can write 
\begin{equation}
\lambda ^{D_F}F_R(p_0;g_R(\lambda ),m_R(\lambda )\lambda ^{-1},\alpha
_R(\lambda ),\mu _0)\mid _{p_0^2=\mu _0^2}=F_R^{(0)}(p;g_R(\lambda
),m_R(\lambda ),\alpha _R(\lambda ))  \eqnum{8.7}
\end{equation}
where the renormalized coupling constant, mass and gauge parameter in the
function $F_R^{(0)}(p,...)$ become the running ones. With the expression
given in Eq. (8.7), Eq. (8.5) will finally be written in the form [20] 
\begin{eqnarray}
F_R(p;g_R,m_R,\alpha _R)=e^{\int_1^\lambda \frac{d\lambda }\lambda \gamma
_F(\lambda )}F_R^{(0)}(p;g_R(\lambda ),m_R(\lambda ),\alpha _R(\lambda )). 
\eqnum{8.8}
\end{eqnarray}

For a gauge field theory, the anomalous dimensions shown in Eq. (8.8) are
all cancelled out in S-matrix elements. To show this point more
specifically, let us take the two-nucleon scattering taking place via $%
\omega $ meson exchanges as an example. The exact matrix element for the
two-nucleon scattering can be written out from the well-known reduction
formula which establishes the relation between an on-mass-shell S-matrix
element and the corresponding off-mass-shell connected Green's function [29,
36]. A connected Green's function may conveniently be derived from the
generating functional $W$ for connected Green's functions as mentioned in
Sec. IV. For the nucleon-nucleon scattering, the S-matrix element is related
to the following four-point connected Green's function 
\begin{equation}
G_c(x_1,x_2;y_1,y_2)=\left\langle 0^{+}\left| T[\psi (x_1)\psi (x_2)%
\overline{\psi }(y_1)\overline{\psi }(y_2)]\right| 0^{-}\right\rangle _C 
\eqnum{8.9}
\end{equation}
where the subscript $C$ marks the connectivity of the Green's function.
According to the familiar procedure of irreducible decomposition [29-31],
the connected Green's function in Eq. (8.9) can be decomposed into three
one-particle irreducible ones as represented graphically in Figs. (4a)-(4c).
In each of the diagrams, there are four external legs which represent the
full of-mass-shell nucleon propagators. These propagators will be converted
to the full on-mass-shell nucleon wave functions by the reduction formula.
The shaded blobs in the diagrams stand for the exact proper (one-particle
irreducible) vertices. Let us first concentrate our attention on the
diagrams in Figs. (4a) and (4b). These two diagrams represent the
two-nucleon scattering taking place in the t-channel via a $\omega $ meson
exchange. The wavy line with a white blob in the figures denotes the full $%
\omega $ meson propagator. Considering the well-known fact that a S-matrix
element expressed in terms of unrenormalized quantities is equal to that
represented by the corresponding renormalized quantities, the scattering
amplitude given by Figs. (4a) and (4b) may be written as [20, 29, 36] 
\begin{equation}
\begin{tabular}{l}
$T_{fi}^{(1)}=\overline{u}_R^\sigma (q_1)\Gamma _R^\mu (q_1,p_1)u_R^\alpha
(p_1)iD_{\mu \nu }^R(k)\overline{u}_R^\rho (q_2)\Gamma _R^\nu
(q_2,p_2)u_R^\beta (p_2)$ \\ 
$-\overline{u}_R^\rho (q_2)\Gamma _R^\mu (q_2,p_1)u_R^\alpha (p_1)iD_{\mu
\nu }^R(k)\overline{u}_R^\sigma (q_1)\Gamma _R^\nu (q_1,p_2)u_R^\beta (p_2)$%
\end{tabular}
\eqnum{8.10}
\end{equation}
where $k=q_1-p_1=p_2-q_2$, $u_R^\alpha (p)$, $\Gamma _R^\mu (q_i,p_i)$ and $%
iD_{\mu \nu }^R(k)$ are the nucleon wave function, the proper vectorial
vertex and the $\omega $ meson propagator respectively which are all
renormalized. The renormalization constants of the wave function, the
propagator and the vertex are denoted by $\sqrt{Z_2}$, $Z_3$ and $Z_\Gamma $
respectively. The constant $Z_\Gamma $ is defined by 
\begin{equation}
Z_\Gamma =Z_2^{-1}Z_3^{-\frac 12}  \eqnum{8.11}
\end{equation}
because the vertex in Eq. (8.10) is now defined by containing a vectorial
coupling constant $g_v^R$ multiplied with an imaginary number $i$ in it. The
renormalized coupling constant is defined as 
\begin{equation}
g_v^R=\frac{Z_2\sqrt{Z_3}}{Z_1}g_v  \eqnum{8.12}
\end{equation}

On the basis of the formula given in Eq. (8.8), the renormalized nucleon
wave function, meson propagator and vertex can be represented in the forms
as shown separately in the following. For the nucleon wave function, we have 
\begin{equation}
u_R^\alpha (p)=e^{\int_1^\lambda \frac{d\lambda }\lambda \gamma _w(\lambda
)}u_{R\alpha }^{(0)}(p)  \eqnum{8.13}
\end{equation}
where 
\begin{equation}
u_{R\alpha }^{(0)}(p)=\left( \frac{E+M_R(\lambda )}{2M_R(\lambda )}\right)
^{1/2}\left( 
\begin{array}{c}
1 \\ 
\frac{\overrightarrow{\sigma }.\overrightarrow{p}}{E+M_R(\lambda )}
\end{array}
\right) \varphi _\alpha (\overrightarrow{p})  \eqnum{8.14}
\end{equation}
is the renormalized wave function which formally is the same as the free
wave function, but, the $M_R(\lambda )$ in it is a running mass and 
\begin{equation}
\gamma _w=\frac 12\mu \frac d{d\mu }\ln Z_2  \eqnum{8.15}
\end{equation}
is the anomalous dimension of the nucleon wave function.

For the renormalized $\omega $ meson propagator, we can write 
\begin{equation}
iD_{\mu \nu }^R(k)=e^{\int_1^\lambda \frac{d\lambda }\lambda \gamma
_3(\lambda )}iD_{\mu \nu }^{R(0)}(k)  \eqnum{8.16}
\end{equation}
where 
\begin{equation}
iD_{\mu \nu }^{R(0)}(k)=-\frac i{k^2-m_\omega ^R(\lambda )+i\varepsilon }%
[g_{\mu \nu }-(1-\alpha _R(\lambda ))\frac{k_\mu k_\nu }{k^2-\nu
^2+i\varepsilon }]  \eqnum{8.17}
\end{equation}
is the free propagator with $m_\omega ^R(\lambda )$ and $\alpha _R(\lambda )$
in it being the running $\omega $ meson mass and gauge parameter and 
\begin{equation}
\gamma _3(\lambda )=\mu \frac d{d\mu }\ln Z_3  \eqnum{8.18}
\end{equation}
is the anomalous dimension of the propagator.

For the renormalized vertex, it reads 
\begin{equation}
\Gamma _R^\mu (q_i,p_j)=e^{\int_1^\lambda \frac{d\lambda }\lambda \gamma
_v(\lambda )}\Gamma _R^{(0)\mu }(q_i,p_j)  \eqnum{8.19}
\end{equation}
where $i,j=1,2$,

\begin{equation}
\Gamma _R^{(0)\mu }(q_i,p_j)=ig_v^R(\lambda )\gamma ^\mu  \eqnum{8.20}
\end{equation}
is the bare vertex in which $g_v^R(\lambda )$ is the running coupling
constant and 
\begin{equation}
\gamma _v(\lambda )=\mu \frac d{d\mu }\ln Z_v=-\mu \frac d{d\mu }\ln Z_2-%
\frac 12\mu \frac d{d\mu }\ln Z_3  \eqnum{8.21}
\end{equation}
is the anomalous dimension of the vertex here the relation in Eq. (8.11) has
been used.

Upon substituting Eqs. (8.13), (8.16) and (8.19) into Eq. (8.10) and
noticing Eqs. (8.15), (8.18) and (8.21), we find that the anomalous
dimensions in the S-matrix element are all cancelled out with each other. As
a result of the cancellation, we arrive at 
\begin{equation}
\begin{tabular}{l}
$T_{fi}^{(1)}=\overline{u}_{R\sigma }^{(0)}(q_1)\Gamma _R^{(0)\mu
}(q_1,p_1)u_{R\alpha }^{(0)}(p_1)iD_{\mu \nu }^{R(0)}(k)\overline{u}_{R\rho
}^{(0)}(q_2)\Gamma _R^{(0)\nu }(q_2,p_2)u_{R\beta }^{(0)}(p_2)$ \\ 
$-\overline{u}_{R\rho }^{(0)}(q_2)\Gamma _R^{(0)\mu }(q_2,p_1)u_{R\alpha
}^{(0)}(p_1)iD_{\mu \nu }^{R(0)}(k)\overline{u}_{R\sigma }^{(0)}(q_1)\Gamma
_R^{(0)\nu }(q_1,p_2)u_{R\beta }^{(0)}(p_2)$%
\end{tabular}
\eqnum{8.22}
\end{equation}
This expression clearly shows that the exact t-channel S-matrix element of
the two-nucleon scattering can be represented in the form as given by the
tree diagrams shown in Figs. (5a) and (5b) provided that all the physical
parameters in the matrix elements are replaced by their effective (running)
ones.

Next, let us turn to the diagram in Fig. (4c). In the diagram, the shaded
blob with four amputated external legs represents the nucleon four-line
proper vertex. The direct term of the scattering amplitude given by Fig.
(4c) can be represented in terms of the renormalized quantities as follows: 
\begin{equation}
T_{fi}^{(2)}=\overline{u}_R^\sigma (q_1)\overline{u}_R^\rho (q_2)\Gamma
_R(p_1,p_2;q_1,q_2)u_R^\alpha (p_1)u_R^\beta (p_2)  \eqnum{8.23}
\end{equation}
In accordance with Eq. (8.8), the renormalized vertex $\Gamma
_R(p_1,p_2;q_1,q_2)$ is of the form 
\begin{equation}
\Gamma _R(p_1,p_2;q_1,q_2)=e^{\int_1^\lambda \frac{d\lambda }\lambda \gamma
_\Gamma (\lambda )}\Gamma _R^{(0)}(p_1,p_2;q_1,q_2)  \eqnum{8.24}
\end{equation}
where $\gamma _\Gamma (\lambda )$ is the anomalous dimension of the vertex
which is determined by the renomalization constant $Z_\Gamma =Z_2^{-2}$
(which is the inverse of the renormalization constant of the nucleon
four-point Green's function). According to Eq. (8.3), 
\begin{equation}
\gamma _\Gamma (\lambda )=\mu \frac d{d\mu }\ln Z_\Gamma =-2\mu \frac d{d\mu 
}\ln Z_2  \eqnum{8.25}
\end{equation}
Substituting Eqs. (8.13) and (8.24) into Eq. (8.23) and noticing Eqs. (8.15)
and (8.25), we also find that the anomalous dimensions are all cancelled
out. Thus, we have 
\begin{equation}
T_{fi}^{(2)}=\overline{u}_{R\sigma }^{(0)}(q_1)\overline{u}_{R\rho
}^{(0)}(q_2)\Gamma _R^{(0)}(p_1,p_2;q_1,q_2)u_{R\alpha }^{(0)}(p_1)u_{R\beta
}^{(0)}(p_2)  \eqnum{8.23}
\end{equation}
As mentioned in Eq. (8.8), the vertex $\Gamma _R^{(0)}(p_1,p_2;q_1,q_2)$ is
given at $p_{i0}^2=q_{i0}^2=\mu ^2$ ($i=1,2$) and the physical parameters in
it are all running ones. Since the unrenormalized vertex $\Gamma
(p_1,p_2;q_1,q_2)$ is not fundamental, it has a complicated structure,
containing a series of tree and loop diagrams [36]. The expression of the
vertex $\Gamma _R^{(0)}(p_1,p_2;q_1,q_2)$ can be determined by the
perturbation method. Unlike the loop-expansion, the perturbation series of
the S-matrix usually is expanded in powers of the coupling constant $g_v$.
The lowest order approximation of the vertex $\Gamma (p_1,p_2;q_1,q_2)$ is
of the order of $g^4$ and contains two terms which are given by the
truncated subdiagrams ( the box and crossed box diagrams) obtained from
Figs. (6e) and (6f) by amputating the external lines. The scattering
amplitude given by the tree diagrams in Figs. (6e) and (6f) and their
corresponding exchanged counterparts are convergent. We may dress these
diagrams by replacing the free wave functions, the free propagators and the
bare vertices with the exact wave functions, the full propagators and the
rigorous proper vertices. In this way, we obtain a series of loop diagrams.
As mentioned before, the dressed wave functions, propagators and vertices in
the S-matrix element can all be replaced by the renormalized ones.
Therefore, they can be expressed in the forms as given in Eqs. (8.13),
(8.16) and (8.19). Due to the cancellation of the anomalous dimensions, we
will obtain an expression of the renormalized scattering amplitude which is
formally the same as that written from the tree diagrams in Figs. (6e) and
(6f) and their exchanged ones. For this reason, the tree diagrams are called
skeletons of the dressed diagrams. There are a series of skeleton diagrams
(or called tree diagrams) of the vertex $\Gamma (p_1,p_2;q_1,q_2)$ such as
the ladder diagrams and some others. But, in practical calculations, it is
only feasible to consider the skeleton diagrams given in lower order
perturbative approximations. We would like to stress that the skeleton
diagrams can all be dressed. The S-matrix elements given by the dressed
diagrams can be written out from the corresponding skeleton (tree) diagrams
provided that the physical parameters are replaced by the solutions of their
RGEs. For other S-matrix elements representing other processes, the
conclusion is completely the same. It is noted that a S-matrix element
evaluated by the $\sigma -\omega $ model is independent of the gauge
parameter as illustrated in the appendix A in the one-loop approximation
(This is the so-called gauge-independence of S-matrix which is implied by
the unitarity of S-matrix elements). This fact indicates that the task of
renormalization for the $\sigma -\omega $ model is reduced to find the
effective coupling constants and the effective masses by solving their RGEs.
These effective quantities completely describe the effect of higher order
loop corrections. As an illustration, the effective coupling constants and
the effective masses given in the one-loop approximation will be derived and
discussed in detail in the next section.

\section{One-loop effective coupling constants and masses}

For the renornalization of the $\sigma -\omega $ model, we need to derive
the effective vectorial coupling constant, the effective scalar coupling
constant, the effective nucleon mass and the effective $\omega $ meson and $%
\sigma $ meson masses. The one-loop expressions of these effective
quantities will be derived and discussed in the following subsections.

\subsection{Effective vectorial coupling constant}

The RGE for the renormalized vectorial coupling constant $g_v^R$ which
appears in the vectorial vertex may be immediately written out from Eq.
(8.2) by setting ${F=g}_v${,} 
\begin{eqnarray}
\mu \frac d{d\mu }g_v^R(\mu )+\gamma _g^v(\mu )g_v^R(\mu )=0.  \eqnum{9.1}
\end{eqnarray}
In view of the definition shown in Eq. (8.1) and the relation in Eq. (6.32),
the renormalization constant defined in Eq. (8.12) will be represented as 
\begin{eqnarray}
Z_g^v=\frac{Z_1}{Z_2Z_3^{\frac 12}}=Z_3^{-\frac 12}  \eqnum{9.2}
\end{eqnarray}
According to the definition in Eq. (8.3), the anomalous dimension ${\gamma }%
_g^v{(\mu )}$ in Eq. (9.1) can be calculated by 
\begin{eqnarray}
\gamma _g^v=\lim_{\varepsilon \to 0}\mu \frac d{d\mu }\ln Z_g^v=-\frac 12%
\lim_{\varepsilon \to 0}\mu \frac d{d\mu }\ln Z_3  \eqnum{9.3}
\end{eqnarray}
where $\varepsilon =2-\frac n2$ with $n$ being the dimension of the space in
which the regularization is performed. Based on the definition denoted in
Eq. (5.19), the renormalization constant $Z_3$ will be given by the
subtraction of the $\omega $ meson vacuum polarization (self-energy)
operator $-i\Pi _{\mu \nu }(k)$. From the one-loop diagram represented in
Fig. (7), one may write [1, 29] 
\begin{equation}
\Pi _{\mu \nu }(k)=-2ig_v^2\int \frac{d^4l}{(2\pi )^4}Tr[\gamma _\mu \frac 1{%
{\bf l}-{\bf k}-M+i\epsilon }\gamma _\nu \frac 1{{\bf l}-M+i\epsilon }] 
\eqnum{9.4}
\end{equation}
where the factor $2$ comes from the fact that the fermion loop can be formed
by both of the proton and neutron loops. By the dimensional regularization
[28-30, 43], the divergent integral shown above is easily calculated. Then,
from the definitions in Eqs. (5.14) and (5.19) , it is not difficult to find
that in the n-dimensional space, the renormalization constant $Z_3$ is
expressed as 
\begin{equation}
\begin{tabular}{l}
$Z_3=1-\Pi (\mu ^2)$ \\ 
$=1+\frac{g_v^2}{2\pi ^2}(4\pi m_g^2)^\varepsilon (2-\varepsilon )\frac{%
\Gamma (1+\varepsilon )}\varepsilon \int_0^1\frac{dxx(x-1)}{[\mu
^2x(x-1)+M^2]^\varepsilon }$%
\end{tabular}
\eqnum{9.5}
\end{equation}
where $m_g$ is a mass introduced to make the coupling constant to be
dimensionless in the n-dimensional space. It is noted here that the factors $%
(4\pi m_g^2)^\varepsilon $ and $\Gamma (1+\varepsilon )$ may all be set to
unity because they do not give an effect on the anomalous dimension when we
set $\varepsilon \rightarrow 0$ in the final step of the calculation for the
anomalous dimension. Inserting Eq. (9.5) into Eq. (9.3), it can be found
that 
\begin{eqnarray}
\gamma _g^v=-\frac{g_v^2}{6\pi ^2}\{1+6\sigma ^2+\frac{12\sigma ^4}{\sqrt{%
1-4\sigma ^2}}\ln \frac{1+\sqrt{1-4\sigma ^2}}{1-\sqrt{1-4\sigma ^2}}\} 
\eqnum{9.6}
\end{eqnarray}
where $\sigma =\frac M\mu $. In this expression, the coupling constant $g_v$
and the nucleon mass $M$ are unrenormalized. In the approximation of order $%
g_v^2$, they can be replaced by the renormalized ones $g_v^R$and $M_R$
because in this approximation, as pointed out in the previous literature
[28], the lowest order approximation of the relation between the $g_v(M)$
and the $g_v^R(M_R)$ is only necessary to be taken into account.
Furthermore, when we introduce the scaling variable $\lambda $ defined by $%
\mu =\mu _0\lambda $ for the renormalization point and set $\mu _0=M_R$
(This can always be done since the $\mu _0$ is fixed, but may be chosen at
will. The above choice amounts to take the renormalization scale parameter
to be the nucleon mass), we have $\sigma =\frac{M_R}{\mu _0\lambda }=\frac 1%
\lambda $. Thus, with the $\gamma _g^v$ expressed in Eq. (9.6) and noticing $%
\mu \frac d{d\mu }=\lambda \frac d{d\lambda }$, Eq. (9.1) may be rewritten
in the form 
\begin{eqnarray}
\lambda \frac{dg_v^R(\lambda )}{d\lambda }=\frac{[g_v^R(\lambda )]^3}{6\pi ^2%
}F_g^v(\lambda )  \eqnum{9.7}
\end{eqnarray}
where 
\begin{eqnarray}
F_g^v(\lambda )=1+\frac 6{\lambda ^2}+\frac{12}{\lambda ^4}f(\lambda ) 
\eqnum{9.8}
\end{eqnarray}
in which 
\begin{equation}
\begin{tabular}{l}
$f(\lambda )=\frac \lambda {\sqrt{\lambda ^2-4}}\ln \frac{\lambda +\sqrt{%
\lambda ^2-4}}{\lambda -\sqrt{\lambda ^2-4}}$ \\ 
$=\{ 
\begin{array}{c}
\frac{2\lambda }{\sqrt{4-\lambda ^2}}\cot ^{-1}\frac \lambda {\sqrt{%
4-\lambda ^2}},\text{ }if\text{ }\lambda \leq 2, \\ 
\frac{2\lambda }{\sqrt{\lambda ^2-4}}\coth ^{-1}\frac \lambda {\sqrt{\lambda
^2-4}},\text{ }if\text{ }\lambda \geq 2.
\end{array}
$%
\end{tabular}
\eqnum{9.9}
\end{equation}
Upon substituting Eqs. (9.8) and (9.9) into Eq. (9.7) and then integrating
Eq. (9.7) by applying the familiar integration formulas, the effective
(running) coupling constant will be found to be 
\begin{eqnarray}
\alpha _R^v(\lambda )=\frac{\alpha _R^v}{1-\frac{4\alpha _R^v}{3\pi }%
G_v(\lambda )}  \eqnum{9.10}
\end{eqnarray}
where ${\alpha _R^v(\lambda )=[g}_R^v(\lambda )]^2/4\pi $, $\alpha
_R^v=\alpha _R^v(1)$ which is the coupling constant that should be
determined by experiment and 
\begin{equation}
\begin{tabular}{l}
$G_v(\lambda )=\int_1^\lambda \frac{d\lambda }\lambda F_g^v(\lambda )$ \\ 
$=2+\sqrt{3}\pi -\frac 2{\lambda ^2}+(1+\frac 2{\lambda ^2})\frac 1\lambda
\varphi (\lambda )$%
\end{tabular}
\eqnum{9.11}
\end{equation}
in which 
\begin{equation}
\begin{array}{c}
\varphi (\lambda )=\sqrt{\lambda ^2-4}\ln \frac 12(\lambda +\sqrt{\lambda
^2-4}) \\ 
=\{ 
\begin{array}{c}
-\sqrt{4-\lambda ^2}\cos ^{-1}\frac \lambda 2,\text{ }if\text{ }\lambda \leq
2, \\ 
\sqrt{\lambda ^2-4}\cosh ^{-1}\frac \lambda 2,\text{ }if\text{ }\lambda \geq
2.
\end{array}
\end{array}
\eqnum{9.12}
\end{equation}

As mentioned before, the variable $\lambda $ is also the scaling parameter
of momenta, $p=\lambda p_0$, and it is convenient to put ${p_0}^2={\mu _0}^2$
so as to apply the boundary condition. Thus, owing to the choice $\mu _0=M_R$%
, we have ${p_0}^2={M_R}^2$ and $\lambda =(\frac{p^2}{M_R^2})^{\frac 12}$.
In this case, it is apparent that when $\lambda =1$, Eq. (9.10) will be
reduced to the result given on the nucleon mass shell with the value $\alpha
_R^v(1)=\alpha _R^v$. The behaviors of the effective coupling constants
obtained in the timelike and spacelike momentum space subtractions are
separately represented in Figs. (8) and (9). The figures show that the
effective coupling constant ${\alpha _R^v(\lambda )}$ has a singularity $%
\lambda _0$ (the Landau pole). The position of the pole strongly depends on
the parameter $\alpha _R^v$. By our numerical test, we find, when the $%
\alpha _R^v$ is getting smaller and smaller, the $\lambda _0$ is getting
larger and larger. If the $\alpha _R^v$ goes to zero, the $\lambda _0$
approaches a value near infinity, similar to the case of QED. While, when
the $\alpha _R^v$ is getting larger and larger, the $\lambda _0$ moves
toward unity; but it can not arrive at unity because $\alpha _R^v(1)=\alpha
_R^v$. In the region [0,1] of $\lambda $, the ${\alpha _R^v(\lambda )}$ has
no singularity to appear. In Fig. (8), there are two lines representing the $%
{\alpha _R^v(\lambda )}$ given in the timelike momentum subtraction: one is
given by taking $\alpha _R^v=1$ and has a singularity at $\lambda \simeq
1.1385$; another is obtained by taking $\alpha _R^v=0.5$ and has a
singularity at $\lambda \simeq 1.3885$. When $\lambda $ goes from $\lambda
_0 $ to zero, the ${\alpha _R^v(\lambda )}$ decreases and tends to zero,
exhibiting an asymptotically free behavior as we met in QED. In Fig. (9),
the two lines represent the ${\alpha _R^v(\lambda )}$ given in the spacelike
momentum subtraction: one line is obtained by taking $\alpha _R^v=1$ and has
a singularity at $\lambda _0\simeq 26.4689$; another is given by $\alpha
_R^v=$ $0.5$ and has a singularity at $\lambda _0\simeq 280.431$. When $%
\lambda $ goes to zero, the ${\alpha _R^v(\lambda )}$ {approaches the
constant} ${\alpha _R^v}$. As one knows, the Landau poles mentioned above
give a limitation of applicability of the one loop renormalization. That is
to say, beyond the region [$0,\lambda _0]$, the ${\alpha _R^v(\lambda )}$ is
meaningless even though in the limit: $\lambda \rightarrow \infty $, the ${%
\alpha _R^v(\lambda )}$ tends to zero from an opposite direction. In
comparison of Fig. (8) with Fig. (9), it is clear to see that the range of
applicability for the ${\alpha _R^v(\lambda )}$ given in the spacelike
momentum subtraction is much larger than the range for the ${\alpha
_R^v(\lambda )}$ given in the timelike momentum subtraction.

\subsection{Effective $\omega $ meson mass}

In Eq. (8.2), when we set $F=m_\omega $, we have a RGE for the renormalized
mass of $\omega $ meson such that 
\begin{eqnarray}
\mu \frac d{d\mu }m_\omega ^R(\mu )+\gamma _m^\omega (\mu )m_\omega ^R(\mu
)=0  \eqnum{9.13}
\end{eqnarray}
From the definitions given in Eq. (8.1) and the first equality in Eq.
(5.22), we find 
\begin{equation}
Z_m^\omega =Z_3^{-\frac 12}=Z_g^v  \eqnum{9.14}
\end{equation}
so that 
\begin{equation}
\gamma _m^\omega (\mu )=\gamma _g^v(\mu )=-\frac{[g_v^R(\lambda )]^2}{6\pi ^2%
}F_g^v(\lambda )  \eqnum{9.15}
\end{equation}
where $F_g^v(\lambda )$ was given in Eq. (9.8). With the above expression,
Eq. (9.13) can be written as 
\begin{equation}
\frac{dm_\omega ^R}{m_\omega ^R}=\frac{2\alpha _R^v(\lambda )}{3\pi }%
F_g^v(\lambda )\frac{d\lambda }\lambda   \eqnum{9.16}
\end{equation}
Integrating the above equation, one gets 
\begin{equation}
m_\omega ^R(\lambda )=m_\omega ^Re^{\frac 2{3\pi }\int_1^\lambda \frac{%
d\lambda }\lambda \alpha _R^v(\lambda )F_g^v(\lambda )}  \eqnum{9.17}
\end{equation}
where $m_\omega ^R=m_\omega ^R(1)$ is the observed $\omega $ meson mass.
This is just the one-loop result of the effective $\omega $ meson mass. If
we take the approximation $\alpha _R^v(\lambda )\simeq \alpha _R^v$, the
above expression becomes 
\begin{equation}
m_\omega ^R(\lambda )\simeq m_\omega ^Re^{\frac 2{3\pi }G_v(\lambda )} 
\eqnum{9.18}
\end{equation}
where $G_v(\lambda )$ was given in Eq. (9.11).

To have an insight into the behavior of the effective masses in Eq. (9.17),
we take the $m_\omega ^R(\lambda )$ given by taking $\alpha _R^v=1$ as an
example. This $m_\omega ^R(\lambda )$ is shown in Fig. (10). In the figure,
the solid line represents the effective mass obtained in the spacelike
momentum subtraction and the dashed line describes the one given in the
timelike momentum subtraction. Comparing Fig. (10) with Figs. ( 8) and (9),
we see that the both effective masses have the same singularities and the
same scopes of applicability as the corresponding effective coupling
constants $\alpha _R^v(\lambda )$. Particularly, the position of the
singularity strongly depends on the choice of $\alpha _R^v$ as the $\alpha
_R^v(\lambda )$ does. When $\lambda $ tends to zero, the $m_\omega
^R(\lambda )$ for the spacelike momentum approaches a nonvanishing value
near the $m_\omega ^R$, while, the $m_\omega ^R(\lambda )$ for the timelike
momentum goes to zero.

\subsection{Effective nucleon mass}

The RGE for the renormalized nucleon mass, according to Eq. (8.2), can be
written as 
\begin{eqnarray}
\mu \frac d{d\mu }M_R(\mu )+\gamma _M(\mu )M_R(\mu )=0  \eqnum{9.19}
\end{eqnarray}
where $\gamma _M(\mu )$ is the anomalous dimension of nucleon mass. In view
of Eqs. (6.25) and (8.3), we have 
\begin{equation}
\gamma _M(\mu )=\mu \frac d{d\mu }\ln Z_M=-\mu \frac d{d\mu }\ln \{1+Z_2[%
\frac AM+(1-\frac \mu M)B]\}  \eqnum{9.20}
\end{equation}
To determine the constants $A,B$ and $Z_2$ in the one-loop approximation, it
is necessary to compute the nucleon self-energy written in Eq. (6.35). By
the dimensional regularization, it is not difficult to derive from Eq.
(6.35) the following expression 
\begin{equation}
\Sigma (p)=({\bf p}-\mu )\Sigma _1(p)+\Sigma _2(p)  \eqnum{9.21}
\end{equation}
where 
\begin{equation}
\begin{tabular}{l}
$\Sigma _1(p)=\frac{g_v^2}{(4\pi )^2}\int_0^1dx\frac{2(x-1)}{\varepsilon
\Theta _v(x)^\varepsilon }+(1-\alpha )\frac{g_v^2}{(4\pi )^2}%
\{\int_0^1dx\int_0^1dyy\{\frac{(1+3xy)}{\varepsilon \Theta
_v(x,y)^\varepsilon }+$ \\ 
$\frac 1{\Theta _v(x,y)}x^2y^2[(1-xy)[p^2+\mu ({\bf p}+\mu )]+M({\bf p}+\mu
)]\}-\frac 12\}$ \\ 
$+\frac{g_s^2}{(4\pi )^2}\int_0^1dx\frac{x-1}{\varepsilon \Theta
_s(x)^\varepsilon }$%
\end{tabular}
\eqnum{9.22}
\end{equation}
and 
\begin{equation}
\begin{tabular}{l}
$\Sigma _2(p)=\frac{g_v^2}{(4\pi )^2}\int_0^1dx\frac{2[(x-1)\mu +2M]}{%
\varepsilon \Theta _v(x)^\varepsilon }+(1-\alpha )\frac{g_v^2}{(4\pi )^2}%
\{\int_0^1dx\int_0^1dyy$ \\ 
$\times \{\frac{(1+3xy)\mu -2M}{\varepsilon \Theta _v(x,y)^\varepsilon }+%
\frac 1{\Theta _v(x,y)}\mu ^2x^2y^2[(1-xy)\mu +M]\}+\frac 12(M-\mu )\}$ \\ 
$+\frac{g_s^2}{(4\pi )^2}\int_0^1dx\frac{(x-1)\mu -M}{\varepsilon \Theta
_s(x)^\varepsilon }$%
\end{tabular}
\eqnum{9.23}
\end{equation}
where 
\begin{equation}
\begin{tabular}{l}
$\Theta _v(x)=p^2x(x-1)+M^2x+m_\omega ^2(1-x)$ \\ 
$\Theta _v(x,y)=p^2xy(xy-1)+M^2xy+m_\omega ^2[(1-x)y+\alpha (1-y)]$ \\ 
$\Theta _s(x)=p^2x(x-1)+M^2x+m_\sigma ^2(1-x).$%
\end{tabular}
\eqnum{9.24}
\end{equation}
From the definition $A=\Sigma (\mu )$ written in Eq. (6.17) and the
expressions in Eqs, (9.21)-(9.23), we find 
\begin{equation}
A=A_1+A_2+A_3  \eqnum{9.25}
\end{equation}
where 
\begin{equation}
A_1=\frac{g_v^2}{(4\pi )^2}\int_0^1dx\frac{2[(x-1)\mu +2M]}{\varepsilon
\Omega _v(x)^\varepsilon },  \eqnum{9.26}
\end{equation}
\begin{equation}
\begin{tabular}{l}
$A_2=(1-\alpha )\frac{g_v^2}{(4\pi )^2}\{\int_0^1dx\int_0^1dyy\{\frac{%
(1+3xy)\mu -2M}{\varepsilon \Omega _v(x,y)^\varepsilon }$ \\ 
$+\frac 1{\Omega _v(x,y)}x^2y^2[(1-xy)\mu ^3+M\mu ^2]\}+\frac 12(M-\mu )\}$%
\end{tabular}
\eqnum{9.27}
\end{equation}
and 
\begin{equation}
A_3=\frac{g_s^2}{(4\pi )^2}\int_0^1dx\frac{(x-1)\mu -M}{\varepsilon \Omega
_s(x)^\varepsilon }  \eqnum{9.28}
\end{equation}
in which 
\begin{equation}
\begin{tabular}{l}
$\Omega _v(x)=\mu ^2x(x-1)+M^2x+m_\omega ^2(1-x)$ \\ 
$\Omega _v(x,y)=\mu ^2xy(xy-1)+M^2xy+m_\omega ^2[(1-x)y+\alpha (1-y)]$ \\ 
$\Omega _s(x)=\mu ^2x(x-1)+M^2x+m_\sigma ^2(1-x).$%
\end{tabular}
\eqnum{9.29}
\end{equation}
The constant $B$ appearing in Eqs. (6.21) and (6.25), according to Eq.
(6.19), ought to be computed by 
\begin{equation}
B=({\bf p-}\mu )^{-1}[\Sigma (p)-A]\mid _{{\bf p=}\mu }.  \eqnum{9.30}
\end{equation}
On inserting Eqs. (9.21)-(9.29) into Eq. (9.30) and employing the formula 
\begin{equation}
\frac 1{a^\varepsilon }-\frac 1{b^\varepsilon }=\int_0^1dx\frac{\varepsilon
(b-a)}{[ax+b(1-x)]^{1+\varepsilon }},  \eqnum{9.31}
\end{equation}
one may derive 
\begin{equation}
B=B_1+B_2+B_3  \eqnum{9.32}
\end{equation}
where 
\begin{equation}
B_1=\frac{g_v^2}{(4\pi )^2}\int_0^1dx\{\frac{2(x-1)}{\varepsilon \Omega
_v(x)^\varepsilon }-\frac 4{\Omega _v(x)}x(x-1)[(x-1)\mu ^2+2M\mu ]\}, 
\eqnum{9.33}
\end{equation}
\begin{equation}
\begin{tabular}{l}
$B_2=(1-\alpha )\frac{g_v^2}{(4\pi )^2}\{\int_0^1dx\int_0^1dyy\{\frac{1+3xy}{%
\varepsilon \Omega _v(x)^\varepsilon }+\frac 1{\Omega _v(x)}xy[\mu
^2(2+7xy-9x^2y^2)$ \\ 
$-2M\mu (2-3xy)]+\frac 2{\Omega _v(x)^2}x^3y^3(1-xy)[(1-xy)\mu ^4+M\mu ^3]\}-%
\frac 12\}$%
\end{tabular}
\eqnum{9.34}
\end{equation}
and 
\begin{equation}
B_3=\frac{g_s^2}{(4\pi )^2}\int_0^1dx\{\frac{(x-1)}{\varepsilon \Omega
_s(x)^\varepsilon }-\frac 2{\Omega _s(x)}x(x-1)[(x-1)\mu ^2-M\mu ]\}. 
\eqnum{9.35}
\end{equation}
The terms expressed in Eqs. (9.27) and (9.34) are dependent on the gauge
parameter $\alpha $ and look more complicated. However, as demonstrated in
the Appendix A, the S-matrix elements evaluated in the $\sigma -\omega $
model are gauge-independent. Therefore, for simplicity, these terms will not
be taken into account later on. This means that we limit ourself to working
in the Feynman gauge.

When Eqs. (9.25) and (9.32) with the expressions given in Eqs. (9.26),
(9.28), (9.33) and (9.35) are substituted into Eq. (9.20), noticing that $%
Z_2\simeq 1$ should be taken in Eq. (9.20) in the approximation of order $%
g_i^2$, one may find an explicit expression of the anomalous dimension $%
\gamma _M(\mu )$ through a tedious calculation 
\begin{equation}
\gamma _M(\lambda )=\gamma _M^{(1)}(\lambda )+\gamma _M^{(2)}(\lambda ) 
\eqnum{9.36}
\end{equation}
where $\gamma _M^{(1)}(\lambda )$ is derived from the constants in Eqs.
(9.26) and (9.33), while $\gamma _M^{(2)}(\lambda )$ is given by the
constants in Eqs. (9.28) and (9 35). The expressions of the $\gamma
_M^{(1)}(\lambda )$ and $\gamma _M^{(2)}(\lambda )$ are separately described
in the following. For the anomalous dimension $\gamma _M^{(1)}(\lambda )$,
we have 
\begin{equation}
\gamma _M^{(1)}(\lambda )=\frac{\alpha _R^v}\pi \Sigma _v(\lambda ) 
\eqnum{9.37}
\end{equation}
in which 
\begin{equation}
\Sigma _v(\lambda )=\xi _0^v(\lambda )+\sum_{i=1}^4\xi _i^v(\lambda
)J_i(\lambda )  \eqnum{9.38}
\end{equation}
where the functions $\xi _i^v(\lambda )$ are 
\begin{equation}
\xi _0^v(\lambda )=\frac 12(3+\lambda )-\frac 2\lambda +\frac{1-\beta ^2}{%
\lambda ^2}(1-\lambda ),  \eqnum{9.39}
\end{equation}
\begin{equation}
\begin{array}{c}
\xi _1^v(\lambda )=\frac 2{\lambda ^4}[2(1-\beta ^2)^2-3\beta ^2\lambda
^2](1-\frac 1\lambda ) \\ 
+\frac 1{\lambda ^6}[2(1-\beta ^2)^3-5\beta ^2(1-\beta ^2)\lambda ^2-\beta
^2\lambda ^4](1-\lambda ),
\end{array}
\eqnum{9.40}
\end{equation}
\begin{equation}
\begin{array}{c}
\xi _2^v(\lambda )=\frac 1{\lambda ^4}[\beta ^2\lambda ^2-3(1-\beta
^2)^2](1-\lambda )-\frac{6(1-\beta ^2)}{\lambda ^2}(1-\frac 1\lambda ),
\end{array}
\eqnum{9.41}
\end{equation}
\begin{equation}
\begin{array}{c}
\xi _3^v(\lambda )=\frac{4\beta ^2}{\lambda ^6}[\beta ^2\lambda ^2-(1-\beta
^2)^2](1-\frac 1\lambda ) \\ 
+\frac{2\beta ^2}{\lambda ^8}[\beta ^2\lambda ^4+2(1-\beta ^2)\beta
^2\lambda ^2-(1-\beta ^2)^3](1-\lambda )
\end{array}
\eqnum{9.42}
\end{equation}
and 
\begin{equation}
\begin{array}{c}
\xi _4^v(\lambda )=\frac 4{\lambda ^6}[(1-\beta ^4)\lambda ^2-(1-\beta
^2)^3](1-\frac 1\lambda ) \\ 
-\frac 2{\lambda ^8}[(1-\beta ^2)^4-(1-\beta ^2)^2(1+2\beta ^2)\lambda
^2+\beta ^4\lambda ^4](1-\lambda )
\end{array}
\eqnum{9.43}
\end{equation}
with $\beta =m_\omega /M$ and the functions $J_i(\lambda )$ are given by the
integrals shown below. With defining $a=(1-\beta ^2)\lambda ^{-2}$ and $%
b=\beta ^2\lambda ^{-2}$, we can write 
\begin{equation}
\begin{array}{c}
J_1(\lambda )=\int_0^1dx\frac 1{x(x-1)+ax+b}=\frac{\lambda ^2}{\sqrt{%
q(\lambda )}}\ln \frac{\lambda ^2-1-\beta ^2-\sqrt{q(\lambda )}}{\lambda
^2-1-\beta ^2+\sqrt{q(\lambda )}}
\end{array}
\eqnum{9.44}
\end{equation}
where 
\begin{equation}
q(\lambda )=\lambda ^4-2(1+\beta ^2)\lambda ^2+(1-\beta ^2)^2,  \eqnum{9.45}
\end{equation}
\begin{equation}
\begin{array}{c}
J_2(\lambda )=\int_0^1dx\frac x{x(x-1)+ax+b}=-\ln \beta +\frac{\lambda
^2-1+\beta ^2}{2\sqrt{q(\lambda )}}\ln \frac{\lambda ^2-1-\beta ^2-\sqrt{%
q(\lambda )}}{\lambda ^2-1-\beta ^2+\sqrt{q(\lambda )}},
\end{array}
\eqnum{9.46}
\end{equation}
\begin{equation}
\begin{array}{c}
J_3(\lambda )=\int_0^1dx\frac 1{[x(x-1)+ax+b]^2} \\ 
=[(1-\beta ^2)^2-(1+\beta ^2)\lambda ^2]\frac{\lambda ^4}{\beta ^2q(\lambda )%
}-\frac{2\lambda ^6}{q(\lambda )^{3/2}}\ln \frac{\lambda ^2-1-\beta ^2-\sqrt{%
q(\lambda )}}{\lambda ^2-1-\beta ^2+\sqrt{q(\lambda )}}
\end{array}
\eqnum{9.47}
\end{equation}
and 
\begin{equation}
\begin{array}{c}
J_4(\lambda )=\int_0^1dx\frac x{[x(x-1)+ax+b]^2} \\ 
=-(\lambda ^2+1-\beta ^2)\frac{\lambda ^4}{q(\lambda )}-(\lambda ^2-1+\beta
^2)\frac{\lambda ^4}{q(\lambda )^{3/2}}\ln \frac{\lambda ^2-1-\beta ^2-\sqrt{%
q(\lambda )}}{\lambda ^2-1-\beta ^2+\sqrt{q(\lambda )}}.
\end{array}
\eqnum{9.48}
\end{equation}

For the anomalous dimension $\gamma _M^{(2)}(\lambda )$, we can write 
\begin{equation}
\gamma _M^{(2)}(\lambda )=\frac{\alpha _R^s}{2\pi }\Sigma _s(\lambda ) 
\eqnum{9.49}
\end{equation}
in which $\alpha _R^s=(g_s^R)^2/4\pi $ and 
\begin{equation}
\Sigma _s(\lambda )=\xi _0^s(\lambda )+\sum_{i=1}^4\xi _i^s(\lambda
)J_i^0(\lambda )  \eqnum{9.50}
\end{equation}
where the functions $\xi _i^s(\lambda )$ are 
\begin{equation}
\xi _0^s(\lambda )=-\frac 32+\frac \lambda 2+\frac 1{\lambda ^2}-\frac{\beta
_0^2}{\lambda ^2}(1-\lambda )  \eqnum{9.51}
\end{equation}
with $\beta _0=m_\sigma /M$, 
\begin{equation}
\begin{array}{c}
\xi _1^s(\lambda )=\frac 1{\lambda ^6}[2(1-\beta _0^2)^3-5\beta _0^2(1-\beta
_0^2)\lambda ^2-\beta _0^2\lambda ^4](1-\lambda ) \\ 
-\frac 1{\lambda ^4}[2(1-\beta _0^2)^2-3\beta _0^2\lambda ^2](1-\frac 1%
\lambda )
\end{array}
\eqnum{9.52}
\end{equation}
\begin{equation}
\begin{array}{c}
\xi _2^s(\lambda )=\frac 1{\lambda ^4}[\beta _0^2\lambda ^2-3(1-\beta
_0^2)^2](1-\lambda )+\frac{3(1-\beta _0^2)}{\lambda ^2}(1-\frac 1\lambda ),
\end{array}
\eqnum{9.53}
\end{equation}

\begin{equation}
\begin{array}{c}
\xi _3^s(\lambda )=\frac{2\beta _0^2}{\lambda ^6}[(1-\beta _0^2)^2-\beta
_0^2\lambda ^2](1-\frac 1\lambda ) \\ 
+\frac{2\beta _0^2}{\lambda ^8}[\beta _0^2\lambda ^4+2(1-\beta _0^2)\beta
_0^2\lambda ^2-(1-\beta _0^2)^3](1-\lambda )
\end{array}
\eqnum{9.54}
\end{equation}
and 
\begin{equation}
\begin{array}{c}
\xi _4^s(\lambda )=\frac 2{\lambda ^6}[(1-\beta _0^2)^3-(1-\beta
_0^4)\lambda ^2](1-\frac 1\lambda ) \\ 
-\frac 2{\lambda ^8}[(1-\beta _0^2)^4-(1-\beta _0^2)^2(1+2\beta _0^2)\lambda
^2+\beta _0^4\lambda ^4](1-\lambda )
\end{array}
\eqnum{9.55}
\end{equation}
and the functions $J_i^0(\lambda )$ formally are the same as the functions $%
J_i(\lambda )$ except that the parameter $\beta $ in the $J_i(\lambda )$ is
now replaced by $\beta _0$, 
\begin{equation}
J_i^0(\lambda )=J_i(\lambda )\mid _{\beta \rightarrow \beta _0}  \eqnum{9.56}
\end{equation}
Substituting the $\gamma _M(\lambda )$ as expressed in Eqs. (9.36)-(9.56)
into Eq. (9.19) and solving the equation with noticing $\mu d/\mu =\lambda
d/\lambda $, we obtain 
\begin{equation}
M_R(\lambda )=M_Re^{-\int_1^\lambda d\lambda /\lambda \gamma _M(\lambda
)}=M_Re^{-\int_1^\lambda d\lambda /\lambda [\alpha _R^v(\lambda )/\pi \Sigma
_v(\lambda )+\alpha _R^s(\lambda )/2\pi \Sigma _s(\lambda )]}  \eqnum{9.57}
\end{equation}
where $M_R=M_R(1)$ is the observed nucleon mass. The coupling constants in
the above have been taken to be running ones. The $\alpha _R^v(\lambda )$
was given in Eq. (9.10), while, the $\alpha _R^s(\lambda )$ will be derived
in the next subsection.

It would be emphasized that in the timelike momentum space subtraction, the
scaling parameter $\lambda $ is real so that the effective nucleon mass is
real, while, in the spacelike momentum space subtraction, the $\lambda $ is
imaginary so that the effective nucleon mass becomes complex one. In the
latter case, the $\lambda $ in the $\gamma _M(\lambda )$ should be set to be 
$i\lambda .$ Observing the expressions in Eqs. (9.39)-(4.48) and
(8.51)-(9.55), we see that in the both subtractions, the functions $%
J_i(\lambda )$ are always real. The real and imaginary parts of the $\gamma
_M(\lambda )$ are distinguished by the real and imaginary parts of the
functions $\xi _i^v(\lambda )$ and $\xi _i^s(\lambda )$. In Figs. (11)-(12),
we show the behaviors of the effective nucleon masses given by the
expression in Eq. (9.57) for which the coupling constants be taken as the
constants $\alpha _R^v$ and $\alpha _R^s$ for simplicity of computation. The
effective mass $M_R(\lambda )$ obtained in the timelike momentum space
subtraction is exhibited in Fig. (11). In the figure, the dashed, dotted and
solid lines represent the $M_R(\lambda )$ given by taking ($\alpha
_R^v,\alpha _R^s)=(0,0.5)$, $(0.5,0)$ and $(0.5,0.5)$ respectively. From the
figure, we see that in the region [0,1] of $\lambda $, the $M_R(\lambda )$
almost keeps a constant $M_R$. Beyond this region, if only the scalar
coupling is considered, the $M_R(\lambda )$ increases up to a maximum at $%
\lambda _0=4.21$ and then decreases to zero rather rapidly when $\lambda
\rightarrow \infty $. While, if the vectorial coupling enters, the $%
M_R(\lambda )$ increases a little when $\lambda $ goes to a smaller $\lambda
_0$ and then decreases much rapidly down to zero when $\lambda $ varies from 
$\lambda _0$ to $\infty $. The effective mass $M_R(\lambda )$ obtained in
the spacelike momentum space subtraction is shown in Fig. (12). In the
figure, the real and imaginary parts of the $M_R(\lambda )$ are displayed
separately. The dashed, solid and dotted lines in the figure represent the
real and imaginary parts of the $M_R(\lambda )$ which are obtained by taking 
$(\alpha _R^v,\alpha _R^s)=(0.5,0)$, $(0.5,0.5)$ and $(1,1)$ respectively.
The figure indicates that in the region [0,1] of $\lambda $, the real part
of the $M_R(\lambda )$ keeps almost a constant equal to $M_R$, while, the
imaginary part of the $M_R(\lambda )$ is almost zero. When $\lambda $ varies
from unity to infinity, the real part of the $M_R(\lambda )$ at first
increases smoothly and the imaginary part of the $M_R(\lambda )$ decreases,
then, both of them drastically oscillate and damp to zero. The figure also
shows that the stronger the couplings (especially, the vectorial coupling),
the larger is the frequency of the oscillation. The appearance of the
oscillation implies that the $M_R(\lambda )$ is invalid to use in the region
that the oscillation appears.

\subsection{Effective scalar coupling constant}

When setting ${F=g}_s$ in Eq. (8.2), one obtains the RGE for the
renormalized scalar coupling constant $g_s^R$ which is included in the
scalar (nucleon-nucleon--$\sigma $ meson) vertex 
\begin{eqnarray}
\mu \frac d{d\mu }g_s^R(\mu )+\gamma _g^s(\mu )g_s^R(\mu )=0  \eqnum{9.58}
\end{eqnarray}
Analogous to the case of vectorial coupling, the anomalous dimension ${%
\gamma }_g^v{(\mu )}$ determined by

\begin{equation}
\gamma _g^s=\lim_{\varepsilon \to 0}\mu \frac d{d\mu }\ln Z_g^s  \eqnum{9.59}
\end{equation}
should be calculated from the renormalization constant $Z_g^s$ which is
represented as

\begin{eqnarray}
Z_g^s=\frac{Z_1^{\prime }}{Z_2Z_3^{\prime \frac 12}}  \eqnum{9.60}
\end{eqnarray}
where $Z_2$, $Z_3^{\prime }$ and $Z_1^{\prime }$ were defined respectively
in Eqs. (6.21), (7.6) and (7.17). According to these definitions, in the
approximation of order $g^2$ and in the Feynman gauge, The $Z_g^s$ will be
written as 
\begin{equation}
Z_g^s=1+B_1+B_3-L_1^{\prime }-L_3^{\prime }+\Omega _1  \eqnum{9.61}
\end{equation}
where $B_1$ and $B_3$ were represented in Eqs. (9.33) and (9.35)
respectively, $L_1^{\prime }$ and $L_3^{\prime }$ are the parts of the
constant $L^{\prime }$ which can conveniently be determined by the identity
in Eq. (7.22). From the identity and the representations written in Eqs.
(9.26) and (9.28), it is easy to get 
\begin{equation}
\begin{array}{c}
L_1^{\prime }=\frac{\partial A_1}{\partial M}=\frac{g_v^2}{4\pi ^2}%
\int_0^1dx\{\frac 1{\varepsilon \Omega _v(x)^\varepsilon }-[x(x-1)\mu +2Mx]%
\frac M{\Omega _v(x)}\} \\ 
L_3^{\prime }=\frac{\partial A_3}{\partial M}=-\frac{g_s^2}{16\pi ^2}%
\int_0^1dx\{\frac 1{\varepsilon \Omega _s(x)^\varepsilon }+[x(x-1)\mu -Mx]%
\frac{2M}{\Omega _s(x)}\}
\end{array}
\eqnum{9.62}
\end{equation}
The one-loop expression of the divergent constant $\Omega _1(\mu ^2)$ in Eq.
(9.61) can be derived from the $\sigma $ meson one-loop self-energy depicted
in Fig. (13). From Fig. (13), it reads 
\begin{equation}
\Omega (q)=-2ig_s^2\int \frac{d^4l}{(2\pi )^4}Tr[\frac 1{({\bf l}-{\bf q}%
-M+i\varepsilon )}\frac 1{({\bf q}-M+i\varepsilon )}]  \eqnum{9.63}
\end{equation}
where the factor 2 also arises from nucleon doublet. By the dimensional
regularization, the above integral is easily calculated and expressed as 
\begin{equation}
\begin{tabular}{l}
$\Omega (q)=\frac{g_s^2}{(4\pi )^2}\int_0^1dx\frac{24[q^2x(x-1)+M^2]}{%
\varepsilon [q^2x(x-1)+M^2]^\varepsilon }$ \\ 
$=\Omega _1(q^2)q^2+\Omega _2(q^2)m_\sigma ^2.$%
\end{tabular}
\eqnum{9.64}
\end{equation}
which gives rise to 
\begin{equation}
\Omega _1(\mu ^2)=\frac{g_s^2}{(4\pi )^2}\int_0^1dx\frac{24x(x-1)}{%
\varepsilon [\mu ^2x(x-1)+M^2]^\varepsilon }  \eqnum{9.65}
\end{equation}
and 
\begin{equation}
\Omega _2(\mu ^2)=\frac{g_s^2}{(4\pi )^2}\int_0^1dx\frac{24M^2/m_\sigma ^2}{%
\varepsilon [\mu ^2x(x-1)+M^2]^\varepsilon }.  \eqnum{9.66}
\end{equation}
On substituting Eq. (9.61) into Eq. (9.59), we get 
\begin{equation}
\gamma _g^s(\lambda )=\gamma _{g1}^s(\lambda )+\gamma _{g2}^s(\lambda
)+\gamma _{g3}^s(\lambda )  \eqnum{9.67}
\end{equation}
where $\gamma _{g1}^s(\lambda )$, $\gamma _{g2}^s(\lambda )$ and $\gamma
_{g3}^s(\lambda )$ are separately defined and described in the following.

For the $\gamma _{g1}^s(\lambda ),$ we have 
\begin{equation}
\gamma _{g1}^s(\lambda )=\mu \frac d{d\mu }(B_1-L_1^{\prime })=\frac{\alpha
_v^R}\pi \Gamma ^v(\lambda )  \eqnum{9.68}
\end{equation}
where 
\begin{equation}
\Gamma ^v(\lambda )=\eta _0^v(\lambda )+\sum_{i=1}^4\eta _i^v(\lambda
)J_i(\lambda )  \eqnum{9.69}
\end{equation}
in which 
\begin{equation}
\begin{array}{c}
\eta _0^v(\lambda )=\frac 32-\frac 3\lambda +\frac{1-\beta ^2}{\lambda ^2}
\\ 
\eta _1^v(\lambda )=\frac 1{\lambda ^6}[2(1-\beta ^2)^3-6(1-\beta
^2)^2\lambda +(1-\beta ^2)(4 \\ 
-5\beta ^2)\lambda ^2+9\beta ^2\lambda ^3-3\beta ^2\lambda ^4] \\ 
\eta _2^v(\lambda )=-\frac 3{\lambda ^4}[(1-\beta ^2)^2-3(1-\beta ^2)\lambda
+(2-\beta ^2)\lambda ^2] \\ 
\eta _3^v(\lambda )=\frac{2\beta ^2}{\lambda ^8}[\beta ^2\lambda ^4-3\beta
^2\lambda ^3-2(1-\beta ^2)^2\lambda ^2 \\ 
+3(1-\beta ^2)^2\lambda -(1-\beta ^2)^3] \\ 
\eta _4^v(\lambda )=\frac 2{\lambda ^8}[(2-\beta ^2)\lambda ^4-3(1-\beta
^4)\lambda ^3-(1-\beta ^2)^2(1 \\ 
-2\beta ^2)\lambda ^2+3(1-\beta ^2)^3\lambda -(1-\beta ^2)^4]
\end{array}
\eqnum{9.70}
\end{equation}
and the functions $J_i(\lambda )$ were given in Eqs. (9.44)-(9.48).

For the $\gamma _{g2}^s(\lambda ),$ we can write 
\begin{equation}
\gamma _{g2}^s(\lambda )=\mu \frac d{d\mu }(B_3-L_3^{\prime })=\frac{%
(g_s^R)^2}{8\pi ^2}\Gamma _1^s(\lambda )  \eqnum{9.71}
\end{equation}
where 
\begin{equation}
\Gamma _1^s(\lambda )=\eta _0^s(\lambda )+\sum_{i=1}^4\eta _i^s(\lambda
)J_i^0(\lambda )  \eqnum{9.72}
\end{equation}
in which 
\begin{equation}
\begin{array}{c}
\eta _0^s(\lambda )=-\frac 32+\frac{1-\beta _0^2}{\lambda ^2} \\ 
\eta _1^s(\lambda )=\frac 1{\lambda ^6}[2(1-\beta _0^2)^3-(1-\beta
_0^2)(2+5\beta _0^2)\lambda ^2] \\ 
\eta _2^s(\lambda )=\frac 3{\lambda ^4}[\lambda ^2-(1-\beta _0^2)^2] \\ 
\eta _3^s(\lambda )=\frac{2\beta _0^2}{\lambda ^8}[\beta _0^2\lambda
^4+(1-\beta _0^2)(1+2\beta _0^2)\lambda ^2-(1-\beta _0^2)^3] \\ 
\eta _4^s(\lambda )=-\frac 2{\lambda ^8}[(1+\beta _0^4)\lambda ^4-2(1-\beta
_0^2)^2(1+\beta _0^2)\lambda ^2+(1-\beta _0^2)^4]
\end{array}
\eqnum{9.73}
\end{equation}
and the functions $J_i^0(\lambda )$ were defined in Eq. (9.56).

For the $\gamma _{g3}^s(\lambda )$, By virtue of the expression given in Eq.
(9.65), one can get 
\begin{equation}
\gamma _{g3}^s(\lambda )=\mu \frac d{d\mu }\Omega _1=-\frac{(g_s^R)^2}{8\pi
^2}\Gamma _2^s(\lambda )  \eqnum{9.74}
\end{equation}
where 
\begin{equation}
\Gamma _2^s(\lambda )=2[1+\frac 6{\lambda ^2}-\frac{12}{\lambda ^3\sqrt{%
\lambda ^2-4}}\ln \frac{\lambda -\sqrt{\lambda ^2-4}}{\lambda +\sqrt{\lambda
^2-4}}]  \eqnum{9.75}
\end{equation}

Based on the anomalous dimension $\gamma _g^s(\lambda )$ given in Eqs.
(9.67), (9.68), (9.71) and (9.74), the RGE in Eq. (9.58) may be represented
in the form 
\begin{equation}
\frac{dg_s^R(\lambda )}{d\lambda }+P(\lambda )g_s^R(\lambda )+Q(\lambda
)(g_s^R(\lambda ))^3=0  \eqnum{9.76}
\end{equation}
where 
\begin{equation}
P(\lambda )=\frac{\alpha _R^v(\lambda )}{\pi \lambda }\Gamma ^v(\lambda ),%
\text{ }Q(\lambda )=\frac 1{8\pi ^2\lambda }[\Gamma _1^s(\lambda )-\Gamma
_2^s(\lambda )]  \eqnum{9.77}
\end{equation}
In the above equation, the $\alpha _R^v(\lambda )$ is a known quantity as
given in Eq. (9.10). So, Eq. (9.76) is the equation used to determine the
unknown quantity $g_s^R(\lambda )$ only. To solve the nonlinear equation, we
may set 
\begin{equation}
g_s^R(\lambda )=u(\lambda )^{-1/2}  \eqnum{9.78}
\end{equation}
which leads Eq. (9.76) to a linear equation obeyed by the function $%
u(\lambda )$ 
\begin{equation}
\frac{du(\lambda )}{d\lambda }-2P(\lambda )u(\lambda )-2Q(\lambda )=0 
\eqnum{9.79}
\end{equation}
When setting $Q(\lambda )=0$, we obtain a homogeneous equation whose
solution is 
\begin{equation}
u(\lambda )=u(1)e^{2\int_1^\lambda d\lambda P(\lambda )}  \eqnum{9.80}
\end{equation}
In order to seek the solution of Eq. (9.79), we assume 
\begin{equation}
u(\lambda )=v(\lambda )e^{2\int_1^\lambda d\lambda P(\lambda )}  \eqnum{9.81}
\end{equation}
where $v(\lambda )$ is an unknown function needs to be determined from Eq.
(9.79). Inserting Eq. (9.81) into Eq. (9.79), we get 
\begin{equation}
\frac{dv(\lambda )}{d\lambda }=2Q(\lambda )e^{-2\int_1^\lambda d\lambda
P(\lambda )}  \eqnum{9.82}
\end{equation}
Integrating the above equation, one obtains 
\begin{equation}
v(\lambda )=u(1)+2\int_1^\lambda d\lambda Q(\lambda )e^{-2\int_1^\lambda
d\lambda P(\lambda )}  \eqnum{9.83}
\end{equation}
Combining the expressions in Eqs. (9.78), (9.81) and (9.83), the solution of
Eq. (9.76) is finally given in the form

\begin{equation}
\alpha _R^s(\lambda )=\frac{\alpha _R^sK(\lambda )}{1+\alpha _R^s/\pi
G_s(\lambda )}  \eqnum{9.84}
\end{equation}
where $\alpha _R^s(\lambda )=(g_s^R(\lambda ))^2/4\pi $, $\alpha _R^s=\alpha
_R^s(1)$ which is a parameter needed to be determined by fitting the
experimental data, 
\begin{equation}
K(\lambda )=e^{-\frac 2\pi \int_1^\lambda d\lambda /\lambda \alpha
_R^v(\lambda )\Gamma ^v(\lambda )}  \eqnum{9.85}
\end{equation}
and 
\begin{equation}
G_s(\lambda )=\int_1^\lambda d\lambda /\lambda \Gamma ^s(\lambda )K(\lambda )
\eqnum{9.86}
\end{equation}
in which 
\begin{equation}
\Gamma ^s(\lambda )=\Gamma _1^s(\lambda )-\Gamma _2^s(\lambda )  \eqnum{9.87}
\end{equation}

The behaviors of the effective coupling constant $\alpha _R^s(\lambda )$
obtained in the timelike and spacelike momentum subtractions are separately
displayed in Figs. (14) and (15). For timelike momenta, the $\alpha
_R^s(\lambda )$ is real. In Fig. (14), there are four lines representing
this $\alpha _R^s(\lambda )$ which are given by four groups of the
parameters $(\alpha _R^v,\alpha _R^s)=(0,1)$, $(0.5,1)$, $(0,0.2)$ and $%
(0.5,0.2)$ respectively. The figure indicates that the $\alpha _R^s(\lambda
) $ has a Landau pole $\lambda _0.$The poles for the four lines are
respectively located about at $\lambda _0\approx 1.07523$ (for the first
lines), $1.8237$ and $2.3967$. Clearly, the $\lambda _0$ is larger if the
both parameters $(\alpha _R^v,\alpha _R^s)$ are smaller. In particular, the
pole moves to the point near infinity when the both parameters $(\alpha
_R^v,\alpha _R^s)$ tend to zero. When $\lambda $ goes from $\lambda _0$ to
zero and infinity, each $\alpha _R^s(\lambda )$ in Fig. (14) abruptly falls
to zero from the opposite directions. For spacelike momenta, The behavior of
the effective coupling constant is described by the lines in Fig. (15). The
two lines in Figs (15a) and (15b) are given by considering the scalar
coupling only with taking $\alpha _R^s=1$ and $0.2$ respectively. In this
case, the $\alpha _R^s(\lambda )$ is real and has a pole $\lambda _0$. the
poles for the aforementioned lines are located respectively at $\lambda
_0=5.725$ and $234.6$. However, when the vectorial coupling is included, the 
$\alpha _R^s(\lambda )$ becomes complex. In this case, the pole disappears;
instead, there is a maximum to appear as shown in Figs (15c) and (15d). The
lines in Figs (15c) and (15d) are given by taking $(\alpha _R^v,\alpha
_R^s)=(0.5,1)$ and $(0.5,0.2)$ respectively. In the figures, the real part
of the $\alpha _R^s(\lambda )$ is represented by the solid line and the
imaginary part by the dashed line. From Fig. (15c), we see that the real and
imaginary parts of the $\alpha _R^s(\lambda )$ have sharp peaks
corresponding to the pole of the upper line. The peaks are manifested more
clearly by the right amplified lines. Fig. (15d) exhibits that either the
real part or the imaginary part varies rather smoothly due to the decrement
of the parameters $(\alpha _R^v,\alpha _R^s)$ and the larger effect of the
vectorial coupling. It is noted that when $\lambda \rightarrow 0$, the $%
\alpha _R^s(\lambda )$ reaches a constant, while, in the limit of $\lambda
\rightarrow \infty $, the $\alpha _R^s(\lambda )$ goes to zero. In
particular, The behaviors of the $\alpha _R^s(\lambda )$ tell us that the
smaller the parameters $(\alpha _R^v,\alpha _R^s)$, the larger will be the
range of applicability and the $\alpha _R^s(\lambda )$ for the spacelike
momenta have a larger range of applicability than that for the timelike
momenta.

\subsection{Effective $\sigma $ meson mass}

In accordance with Eq. (8.2), the RGE for the renormalized $\sigma $ meson
mass is 
\begin{equation}
\lambda \frac d{d\lambda }m_\sigma ^R(\lambda )+\gamma _m^\sigma (\lambda
)m_\sigma ^R(\lambda )=0  \eqnum{9.88}
\end{equation}
where 
\begin{equation}
\gamma _m^\sigma (\lambda )=\mu \frac d{d\mu }\ln Z_m^\sigma .  \eqnum{9.89}
\end{equation}
From the definitions given in Eqs. (7.6) and (7.9), it is found that in the
approximation of order $g_s^2$, the renormalization constant $Z_m^\sigma $
can be written as 
\begin{equation}
Z_m^\sigma =1-\frac 12[\Omega _1(\mu ^2)+\Omega _2(\mu ^2)]  \eqnum{9.90}
\end{equation}
The one-loop expressions of the divergent constants $\Omega _1(\mu ^2)$ and $%
\Omega _2(\mu ^2)]$ were given in Eqs. (9.65) and (9.66). With these
expressions, the renormalization constants $Z_m^\sigma $ in Eq. (9.90) can
explicitly be written out. Use of this renormalization constant in Eq.
(9.89) yields the $\sigma $ meson mass anomalous dimension as follows: 
\begin{equation}
\gamma _m^\sigma =-\frac{g_s^2}{4\pi }G_s(\lambda )  \eqnum{9.91}
\end{equation}
where 
\begin{equation}
G_s(\lambda )=\frac 6\pi [\frac 16-\frac 1{\beta _0^2}+\frac 1{\lambda ^2}-%
\frac 2\lambda (\frac 1{\lambda ^2}-\frac 1{\beta _0^2})\eta (\lambda )] 
\eqnum{9.92}
\end{equation}
in which $\beta _0=\frac{m_\sigma }M$ and 
\begin{equation}
\eta (\lambda )=\frac 1{\sqrt{\lambda ^2-4}}\ln \frac{\lambda -\sqrt{\lambda
^2-4}}{\lambda +\sqrt{\lambda ^2-4}}  \eqnum{9.93}
\end{equation}
Substituting the above anomalous dimension into Eq. (9.88) and solving the
equation, we obtain an expression of the effective $\sigma $ meson mass such
that 
\begin{equation}
m_\sigma ^R(\lambda )=m_\sigma ^Re^{\int_1^\lambda \frac{d\lambda }\lambda
\alpha _R^s(\lambda )G_s(\lambda )}  \eqnum{9.94}
\end{equation}
where $m_\sigma ^R=m_\sigma ^R(1)$ is a mass parameter which needs to be
determined by experiment. If the coupling constant $\alpha _R^s(\lambda )$
is approximately taken to be a constant $\alpha _R^s$, the integral over $%
\lambda $ can easily be calculated. In this case, we have 
\begin{equation}
m_\sigma ^R(\lambda )=m_\sigma ^Re^{S_\sigma (\lambda )}  \eqnum{9.95}
\end{equation}
where 
\begin{equation}
\begin{tabular}{l}
$S_\sigma (\lambda )=\frac{2\alpha _R^s}\pi [1-\frac 1{\lambda ^2}+\frac{%
\sqrt{3}}2(1-\frac 2{\beta _0^2})\pi $ \\ 
$+(\frac 1{\lambda ^2}+\frac 12-\frac 3{\beta _0^2})\frac{\sqrt{\lambda ^2-4}%
}\lambda \ln \frac 12(\lambda +\sqrt{\lambda ^2-4})].$%
\end{tabular}
\eqnum{9.96}
\end{equation}
It is seen from Eq. (9.94) that the behavior of the effective mass $m_\sigma
^R(\lambda )$ is intimately related to property of the effective coupling
constant $\alpha _R^s(\lambda )$. To give a view of the behavior of the $%
m_\sigma ^R(\lambda )$, we take the $m_\sigma ^R(\lambda )$ evaluated from
Eq. (9.94) by taking $(\alpha _R^v,\alpha _R^s)=(1,1)$ as an example. This $%
m_\sigma ^R(\lambda )$ is shown in Fig. (16). In the figure, the dashed line
represents the $m_\sigma ^R(\lambda )$ given in the timelike momentum
subtraction. This $m_\sigma ^R(\lambda )$ is real and has a singularity at $%
\lambda _0\approx 1.09758$ which implies that the range of applicability of
the $m_\sigma ^R(\lambda )$ is less than $\lambda =1$, the nucleon mass
scale. When $\lambda $ tends to zero, the $m_\sigma ^R(\lambda )$ approaches
a value which does not deviate from the constant $m_\sigma ^R$ so much. The $%
m_\sigma ^R(\lambda )$ given in the spacelike momentum subtraction is
complex .The solid line in Fig. (16) represents the real part of the $%
m_\sigma ^R(\lambda )$ which has a maximum near $\lambda _0=5$ which
indicates that the $m_\sigma ^R(\lambda )$ is applicable in a wide region of
[$0,5$]. Similar to the coupling constant $\alpha _R^s(\lambda )$, when the
parameters $(\alpha _R^v,\alpha _R^s)$ are taken to be smaller, either the
pole or the maximum will be shifted to the point of a large $\lambda _0$.

\section{Summary and discussions}

In this paper, it has been argued that the $\sigma -\omega $ model, as a
constrained system, is really of U(1) local gauge symmetry. This enables us
to quantize the $\sigma -\omega $ model by means of the method used for
quantizing the gauge field theory. In particular, the gauge symmetry allows
us, in a consistent way, to derive various W-T identities which provide a
faithful basis for performing the renormalization of the model. As shown in
Sec. V, the W-T identity in Eq. (5.6) satisfied by the $\omega $ meson
propagator and the W-T identity in Eq. (5.13) for the vacuum polarization
operator determine not only the structures of the propagator and the vacuum
polarization operator, but also the renormalization fashion of the
propagator and the $\omega $ meson mass as shown in Eqs. (5.14), (5.16),
(5.21) and (5.22). Especially, the W-T identity in Eq. (6.13) obeyed by the
vertex gives rise to the correct manner of subtraction of the nucleon
self-energy as denoted in Eq. (6.19). As shown in Sec.VI, the subtraction in
Eq. (6.19) leads to the correct representations for the renormalization
constants of nucleon propagator and nucleon mass as shown in Eqs. (6.21) and
(6.25). Moreover, the identity in Eq. (6.13) directly yields the important
relation between the renormalization constants $Z_1$ and $Z_2$ as written in
Eq. (6.32). This relation together with the relation in Eq. (7.22) which
follows from the identity in Eq. (7.21) greatly simplify the calculation of
the renormalization. It would be mentioned here that in some previous works
[7, 15, 17], the subtraction based on the expression $\Sigma (p)=A{\bf p+}BM$
was ever used. This subtraction gives the nucleon propagator renormalization
constant as $Z_2=[1-A(\mu ^2)]^{-1}$ and nucleon mass renormalization
constant as $Z_M=[Z_2(1+B(\mu ^2)]^{-1}$ which are different from the
renormalization constants written in Eqs. (6.21) and (6.25) and therefore
the relation in Eq. (6.32) could not be fulfilled in this case. The
renormalization of the model under consideration is performed in the
mass-dependent momentum space subtraction scheme by the renormalization
group approach. The prominent advantage of the subtraction scheme is that it
naturally leads to the boundary conditions for the renormalized propagators,
the vertices and the wave functions. The boundary conditions allow us to
give an unique determination of the solutions to the renormalization group
equations for the renormalized propagators, vertices and wave functions
without any ambiguity. As claimed in the Introduction, we limit ourself in
this paper to examine the renormalization of the model at zero temperature
by means of the renormalization group method. Since the perturbative series
expanded in the powers of coupling constants is chosen to be the starting
point of this renormalization, the results of the renormalization would be
different from those obtained in the study of nuclear matter by using the
loop expansion and the spectral function methods. Hopefully, the
renormalization procedure described in this paper will be helpful for
applying the renormalization group approach to study the nuclear matter at
finite temperature and finite density.

The procedure of renormalization group method was demonstrated by the
one-loop renormalization in this paper. Since the renormalization exactly
respects the W-T identities, the results obtained are faithful. Especially,
the one-loop effective coupling constants and masses are given in the
rigorous forms as they are derived from the mass-dependent momentum space
subtraction. The subtraction scheme used is, in principle, suitable not only
for high energy, but also for low energy, unlike the minimal subtraction
scheme [28-31] which is only appropriate in the large momentum limit. In
addition, the expressions of the one-loop effective physical quantities
derived in this paper are applicable for the both of timelike momentum
subtraction and spacelike momentum subtraction. As seen from Figs. (8)-(12)
and (14)- (16), the behaviors of the effective quantities given in the
timelike subtraction and the spacelike subtraction are much different from
one another. In which case we should use the results given in the timelike
momentum subtraction or in the spacelike momentum subtraction? The answer to
this question depends on what process is discussed. For example, when we
study the nucleon-nucleon scattering taking place in the t-channel, as
mentioned in the Appendix B, the transfer momentum in the boson propagator
is spacelike. In this case, it is suitable to take the effective coupling
constants and boson masses given in the spacelike momentum subtraction. If
we investigate the nucleon-antinucleon annihilation process which takes
place in the s-channel, since the transfer momentum is timelike, the
effective coupling constants and boson masses given in the timelike momentum
subtraction should be used. The effect of the one-loop renormalization is
examined by the nucleon-nucleon elastic scattering whose differential cross
section given in the order of $g^{2\text{ }}$is described in the Appendix B
and plotted in Fig. 17. In the figure, We only take the differential cross
sections given at the laboratory kinetic energies $T_{lab}=491.9MeV$ and $%
575.5MeV$ as an example. The figure shows that consideration of the one-loop
renormalization requires the coupling constants to be smaller in order to
fit the experimental data. This actually is a general feature of considering
the renormalization effect.

As exhibited in Sec. IX, the one-loop effective physical parameters given in
the timelike momentum subtraction and the ones given in the spacelike
momentum subtraction not only behave differently, but also have different
ranges of applicability because the singularities of the effective
quantities given by the two subtractions appear at the different momenta.
Especially, the positions of the singularities are strongly dependent on the
coupling constants $\alpha _v^R$ and $\alpha _s^R$. The smaller the coupling
constants, the larger are the ranges of applicability. It would be mentioned
that since the propagators written in Eqs. (5.16), (6.12) and (7.2) are
solved from the Dyson equations, the one-loop renormalization actually
contains the contribution given by partially summing up a set of chain loop
diagrams. Just due to the partial summation, as mentioned before, the
coupling constants must be set to be smaller for fitting the experimental
data of the nucleon scattering. To this end, it is natural to ask if and how
the behaviors of the one-loop effective physical parameters can be modified
by considering higher order loop renormalizations? In other words, if the
coupling constants would be more smaller and the ranges of applicability of
the renormalization would be enlarged when the contributions arising from
more higher order loop diagrams are summed up? Obviously, this is an
interesting problem worthy to pursuing further. In addition, we would like
to address that the $\sigma -\omega $ model should be viewed as a
restrictive model in which the $\sigma $ field is introduced as a
phenomenological field. Aside from the $\sigma -\omega $ model, there are
some other models in QHD which are of a certain gauge symmetry. Especially,
the model proposed by Sakurai in the early time [44], in our opinion, is
most promising to describe the nuclear force because in this model,
exchanges of the light mesons, pion and $\rho ho$ meson, dominate the strong
interaction between nucleons. In view of the argument given in Refs. [32-34,
45], the Sakurai's model is a SU(2) gauge field theory which is not only
gauge-invariant, but also renormalizable. Certainly, the renormalization of
this model may be investigated along the same line as described in this
paper. We will discuss this subject in the future. But, it can not be
expected that a perturbative investigation could give an ultimate solution
to the strong interaction. Just as said in Ref. [46, 47], to resolve the
strong interaction, it is adequate to perform a nonperturbative study of the
interaction kernel appearing in the relativistic equation whose closed
expression can be derived by the procedure as described in Refs. [46, 47].

\section{Acknowledgment}

This subject was supported by National Natural Science Foundation of China .

\section{Appendix A: Gauge-independence of S-matrix elements}

The gauge-independence of S-matrix elements computed by a gauge field theory
is a well-known fact. For the $\sigma -\omega $ model, as argued in this
paper, it actually is a U(1) gauge field theory. So, the same conclusion
should hold for the $\sigma -\omega $ model. To convince oneself of this
fact, we take the nucleon-nucleon scattering amplitudes up to the one-loop
approximation as examples to show that the matrix elements given by the $%
\sigma -\omega $ model are surely independent of the gauge parameter $\alpha 
$. The typical Feynman diagrams representing the scattering amplitudes are
depicted in Figs. (5) and (6). In Fig. (6), only the diagrams with the
internal $\omega $ meson line are necessary to be considered.

For the tree diagram in Fig. (5a), the gauge-independence of its S-matrix
element is well-known. In fact, the S-matrix element 
\begin{equation}
S_1=\overline{u}_{s^{\prime }}(q_2)ig_v\gamma _\mu u_{r^{\prime
}}(q_1)iD^{\mu \nu }(k)\overline{u}_s(p_2)ig_v\gamma _\nu u_r(p_1) 
\eqnum{A.1}
\end{equation}
where $u_s(p)$ is the free nucleon wave function can be divided into two
parts according to the decomposition of free $\omega $ meson propagator 
\begin{equation}
D^{\mu \nu }(k)=D_F^{\mu \nu }(k)+D_\alpha ^{\mu \nu }(k)  \eqnum{A.2}
\end{equation}
where 
\begin{equation}
D_F^{\mu \nu }(k)=-\frac{g^{\mu \nu }}{k^2-m_\omega ^2+i\varepsilon } 
\eqnum{A.3}
\end{equation}
which is the propagator given in the Feynman gauge and 
\begin{equation}
D_\alpha ^{\mu \nu }(k)=(1-\alpha )D_\alpha (k^2)k^\mu k^\nu  \eqnum{A.4}
\end{equation}
which is the $\alpha -$dependent part of the propagator in which 
\begin{equation}
D(k^2)=\frac 1{(k^2-m_\omega ^2+i\varepsilon )(k^2-\nu ^2+i\varepsilon )}. 
\eqnum{A.5}
\end{equation}
For the $\alpha -$dependent part of $S_1$, applying the energy-momentum
conservation $k=q_1-q_2=p_2-p_1$ and Dirac equation (${\bf p-}M{\bf )}%
u_s(p)=0$, it is found 
\begin{equation}
\begin{tabular}{l}
$S_1^\alpha =-(1-\alpha )ig_v^2\overline{u}_{s^{\prime }}(q_2){\bf k}%
u_{r^{\prime }}(q_1)\overline{u}_s(p_2){\bf k}u_r(p_1)D_\alpha (k^2)$ \\ 
$=-(1-\alpha )ig_v^2\overline{u}_{s^{\prime }}(q_2){\bf (q}_1-{\bf q}%
_2)u_{r^{\prime }}(q_1)\overline{u}_s(p_2)({\bf p}_2-{\bf p}%
_1)u_r(p_1)D_\alpha (k^2)$ \\ 
$=0.$%
\end{tabular}
\eqnum{A.6}
\end{equation}
This shows that the $\omega $ meson propagator given in the Feynman gauge is
sufficient to use for evaluating the tree diagram matrix element. In the
same way, one may prove that the S-matrix element given by the tree-diagram
in Fig. (5b) is independent of the gauge parameter as well.

Let us focus on the one-loop diagrams in Fig. (6) where only the direct
diagrams are plotted and necessarily to be examined for our purpose. The
gauge-independence of the matrix element of Fig.(6a) which contains a $%
\omega $ meson self-energy in it can easily be proved by using Eq. (A.6).
So, we only need to examine the gauge-independence of Figs. (6b)-(6f). The
matrix element of Fig. (6b) with a vertex correction in it can be written as 
\begin{equation}
S_2^1=M_\mu (q_{1,}q_2)A^\mu (p_1,p_2)  \eqnum{A.7}
\end{equation}
where 
\begin{equation}
M_\mu (q_{1,}q_2)=\overline{u}_{s^{\prime }}(q_2)ig_v\gamma ^\nu
u_{r^{\prime }}(q_1)iD_{\nu \mu }(k)  \eqnum{A.8}
\end{equation}
which is gauge-independent as shown in (A.6) and 
\begin{equation}
A^\mu (p_1,p_2)=\int \frac{d^4k}{(2\pi )^4}\overline{u}_s(p_2)ig_v\gamma
_\rho iS_F(p_2-k)ig_v\gamma ^\mu iS_F(p_1-k)ig_v\gamma _\sigma
u_r(p_1)iD^{\rho \sigma }(k).  \eqnum{A.9}
\end{equation}
Replacing $D^{\rho \sigma }(k)$ by the $D_\alpha ^{\rho \sigma }(k)$ shown
in (A.4), we have the following gauge-dependent part of $A^\mu (p_1,p_2)$ 
\begin{equation}
A_\alpha ^\mu (p_1,p_2)=-(1-\alpha )g_v^3\int \frac{d^4k}{(2\pi )^4}%
\overline{u}_s(p_2){\bf k}S_F(p_2-k)\gamma ^\mu S_F(p_1-k){\bf k}%
u_r(p_1)D_\alpha (k^2)  \eqnum{A.10}
\end{equation}
where ${\bf k}=\gamma ^\mu k_\mu $ can be written in the form 
\begin{equation}
{\bf k=}S_F^{-1}(p_i)-S_F^{-1}(p_i-k)  \eqnum{A.11}
\end{equation}
here $i=1,2$. Using this relation and Dirac equation, Eq. (A.10) becomes 
\begin{equation}
A_\alpha ^\mu (p_1,p_2)=-(1-\alpha )g_v^3\overline{u}_s(p_2)\gamma ^\mu
u_r(p_1)J  \eqnum{A.12}
\end{equation}
where 
\begin{equation}
J=\int \frac{d^4k}{(2\pi )^4}D_\alpha (k^2)=\lim_{\varepsilon \rightarrow 0}%
\frac i{(4\pi )^2}\int_0^1\frac{dx}{\varepsilon \{[\alpha +(1-\alpha
)x]m_\omega ^2\}^\varepsilon }  \eqnum{A.13}
\end{equation}
here the last equality is given by the dimensional regularization. This
integral gives a divergent constant without containing any finite number in
it. Therefore, it may completely be cancelled out by a counterterm in a
renormalization program and gives no contribution to the renormalized
S-matrix element. On the other hand, since the integral is independent of
momentum, it would not contribute to the anomalous dimension and hence to
any physical quantity.

For the matrix element of Fig. (6c) which contains a nucleon self-energy, it
may be written as 
\begin{equation}
S_2^2=M_\mu (q_{1,}q_2)B^\mu (p_1,p_2)  \eqnum{A.14}
\end{equation}
where 
\begin{equation}
B^\mu (p_1,p_2)=\int \frac{d^4k}{(2\pi )^4}\overline{u}_s(p_2)ig_v\gamma
_\rho iS_F(p_2-k)ig_v\gamma _\sigma iS_F(p_1+q)ig_v\gamma ^\mu
u_r(p_1)iD^{\rho \sigma }(k)  \eqnum{A.15}
\end{equation}
in which the $\alpha -$dependent part is of the form 
\begin{equation}
B_\alpha ^\mu (p_1,p_2)=-(1-\alpha )g_v^3\int \frac{d^4k}{(2\pi )^4}%
\overline{u}_s(p_2){\bf k}S_F(p_2-k){\bf k}S_F(p_1+q)\gamma ^\mu
u_r(p_1)D_\alpha (k^2)  \eqnum{A.16}
\end{equation}
By employing the relation in Eq. (A.11) and Dirac equation, one may find 
\begin{equation}
B_\alpha ^\mu (p_1,p_2)=(1-\alpha )g_v^3\overline{u}_s(p_2)\gamma _\nu
S_F(p_2)\gamma ^\mu u_r(p_1)J_2^\nu =0  \eqnum{A.17}
\end{equation}
This is because the integral in it vanishes 
\begin{equation}
J_2^\nu =\int \frac{d^4k}{(2\pi )^4}k^\nu D_\alpha (k^2)=0  \eqnum{A.18}
\end{equation}
due to that the integrand is an odd function. Similarly, the $\alpha -$%
dependent part of Fig. (6d) can also be proved to give no contribution to
the S-matrix element.

Let us turn to Figs.(6e) and (6f). The matrix elements of the both figures
can be respectively represented as 
\begin{equation}
\begin{tabular}{l}
$S_2^3=g_v^4\int \frac{d^4k}{(2\pi )^4}\overline{u}_{s^{\prime }}(q_2)\gamma
^\mu S_F(q_1-k)\gamma ^\nu u_{r^{\prime }}(q_1)$ \\ 
$\times \overline{u}_s(p_2)\gamma ^\rho S_F(p_1+k)\gamma ^\sigma
u_r(p_1)D_{\mu \rho }(q-k)D_{\nu \sigma }(k)$%
\end{tabular}
\eqnum{A.19}
\end{equation}
and 
\begin{equation}
\begin{tabular}{l}
$S_2^4=g_v^4\int \frac{d^4k}{(2\pi )^4}\overline{u}_{s^{\prime }}(q_2)\gamma
^\mu S_F(q_1-k)\gamma ^\nu u_{r^{\prime }}(q_1)$ \\ 
$\times \overline{u}_s(p_2)\gamma ^\rho S_F(p_2-k)\gamma ^\sigma
u_r(p_1)D_{\mu \sigma }(q-k)D_{\nu \rho }(k)$%
\end{tabular}
\eqnum{A.20}
\end{equation}
where $q=p_2-p_1=q_1-q_2$. Their $\alpha -$dependent parts are denoted by $%
S_{2\alpha }^3$ and $S_{2\alpha }^4$. By making use of the relation in
(A.11) and the relation $q-k=(q_1-k)-q_2=p_2-(k+p_1)$ as well as Dirac
equation, it is easy to find 
\begin{equation}
\begin{tabular}{l}
$S_{2\alpha }^3=-2(1-\alpha )g_v^4\overline{u}_{s^{\prime }}(q_2)\gamma ^\mu
u_{r^{\prime }}(q_1)\overline{u}_s(p_2)\gamma _\mu u_r(p_1)\int \frac{d^4k}{%
(2\pi )^4}\frac{D_\alpha (k^2)}{(q-k)^2-m_\omega ^2+i\varepsilon }$ \\ 
$-(1-\alpha )^2g_v^4\overline{u}_{s^{\prime }}(q_2)\gamma ^\mu u_{r^{\prime
}}(q_1)\overline{u}_s(p_2)\gamma ^\nu u_r(p_1)\int \frac{d^4k}{(2\pi )^4}%
k_\mu k_\nu D_\alpha [(q-k)^2]D_\alpha (k^2)$ \\ 
$=-S_{2\alpha }^4$%
\end{tabular}
\eqnum{A.21}
\end{equation}
which gives $S_{2\alpha }^3+S_{2\alpha }^4=0$ so that the sum of $S_2^3$ and 
$S_2^4$ is independent of the gauge parameter.

\section{Appendix B: Cross section of nucleon-nucleon scattering}

To illustrate the effect of the renormalization described in this paper, we
evaluate the cross section of the nucleon-nucleon elastic scattering. Here
we limit ourself to first consider the cross section given in the
approximation of order $g^2$. In this approximation, only the tree diagrams
denoted in Fig. (5) are concerned. From these diagrams, in the center of
mass frame, the differential cross section is easily calculated and
represented as follows: 
\begin{equation}
\frac{d\sigma }{d\Omega (\theta ,\varphi )}=\frac 1S[\alpha _v^2T_v+\alpha
_s^2T_s-\alpha _v\alpha _sT_{vs})  \eqnum{B.1}
\end{equation}
where $S$ $=4(p^2+M^2)$ with $p$ being the nucleon momentum is the squared
total energy of the system, $T_v$ is contributed from the $\omega $ meson
exchange interaction, $T_s$ is given by the $\sigma $ meson exchange
interaction and $T_{vs}$ is the crossed term related to both of the $\omega $
meson and $\sigma $ meson exchanges. They are separately represented as
follows: 
\begin{equation}
T_v=\frac{R_1^v}{(\Delta _1^v)^2}+\frac{R_2^v}{(\Delta _2^v)^2}+(-1)^{1+I}%
\frac{R_3^v}{\Delta _1^v\Delta _2^v}  \eqnum{B.2}
\end{equation}
where 
\begin{equation}
R_1^v=M^4+2p^2M^2\cos \theta +2M^2p^2+2p^4+2p^4\cos ^4\theta /2,  \eqnum{B.3}
\end{equation}
\begin{equation}
R_2^v=M^4-2p^2M^2\cos \theta +2M^2p^2+2p^4+2p^4\sin ^4\theta /2,  \eqnum{B.4}
\end{equation}
\begin{equation}
R_3^v=16(p^4-M^4),  \eqnum{B.5}
\end{equation}
\begin{equation}
\Delta _1^v=4p^2\sin ^2\theta /2+m_\omega ^2,  \eqnum{B.6}
\end{equation}
and 
\begin{equation}
\Delta _2^v=4p^2\cos ^2\theta /2+m_\omega ^2.  \eqnum{B.7}
\end{equation}
\begin{equation}
T_s=\frac{R_1^s}{(\Delta _1^s)^2}+\frac{R_2^s}{(\Delta _2^s)^2}+(-1)^I\frac{%
R_3^s}{\Delta _1^s\Delta _2^s},  \eqnum{B.8}
\end{equation}
where 
\begin{equation}
R_1^s=4(4p^2\sin ^2\theta /2+M^2),  \eqnum{B.9}
\end{equation}
\begin{equation}
R_2^s=4(4p^2\cos ^2\theta /2+M^2)^2,  \eqnum{B.10}
\end{equation}
\begin{equation}
R_3^s=2[2M^2(p^2+M^2)+p^4(\sin ^4\theta /2+\cos ^4\theta /2],  \eqnum{B.11}
\end{equation}
\begin{equation}
\Delta _1^s=4p^2\sin ^2\theta /2+m_\sigma ^2,  \eqnum{B.12}
\end{equation}
and 
\begin{equation}
\Delta _2^s=4p^2\cos ^2\theta /2+m_\sigma ^2.  \eqnum{B.13}
\end{equation}
\begin{equation}
T_{vs}=\frac{R_1^{vs}}{\Delta _1^v\Delta _1^s}+\frac{R_2^{vs}}{\Delta
_2^v\Delta _2^s}+(-1)^I\frac{R_3^{vs}}{\Delta _1^v\Delta _2^s}+(-1)^I\frac{%
R_4^{vs}}{\Delta _2^v\Delta _1^s},  \eqnum{B.14}
\end{equation}
where 
\begin{equation}
R_1^{vs}=4M^2[(M^2+2p^2)^2+(M^2+2p^2\cos ^2\theta /2)^2],  \eqnum{B.15}
\end{equation}
\begin{equation}
R_2^{vs}=4M^2[(M^2+2p^2)^2+(M^2+2p^2\sin ^2\theta /2)^2],  \eqnum{B.16}
\end{equation}
\begin{equation}
R_3^{vs}=4[2(M^2+p^2\cos ^2\theta /2)^2-M^2(M^2+p^2+p^2\sin ^2\theta /2)] 
\eqnum{B.17}
\end{equation}
and 
\begin{equation}
R_4^{vs}=4[2(M^2+p^2\sin ^2\theta /2)^2-M^2(M^2+p^2+p^2\cos ^2\theta /2)]. 
\eqnum{B.18}
\end{equation}
In the above, ($\theta ,\varphi )$ are the scattering angles, $I$ is the
isospin of the two-nucleon system, $\Delta _k^v$ and $\Delta _k^s$ ($k=1,2$)
are respectively given by the $\omega $ meson and $\sigma $ meson
propagators and $R_i^\alpha $ ($\alpha =v,s,i=1,2,3$) are the functions
coming from the nucleon spinor matrix elements. In Eqs. (B.2), (B.8) and
(B.14), the isospin-related terms arise from the exchanged diagram, while,
the remaining terms represent the contribution of the direct diagram.

To consider the renormalization effect on the two-nucleon scattering, as
mentioned in Sec. VIII, we may directly replace the coupling constant $%
\alpha _v$ and $\alpha _s$ in Eq. (B.1) by their effective ones $\alpha
_R^v(\lambda )$ and $\alpha _R^s(\lambda )$ and the particle masses $M$, $%
m_\omega $ and $m_\sigma $ appearing in the propagators and the functions $%
R_i^\alpha $ by their effective counterparts $M_R(\lambda )$, $m_\omega
^R(\lambda )$ and $m_\sigma ^R(\lambda )$. To this end, it should be noted
that the nucleon spinors in the S-matrix element under consideration are on
the mass-shell, satisfying the free nucleon Dirac equation. The momenta $p_i$
in the spinors are timelike because they meet the relation $p^2=M^2$ where $%
M $ is real. For the renormalized spinor wave function shown in Eq. (8.14),
as easily verified, it also satisfies the Dirac equation and the momentum in
the spinor fulfills the relation $p(\lambda )^2=M_R(\lambda )^2$ where $%
p(\lambda )=(E(\lambda ),\vec p)$ with $E(\lambda )=(\vec p^2+M_R(\lambda
)^2)^{1/2}$. Therefore, for the nucleon-nucleon scattering, it is adequate
to take the effective nucleon mass given in the timelike momentum space
subtraction. While, the momenta in the $\omega $ meson and $\sigma $ meson
propagators, as one knows, are off-shell and spacelike in the t-channel
scattering. Therefore, it is appropriate to take the effective coupling
constants $\alpha _R^v(\lambda )$ and $\alpha _R^s(\lambda )$ and the
effective meson masses $m_\omega ^R(\lambda )$ and $m_\sigma ^R(\lambda )$
given in the spacelike momentum space subtraction. In this section, we only
examine the effect of the one-loop renormalization on the two-proton
scattering by using the effective coupling constants and masses presented in
Sec. IX. The differential cross sections given at the laboratory kinetic
energies $T_{lab}=491.9MeV$ and $575.5MeV$ are shown in Fig. (17a) and
(17b). In the figures, the solid lines and the dashed lines represent
respectively the calculated results with and without considering the
renormalization effect and the experimental data are taken from Ref. [48].
As shown in Fig. (17a) and (17b), in the case without considering the
renormalization effect, the theoretical parameters are taken to be $%
M_R=938MeV,\alpha _\omega ^R=1.1,\alpha _\sigma ^R=1.4,m_\omega ^R=782MeV$
and $m_\sigma ^R=580MeV$; while in the case of considering the
renormalization effect, the parameters must be taken to be $%
M_R=938MeV,\alpha _\omega ^R=0.55,\alpha _\sigma ^R=0.62,m_\omega ^R=782MeV$
and $m_\sigma ^R=670MeV$. It is clearly seen from the figures that
consideration of the renormalization has an effect that to fit the
experimental data, the coupling constants must be set to be smaller.

\section{References}

[1] J. D. Walecka, Ann. Phys. (N. Y) {\bf 83, }491 (1974).

[2] T. Matsui and B. D. Serot, Ann. Phys. {\bf 114}, 107 (1982).

[3] B. D. Serot and J. D. Walecka, Adv. Nucl. Phys. {\bf 16, }1 (1986).

[4] B. D. Serot, Rep. Prog. Phys. {\bf 55}, 1855 (1992).

[5]{\bf \ }B. D. Serot and J. D. Walecka, Int. J. Mod. Phys. E {\bf 6}, 515
(1997).

[6] C. J. Horowith and B. D. Serot, Nucl. Phys. A {\bf 399,} 529 (1983).

[7] A. F. Bielajew, Nucl. Phys. A {\bf 404, }428{\bf \ (1983); Ann. 156, }%
215 {\bf (1984).}

[8] R. J. Perry, Phys. Lett. B {\bf 199}, 489 (1987).

[9] T. D. Cohen, M. K. Banerjee and C. Y. Ren, Phys. Rev. C {\bf 36}, 1653
(1987).

[10] R. J. Furnstahl and C. J. Horowith, Nucl. Phys. A {\bf 485}, 632 (1988).

[11] K. Wehrberger, R. Wittman and B. D. Serot, Phys. Rev. C {\bf 42}, 2680
(1990).

[12] R. J. Furnstahl, R. J. Perry and B. D. Serot, Phys. Rev. C {\bf 40},
321, (1989); {\bf 404} (E) (1990).

[13] M. P. Allendes and B. D. Serot, Phys. Rev. C {\bf 45}, 2975 (1992).

[14] G. Krein, M. Nielsen, R. D. Puff and L. Wilets, Phys. Rev. C 47, 2485
(1993).

[15] B. D. Serot and H. B Tang, Phys. Rev. C {\bf 51}, 969 (1995).

[16] R. J. Furnstahl, B. D. Serot and H.-B. Tang, Nucl. Phys. A {\bf 615},
441 (1997); A {\bf 640, 505(E) (1998)}.

[17] S. S. Wu and Y. J. Yao, Eur. Phys. J. A {\bf 3} 49 (1998).

[18] H. W. Hammer and R. J. Furnstahl, Nucl. Phys. A {\bf 678}, 277 (2000).

[19] R. J. Furnstahl and B. D. Serot, Comments on Modern Physics {\bf 2}, A 
{\bf 23} (2000).

[20] J. C. Su, X. X. Yi and Y. H. Cao, J. Phys. G: Nucl. Part. Phys. {\bf %
25, }2325 (1999); J. C. Su, L. Shan and Y. H. Cao, Commun. Theor. Phys. {\bf %
36}, 665 (2000).

[21] W. A. Bardeen, A. J. Buras, D. W. Duke and T. Muta, Phys. Rev. D {\bf 18%
}, 3998 (1978); W. A. Bardeen and R. A. J. Buras, Phys. Rev. D {\bf 20}, 166
(1979).

[22] W. Celmaster and R. J. Gonsalves, Phys. Rev. Lett. {\bf 42}, 1435
(1979); Phys. Rev. D {\bf 20}, 1420 (1979);

W. Celmaster and D. Sivers, Phys. Rev. D {\bf 23}, 227 (1981).

[23] E. Braaten and J.P. Leveille, Phys. Rev. D {\bf 24}, 1369 (1981).

[24] S. N. Gupta and S. F. Radford, Phys. Rev. D {\bf 25}, 2690 (1982); J.
C. Collins and A. J. Macfarlane, Phys. Rev. D {\bf 10}, 1201 (1974).

[25] C. G. Callan, Phys. Rev. D {\bf 2}, 1541 (1970);

[26] K. Symanzik, Commun, Math. Phys. {\bf 18}, 227 (1970).

[27] S. Weinberg, Phys. Rev. D {\bf 8}, 3497 (1973).

[28] J. C. Collins and A. J. Macfarlane, Phys. Rev. D {\bf 10}, 1201 (1974).

[29] C. Itzykson and \ J-B. Zuber, Quantum \ Field Theory, McGraw-Hill, New
York, 1980.

[30] D. Bailin and A. Love, Introduction to Gauge field Theory, Mid-Couty
Press, London (1986); Institute of Physics Publishing, Bristol and
Philadelphia, 1994.

[31] S. Weinberg, The Quantum of Fields, Vol. I, Vol. II, Cambridge
University Press, 1995.

[32] J. C. Su, Nuovo Cimento Soc. Ital. Fis., B {\bf 117}, 203 (2002).

[33] J. C. Su, Proceedings of the Fifth International Conference on Symmetry
in the Nonlinear Mathematical Physics, p.965, Kyiv, Ukraine (2003).

[34] J. C. Su and J. X. Chen, Phys. Rev. D {\bf 69}, 076002 (2004).

{[35]} C. Becchi, A. Rouet and R. Stora, Phys. Lett. B {\bf 52} (1974) 344;
Commun. Math. Phys. {\bf 42} (1975) 127; I. V. Tyutin, Lebedev Preprint {\bf %
39} (1975).

[36] D. Lurie, Particles and Fields, Interscience Publishers, a divison of
John Viley \& Sons, New York, 1968.

[37] R. J. Furnstahl and B. D. Serot, Phys. Rev. C {\bf 44}, 2141 (1991).

[38] J. C. Ward, Phys. Rev. {\bf 77} (1950) 2931; Y. Takahashi, Nuovo
Cimento {\bf 6}, (1957) 370.

[39] J. C.Taylor, Nucl. Phys. B {\bf 33}, 436 (1971); A. A. Slavnov, Theor.
Math. Phys. {\bf 10}, 99 (1972).

[40] J. C. Su, J. Phys. G: Nucl. Part. Phys. {\bf 27}, 1493 (2001).

[41] L. D. Faddeev, Theor. Math. Phys. {\bf 1}, 1 (1970); L. D. Faddeev and
A. A. Slavnov, Gauge Fields: Introduction to Quantum Theory, The Benjamin
Commings Publishing Company, Inc. (1980).

[42] L. D. Faddeev and V. N. Popov, Phys. Lett. B {\bf 25}, 29 (1967).

[43] G.'t Hooft, Nucl. Phys. B {\bf 61}, 455 (1973).

[44] J. J. Sakurai, Ann. Phys. {\bf 11}, 1 (1960).

[45] J. C. Su, hep-th/9805193, hep-th/980519.

[46] J. C. Su, Commun. Theor. Phys. {\bf 38}, 433 (2002).

[47] J. C. Su, ''Dirac-Schr\"odinger equation for quark-antiquark bound
states and derivation of its interaction kernel'', will soon apprear in J.\
Phys. G. Nucl. Part. Phys.

[48] A. J. Simon, et.el., Phys. Rev. C 48, 662 (1993).

\section{Figure captions}

Fig.(1): The one-loop nucleon self-energy in the $\sigma -\omega $ model.
The solid, wavy line and dashed lines represent the free nucleon, $\omega $
meson and $\sigma $ meson propagators respectively.

Fig. (2): The one-loop vectorial vertices in the $\sigma -\omega $ model.
The lines represent the same as in Fig. (1).

Fig. (3): The one-loop scalar vertices in the $\sigma -\omega $ model. The
lines mark the same as in Fig. (1).

Fig. (4): The diagrams represent the nucleon four-point one-particle
irreducible Green's function. The solid line with a white blob represents
the full nucleon propagator. The wavy line with a white blob denotes the
full $\omega $ meson propagator, The shaded blobs represent the proper
vertices.

Fig. (5): The tree diagrams of nucleon-nucleon scattering. The first two
diagrams represent the interaction generated by the $\omega $ meson
exchange. The remaining two diagrams represent the interaction mediated by
the $\sigma $ meson exchange.

Fig. (6): Some two-nucleon one-loop Feynman diagrams which are chosen to
demonstrate the gauge-independence of the nucleon scattering matrix elements.

Fig. (7): the one-loop diagram of the effective $\omega $ meson self-energy.
The solid line marks the free nucleon propagator and the wavy line denotes
the free $\omega $ meson propagator.

Fig. (8): The effective one-loop vectorial coupling constants ${\alpha
_R^v(\lambda )}$ given by the timelike momentum space subtraction. The solid
line represents the coupling constant given by taking ${\alpha _R^v=0.5}$.
The dashed line denotes the coupling constant given by ${\alpha _R^v=1}$.

Fig. (9): The effective one-loop vectorial coupling constants ${\alpha
_R^v(\lambda )}$ given by the spacelike momentum space subtraction. The
solid line represents the coupling constant given by taking ${\alpha _R^v=0.5%
}$. The dashed line denotes the coupling constant given by ${\alpha _R^v=1}$.

Fig. (10): The effective one-loop $\omega $ meson masses $m_\omega
^R(\lambda )$ given by taking ${\alpha _R^v=1}$. The solid line and the
dashed line represent the effective masses obtained in the spacelike
momentum subtraction and the timelike momentum respectively.

Fig. (11): The effective one-loop nucleon masses $M_R(\lambda )$ obtained in
the timelike momentum subtraction$.$ The solid, dashed and dotted lines
represent the effective masses given by $(\alpha _R^v,\alpha _R^s)=(0.5,0)$, 
$(0,0.5)$ and $(0.5,0.5)$ respectively.

Fig. (12): The effective one-loop nucleon masses $M_R(\lambda )$ obtained in
the spacelike momentum subtraction. The dashed, solid and dotted lines
represent the effective masses given by taking $(\alpha _R^v,\alpha
_R^s)=(0.5,0)$, $(0.5,0.5)$ and $(1,1)$ respectively. The upper figure
describes the real part of the $M_R(\lambda ).$ Another figure shows the
imaginary part of the $M_R(\lambda )$.

Fig. (13): The $\sigma $ meson one-loop self-energy. The solid line
represents the free nucleon propagator and the dashed line denotes the free $%
\sigma $ meson propagator.

Fig. (14): The effective one-loop scalar coupling constants ${\alpha
_R^s(\lambda )}$ obtained in the timelike momentum subtraction. The dashed
and solid lines on the left represents the effective coupling constants
given by $(\alpha _R^v,\alpha _R^s)=(0,1)$ and $(0.5,1)$. The dashed and
solid lines on the right denote the effective coupling constants given by $%
(\alpha _R^v,\alpha _R^s)=(0,0.2)$ and $(0.5,0.2)$.

Fig. (15): The effective one-loop scalar coupling constants ${\alpha
_R^s(\lambda )}$ obtained in the spacelike momentum subtraction. The four
lines represent the effective coupling constants given by taking $(\alpha
_R^v,\alpha _R^s)=(0,1)$, $(0,0.2)$, $(0.5,1)$ and $(0.5,0.2)$ respectively.
The solid and dashed lines denote the real parts and the imaginary parts of
the coupling constants respectively.

Fig. (16): The effective one-loop $\sigma $ meson masses $m_\sigma
^R(\lambda )$ obtained by taking $(\alpha _R^v,\alpha _R^s)=(1,1)$. The
solid line represents the effective mass given in the spacelike momentum
subtraction, The dashed line shows the real part of the $m_\sigma ^R(\lambda
)$ given in the timelike momentum subtraction.

Fig. (17): The two proton elastic differential cross sections given at the
laboratory kinetic energies $T_{lab}=491.9MeV$ and $575.5MeV$. The black
squares show the experimental data. The solid lines represent the
theoretical values calculated by considering the one-loop renormalization
effect. The dashed lines represent the theoretical values without
considering the one-loop renormalization effect.

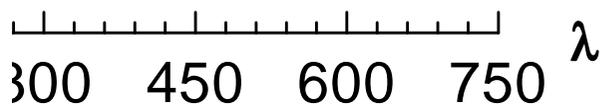
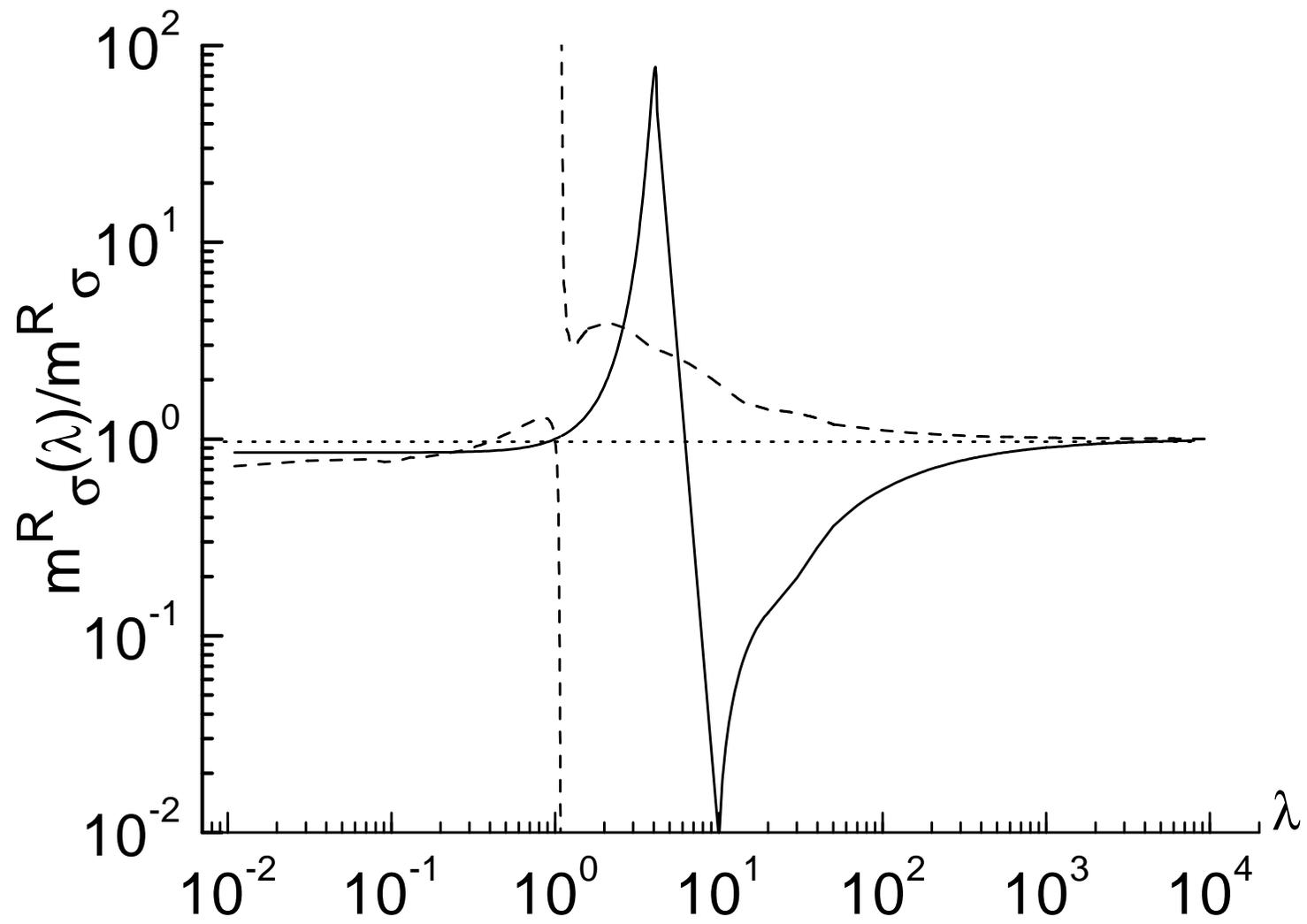
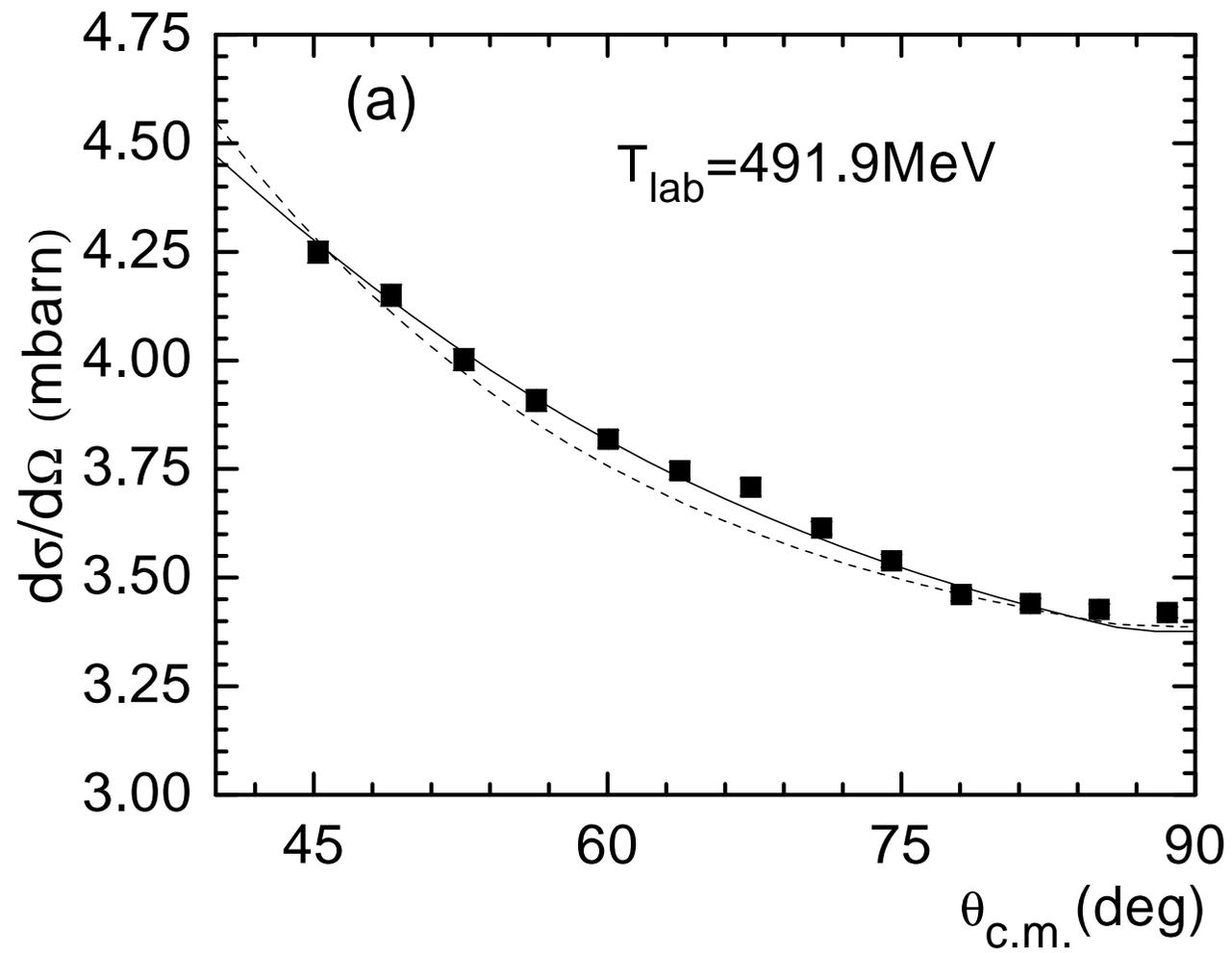
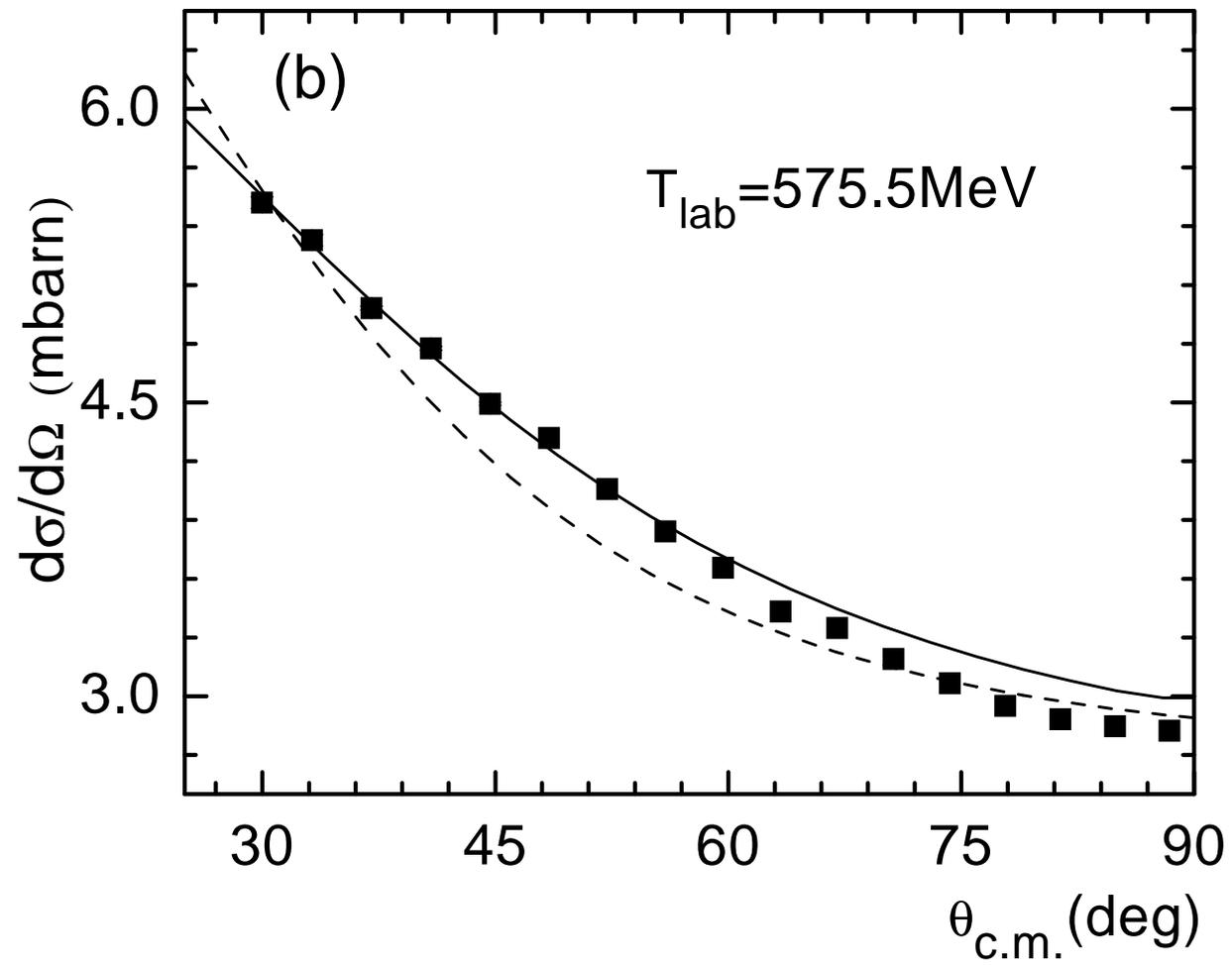
\end{document}